\documentclass[preprint,12pt]{elsarticle}

\makeatletter
\def\ps@pprintTitle{%
    \let\@oddhead\@empty
    \let\@evenhead\@empty
    \def\@oddfoot{\footnotesize
         {} \hfill {April 2024}}%
    \let\@evenfoot\@oddfoot
    }
\makeatother

\usepackage{amssymb}
\usepackage[fleqn]{amsmath}
\usepackage{amsmath,amsthm,bm} 
\usepackage{amsfonts}
\usepackage{mathtools}
\usepackage{fdsymbol}
\usepackage{mathrsfs}
\usepackage{dutchcal} 

\usepackage[colorlinks=true, allcolors=blue]{hyperref}
\usepackage[margin=2.5cm,includefoot]{geometry} 
\usepackage{caption}
\usepackage{subcaption}
\usepackage{booktabs}  
\usepackage{longtable} 
\usepackage{multirow}  
\graphicspath{ {./Figures/} } 

\usepackage{natbib}
\hypersetup{pdfauthor={Zihao Liu, Mohsen Ebrahimzadeh Hassanabadi, Daniel Dias-da-Costa}}

\begin{document}
\begin{frontmatter}



\title{A recursive smoothing method for input and state estimation of vibrating structures}


\author[a]{Zihao Liu}
\author[a]{Mohsen Ebrahimzadeh Hassanabadi}
\author[a]{Daniel Dias-da-Costa\corref{cor1}}
\ead{daniel.diasdacosta@sydney.edu.au}
\cortext[cor1]{Corresponding author}

\address[a]{School of Civil Engineering, The University of Sydney, Sydney, NSW 2006, Australia}

\begin{abstract}
\noindent Recursive Bayesian filters have been widely deployed in structural system identification where output-only filters are of higher practicality. Unfortunately, the estimation obtained by instantaneous system inversion via filters can be compromised by an ill-conditionedness of the system, which is a consequence of the architecture of the sensor network. To significantly reduce the ill-conditioning and increase the robustness to available networks, a new recursive smoothing algorithm is proposed for simultaneous input and state estimation of linear systems. Unlike the existing minimum-variance unbiased (MVU) smoothing methods that are restricted to either systems with no direct feedthrough or systems with a full-rank feedforward matrix, the proposed smoothing algorithm is universally applicable to linear systems with and without direct feedthrough as well as those with a rank-deficient feedforward matrix. The proposed smoothing method does not assume any prior knowledge of the statistical characteristics or evolutionary model pertaining to the input. A different indexing of the discrete-time input leads to a distinct linear algebra from the existing MVU smoothing methods. An eight-storey shear frame and the Taipei 101 tower in Taiwan are used as case studies, and a thorough comparison is established with the Augmented Kalman Filter, MVU filters and MVU smoothing methods. It is shown that the incorporation of singular value truncation for system inversion can result in a noticeable improvement in the estimation. Moreover, across various sensor networks and in the presence of a rank-deficient feedforward matrix, the proposed method could achieve at least $67$\% noise reduction compared to other filters and at least $30$\% improvement compared to other smoothing methods.

\end{abstract}


\end{frontmatter}

\section{Introduction}
\label{section1}
\noindent The Kalman filter~\cite{Kalman1960} is one of the most popular recursive Bayesian algorithms in civil engineering for estimating the state of linear systems and has been utilised for various applications. This includes estimating the state of structures from multiple sampling rate measurements~\cite{Zhu2023} and predicting fatigue in metallic structures by estimating the power spectral density of stresses at unmeasured locations~\cite{Papadimitriou2011}. One significant limitation of the Kalman filter is the requirement of some prior knowledge of the input to the system, such as the dynamic load, to be taken as known or assumed as a zero-mean white noise. To address this, \citet{LourensReyndersEATAL2012} proposed the Augmented Kalman Filter (AKF), which concatenates the unknown input and the state vector to jointly estimate both the state and input. The AKF was implemented to localise collision-introduced impact forces in truss structures~\cite{Saleem2019}, and identify hysteresis~\cite{Wang2021}. The requirement of measuring displacement and velocity could be relaxed by the Dual Kalman Filter developed by \citet{EftekharAzam2015} to estimate the state and input of linear systems with acceleration-only observations. Several nonlinear extensions were proposed for parameter identification and damage detection that expand beyond structural systems with time-varying dynamic characteristics. For example, the Extended Kalman Filter was deployed to detect crack propagation in boom structures~\cite{Cazzulani2013}, and identify modal properties of structures with tuned mass dampers~\cite{Roffel2014}. Furthermore, the input augmentation in the Extended Kalman Filter enables the combined state, input, and parameter estimation~\cite{Naets2015, EbrahimzadehHassanabadi2020}. In addition to the Extended Kalman Filter, another nonlinear approach -- the Unscented Kalman Filter -- was utilised for highly nonlinear problems including the identification of the Bouc–Wen hysteresis model~\cite{Wu2007}, and vibration control of duffing oscillator and quarter-car model~\cite{Dertimanis2021}. In recent years, there has been a sustained growth in the application of Kalman-type recursive Bayesian methods in civil and structural engineering, combined with other techniques, such as computer vision~\cite{Cha2017, Ma2022} and artificial neural networks~\cite{Lin2022, Zhou2022}.

Although Kalman-type filters are frequently used for estimating the state and input of a structural system, they suffer from a significant limitation: input estimation requires prior statistical characteristics of the input which is often unknown, or an assumption of a fictitious input model. Consequently, Kalman-type filters need a refined tuning of the modelling error for input estimation, which can lead to a complex process. This limitation for output-only estimations motivated the development of Minimum-Variance Unbiased Filters (MVUFs) that do not rely on statistical information and fictitious models of unknown input. \citet{Gillijns2007a} developed an MVUF for simultaneous state and input estimation based on the pioneering method proposed by \citet{Kitanidis1987}. This filter will be referred to as MVUF-NDF in the remainder of this paper. Unlike the AKF, which concatenates the unknown input and the state vectors, the MVUF-NDF uses the least squares method to estimate the input from the innovation. The MVUF-NDF is limited to linear systems with no direct feedthrough, meaning that no acceleration is measured in the system. This limitation can be solved with a separate filter, referred to as MVUF-DF, proposed specifically for systems with direct feedthrough~\cite{Gillijns2007b} and requiring a full-rank feedforward matrix. This means that the MVUF-DF is only applicable to systems in which the number of acceleration measurements is not less than the number of unknown inputs. \citet{LourensPapadimitriouETAL2012} proposed an enhancement to avoid the numerical issue caused by using a Reduced-Order Model (ROM) with a large number of acceleration sensors. This was achieved by truncating the singular value decomposition. The MVUF-DF was further extended to nonlinear systems by \citet{Wan2018} to enable parameter identification. However, the MVUFs are constrained by the feedforward matrix; the MVUF-DF can only observe systems with a full-rank feedforward matrix, while the MVUF-NDF is exclusively applicable to systems without direct feedthrough. Therefore, there is a lack of MVUF methods that are universally applicable to linear systems with and without direct feedthrough as well as those with a rank-deficient feedforward matrix.~\citet{EbrahimzadehHassanabadi2023} proposed a Universal Filter (UF) to cope with this restriction by relaxing the restrictive rank condition requirements for the feedforward matrix. 

The instantaneous state and input estimation methods can only use the data observed at a single timestep. As a result, the system inversion may become ill-conditioned, which can adversely impact the estimation quality. The ill-conditionedness can stem from several causes, including a limited number of sensors, displacement-only and strain-only observations, and non-collocated sensor networks. To remedy this issue of instantaneous estimation methods, smoothing algorithms extend the observation window. By doing so, smoothing algorithms can benefit from extra sensory information ahead for estimating the state and input, therefore leading to an improved quality of the estimation. \citet{Maes2018} proposed a smoothing algorithm for systems with direct feedthrough, termed as MVUS-DF herein, which is a generalisation of the MVUF-DF. Extending the MVUF-NDF, \citet{EbrahimzadehHassanabadi2022} developed a smoothing algorithm for systems with no acceleration measurements, referred to as MVUS-NDF in this paper; a recursive method was proposed in this work to efficiently calculate cross-covariances. However, in alignment with the aforementioned MVUFs~\cite{Gillijns2007a, Gillijns2007b}, their smoothing methods~\cite{Maes2018, EbrahimzadehHassanabadi2022} are also restricted by the rank condition of the feedforward matrix of the observed system.

\begin{figure}[ht]
    \centering
    \includegraphics[trim={1.5cm 6.5cm 1.5cm 6.5cm}, clip, width=\textwidth]{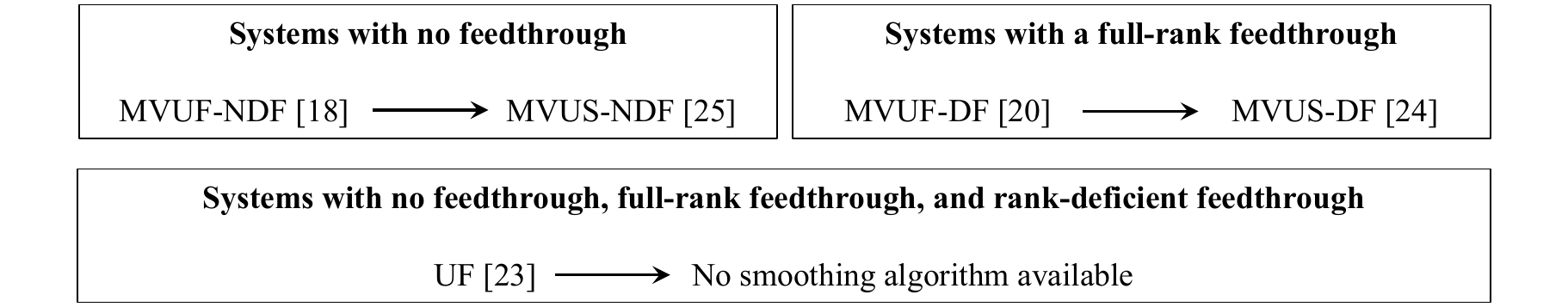}
    \caption{Current MVUFs and their smoothing counterparts.}
    \label{Intro_visual}
\end{figure}

Fig.~\ref{Intro_visual} presents the aforementioned MVUFs and their smoothing counterparts. Considering the limitations of the current MVUS methods, and the need for a universal smoothing method, this paper proposes a novel recursive smoothing algorithm to simultaneously estimate the input and state of a linear structural system. The developed algorithm does not rely on any presumptions regarding prior information on input statistics or its evolution model. Unlike the existing MVUS algorithms that are either limited to the systems with no direct feedthrough or systems with direct feedthrough, the proposed smoothing algorithm in this paper is universally applicable to linear systems with and without direct feedthrough and ones with a rank-deficient feedforward matrix. Furthermore, the proposed smoothing algorithm uses weighted least squares for input estimation. This provides an advantage over Kalman-type methods that require statistical knowledge or assumption on the input model. On the contrary, the developed method does not require any evolution model for input and the input statistics are taken as unknown. It is noteworthy that despite being inspired by the UF, the derivation of error covariance matrices in the smoothing process becomes complex due to the extended observation window. For ease of reference, the proposed smoothing algorithm is named Universal Smoothing (US) hereafter.

In the following, the US is presented in detail. The mathematical derivation of the smoothing and the system equations for linear structural systems are presented in Section~\ref{section2}. Section~\ref{section3} includes numerical validations of the smoothing using a shear frame and a high-rise building. Finally, the conclusions are drawn in Section~\ref{section4}.

\section{Derivation of the Universal Smoothing}
\label{section2}
\noindent In this section, derivation of the US is presented. The derivation begins with extending the observation equation of the state-space model of a linear system for the purpose of smoothing, which is followed by an overview of the smoothing steps. The input gain matrix that minimises the variance of the input estimation is obtained using the weighted least squares method, whereas the state gain is obtained by minimising the trace of the error covariance of the state estimation. Precise solutions to the propagations of estimation errors and corresponding covariances are also obtained. Finally, a discrete-time recursive solution of a Multi-Degree-of-Freedom (MDOF) structural system is derived using the matrix exponential method.

\subsection{Equations of the system}
\label{section2.1}

\noindent The discrete-time process equation of a linear structural system subjected to dynamic loads can be defined by: 
\begin{equation} \label{eq1} 
    \mathbf{x}_k=\mathbf{A}_{k-1}\mathbf{x}_{k-1}+\mathbf{G}_{k-1}\mathbf{p}_k+\mathbf{w}_{k-1},
\end{equation}
\noindent where $\mathbf{x}_k\in\mathbb{R}^n$ is the state vector, $\mathbf{p}_k\in\mathbb{R}^m$ is the dynamic input to the system, $\mathbf{A}_k\in\mathbb{R}^{n\times n}$ and $\mathbf{G}_k\in\mathbb{R}^{n\times m}$ are state and input matrices, respectively. The modelling error is expressed by a zero-mean white noise $\mathbf{w}_k\in\mathbb{R}^n$. 
A zero-order-hold assumption, which is the same as the one used in the UF, has been made for Eq.~\eqref{eq1} such that the state vector $\mathbf{x}_k$ and the input vector $\mathbf{p}_k $ are at the same timestep $k$. It is crucial to note that the zero-order-hold assumption taken herein is different from the one for the MVUF-based methods \cite{Gillijns2007a, Gillijns2007b, Maes2018, EbrahimzadehHassanabadi2022} which has one timestep lag in the subscript of the input in the process equation, i.e., $\mathbf{p}_{k-1}$; an important consequence of shifting the input subscript in the process equation, is that the derivations in \cite{Gillijns2007a, Gillijns2007b, Maes2018, EbrahimzadehHassanabadi2022} become inconsistent with the system defined in this paper, therefore, their subsequent system inversion algorithms are not applicable. The MVU input-state estimation of such a system focused here requires a different linear algebra for the MVU joint input-state estimation, which was addressed in \cite{EbrahimzadehHassanabadi2023} for a new filter with a distinct formulation.

The observation equation of the structural system reads as:
\begin{equation} \label{eq2}\mathbf{y}_k=\mathbf{C}_k\mathbf{x}_k+\mathbf{H}_k\mathbf{p}_k+\mathbf{v}_k,
\end{equation}
\noindent in which $\mathbf{y}_k\in\mathbb{R}^d$ contains the observed output from the system. The output matrix $\mathbf{C}_k\in\mathbb{R}^{d\times n}$ and feedforward matrix $\mathbf{H}_k\in\mathbb{R}^{d\times m}$ relate the state vector $\mathbf{x}_k$ and input vector $\mathbf{p}_k$ to the observation $\mathbf{y}_k$ at the same timestep. The zero-mean white noise $\mathbf{v}_k\in\mathbb{R}^d$ represents the observation noise. It should be noted that when $\mathbf{H}_k={\mathbf{0}}$, Eq.~\eqref{eq2} describes a system without direct feedthrough, while $\mathbf{H}_k\neq{\mathbf{0}}$ leads to a system with direct feedthrough. 

By introducing an observation window $N$ ($N=1,\ 2,\ \ldots$) in the system described by Eqs.~\eqref{eq1} and \eqref{eq2}, an extended observation equation is formulated below:
\begin{equation} \label{eq3}\mathbcal{y}_k=\mathbcal{C}_k\mathbf{x}_k+\mathbcal{H}_k\mathbcal{p}_k+\mathbcal{D}_k\mathbcal{w}_k+\mathbcal{v}_k,
\end{equation}
\noindent where the extended observation $\mathbcal{y}_k\in\mathbb{R}^{(N+1)d}$, the extended input vector $\mathbcal{p}_k\in\mathbb{R}^{(N+1)m}$, the extended modelling error $\mathbcal{w}_k\in\mathbb{R}^{(N+1)n}$ and the extended observation noise $\mathbcal{v}_k\in\mathbb{R}^{(N+1)d}$ are respectively given by: 
\begin{equation} \label{eq4}
    \mathbcal{y}_k\triangleq\begin{bmatrix}\mathbf{y}_k^T & \cdots & \mathbf{y}_{k+N}^T \\\end{bmatrix}^T,
\end{equation}
\vspace{-0.5cm}
\begin{equation} \label{eq5}
    \mathbcal{p}_k\triangleq\begin{bmatrix}\mathbf{p}_k^T & \cdots & \mathbf{p}_{k+N}^T \\\end{bmatrix}^T,
\end{equation}
\vspace{-0.5cm}
\begin{equation} \label{eq6}
    \mathbcal{w}_k\triangleq\begin{bmatrix}\mathbf{w}_{k-1}^T & \cdots & \mathbf{w}_{k+N-1}^T\\\end{bmatrix}^T,
\end{equation}
\vspace{-0.5cm}
\begin{equation} \label{eq7}
    \mathbcal{v}_k\triangleq\begin{bmatrix}\mathbf{v}_k^T & \cdots & \mathbf{v}_{k+N}^T\\\end{bmatrix}^T,
\end{equation}
\noindent furthermore, the extended output matrix $\mathbcal{C}_k$, extended feedforward matrix $\mathbcal{H}_k$ and extended modelling error  matrix $\mathbcal{D}_k$ can be obtained as:
\begin{equation} \label{eq8}
    \mathbcal{C}_k=
    \begin{bmatrix}\mathbf{C}_k\\
    \mathbf{C}_{k+1}\mathbf{A}_k\\
    \mathbf{C}_{k+2}\mathbf{A}_{k+1}\mathbf{A}_k\\
    \vdots\\
    \mathbf{C}_{k+N}\prod_{i=0}^{N-1}\mathbf{A}_{k+i}\\\end{bmatrix},
\end{equation}
\vspace{-0.5cm}
 \small{
\begin{equation} \label{eq9}
    \mathbcal{H}_k=
    \begin{bmatrix}\mathbf{H}_k&\mathbf{0}&\mathbf{0}&\cdots&\mathbf{0}\\
    \mathbf{0}&\mathbf{C}_{k+1}\mathbf{G}_k+\mathbf{H}_{k+1}&\mathbf{0}&\cdots&\mathbf{0}\\
    \mathbf{0}&\mathbf{C}_{k+2}\mathbf{A}_{k+1}\mathbf{G}_k&\mathbf{C}_{k+2}\mathbf{G}_{k+1}+\mathbf{H}_{k+2}&\cdots&\mathbf{0}\\
    \mathbf{0}&\mathbf{C}_{k+3}\mathbf{A}_{k+2}\mathbf{A}_{k+1}\mathbf{G}_k&\mathbf{C}_{k+3}\mathbf{A}_{k+2}\mathbf{G}_{k+1}&\cdots&\mathbf{0}\\
    \vdots&\vdots&\vdots&\ddots&\vdots\\
    \mathbf{0}&\mathbf{C}_{k+N}\left(\prod_{i=1}^{N-1}\mathbf{A}_{k+i}\right)\mathbf{G}_k&\mathbf{C}_{k+N}\left(\prod_{i=2}^{N-1}\mathbf{A}_{k+i}\right)\mathbf{G}_{k+1}&\cdots&\mathbf{C}_{k+N}\mathbf{G}_{k+N-1}+\mathbf{H}_{k+N}\\\end{bmatrix},
\end{equation}
}
\vspace{-0.5cm}
\normalsize
\begin{equation} \label{eq10}
    \mathbcal{D}_k=
    \begin{bmatrix}\mathbf{0}&\mathbf{0}&\mathbf{0}&\cdots&\mathbf{0}\\
    \mathbf{0}&\mathbf{C}_{k+1}&\mathbf{0}&\cdots&\mathbf{0}\\    \mathbf{0}&\mathbf{C}_{k+2}\mathbf{A}_{k+1}&\mathbf{C}_{k+2}&\cdots&\mathbf{0}\\ \mathbf{0}&\mathbf{C}_{k+3}\mathbf{A}_{k+2}\mathbf{A}_{k+1}&\mathbf{C}_{k+3}\mathbf{A}_{k+2}&\cdots&\mathbf{0}\\
    \vdots&\vdots&\vdots&\ddots&\vdots\\
    \mathbf{0}&\mathbf{C}_{k+N}\prod_{i=1}^{N-1}\mathbf{A}_{k+i}&\mathbf{C}_{k+N}\prod_{i=2}^{N-1}\mathbf{A}_{k+i}&\cdots&\mathbf{C}_{k+N}\\\end{bmatrix}.
\end{equation}

The covariance matrices of modelling error and observation noise are given by:
\begin{equation} \label{eq11}\mathbf{Q}_k\triangleq\mathbb{E}\left[\mathbf{w}_k\mathbf{w}_k^T\right],
\end{equation}
\vspace{-0.5cm}
\begin{equation} \label{eq12}\mathbf{R}_k\triangleq\mathbb{E}\left[\mathbf{v}_k\mathbf{v}_k^T\right].
\end{equation}

\subsection{Smoothing steps}
\label{section2.2}
\noindent As the input is unknown, a biased estimate of the state vector without the contribution of the input can be obtained based on the information from the previous timestep, i.e.: 
\begin{equation} \label{eq13}
    \hat{\bm{\mathbf{\chi}}}_k=\mathbf{A}_{k-1}\hat{\mathbf{x}}_{k-1},
\end{equation}
\noindent and the innovation term $\tilde{\mathbcal{y}}_k$ related to the input estimation can be defined as:
\begin{equation} \label{eq14}
    \tilde{\mathbcal{y}}_k\triangleq\mathbcal{y}_k-\mathbcal{C}_k\hat{\bm{\mathbf{\chi}}}_k.
\end{equation}

Note that the biasedness of the Eq.~\eqref{eq13} is caused by the missing contribution of the input, thus, the innovation term implicitly defines the input estimate. The input gain $\mathbf{M}_k$ minimising the variance of input estimation can be obtained by using weighted least squares, in which case the extended input vector can be estimated by:
\begin{equation} \label{eq15}
\hat{\mathbcal{p}}_k=\mathbf{M}_k\left(\mathbcal{y}_k-\mathbcal{C}_k\hat{\bm{\mathbf{\chi}}}_k\right).
\end{equation}

The unbiased minimum variance input estimate can be extracted from the extended input vector using:
\begin{equation} \label{eq16}
\hat{\mathbf{p}}_k=\bm{\mathbf{\varepsilon}}_m\hat{\mathbcal{p}}_k,
\end{equation}
\noindent where
\begin{equation} \label{eq17}   \bm{\mathbf{\varepsilon}}_m=\begin{bmatrix}\mathbf{I}_m & \bm{\mathbf{0}}_{m\times Nm}\\ \end{bmatrix}.
\end{equation}

Note that the unbiasedness of the input estimation will be proven in Section~\ref{section2.3}. Next, an a priori estimate of the state vector can be obtained by including the contribution of the estimated input $\hat{\mathbf{p}}_k$:
\begin{equation} \label{eq18}
\hat{\mathbcal{x}}_k=\hat{\bm{\mathbf{\chi}}}_k+\mathbf{G}_{k-1}\hat{\mathbf{p}}_k.
\end{equation}

 Finally, an a posteriori estimate of the state vector can be obtained by adding the contribution of the extended observation with an optimal gain matrix $\mathbf{K}_k$ which minimises the error variance of state estimation, i.e.:
\begin{equation} \label{eq19} \hat{\mathbf{x}}_k=\hat{\mathbcal{x}}_k+\mathbf{K}_k\left(\mathbcal{y}_k-\mathbcal{C}_k\hat{\mathbcal{x}}_k-\mathbcal{H}_k\hat{\mathbcal{p}}_k\right).
\end{equation}

\subsection{Input estimation}
\label{section2.3}

\noindent The requirement of unbiased estimation of input and the derivation of the input gain $\mathbf{M}_k$ are presented herein. Firstly, the error of the state estimation can be defined by:
\begin{equation} \label{eq20}
    \tilde{\mathbf{x}}_k\triangleq\mathbf{x}_k-\hat{\mathbf{x}}_k,
\end{equation}
\noindent and introducing $\hat{\mathbf{x}}_{k-1}=\mathbf{x}_{k-1}-\tilde{\mathbf{x}}_{k-1}$ into Eq.~\eqref{eq13} gives:
\begin{equation} \label{eq21}
    \hat{\bm{\mathbf{\chi}}}_k=\mathbf{A}_{k-1}\mathbf{x}_{k-1}-\mathbf{A}_{k-1}\tilde{\mathbf{x}}_{k-1}.
\end{equation}

Together with Eqs.~\eqref{eq14} and \eqref{eq21}, the innovation can be written as:
\begin{equation} \label{eq22}
    \tilde{\mathbcal{y}}_k=\mathbcal{y}_k-\bm{\mathbf{\Gamma}}_k\mathbf{x}_{k-1}+\bm{\mathbf{\Gamma}}_k\tilde{\mathbf{x}}_{k-1},
\end{equation}
\noindent where
\begin{equation} \label{eq23}
    \bm{\mathbf{\Gamma}}_k=\mathbcal{C}_k\mathbf{A}_{k-1}.
\end{equation}

Eq.~\eqref{eq1} can be rearranged to $\mathbf{A}_{k-1}\mathbf{x}_{k-1}=\mathbf{x}_k-\mathbf{G}_{k-1}\mathbf{p}_k-\mathbf{w}_{k-1}$, and substituting it into Eq.~\eqref{eq22} reads:
\begin{equation} \label{eq24}
    \tilde{\mathbcal{y}}_k=\mathbcal{y}_k-\mathbcal{C}_k\left(\mathbf{x}_k-\mathbf{G}_{k-1}\mathbf{p}_k-\mathbf{w}_{k-1}\right)+\bm{\mathbf{\Gamma}}_k\tilde{\mathbf{x}}_{k-1}.
\end{equation}

The extended observation equation can be replaced into Eq.~\eqref{eq24} giving rise to:
\begin{equation} \label{eq25}    \tilde{\mathbcal{y}}_k=\mathbcal{H}_k\mathbcal{p}_k+\mathbf{C}_k\mathbf{G}_{k-1}\mathbf{p}_k+\mathbf{\Gamma}_k\tilde{\mathbf{x}}_{k-1}+\mathbf{D}_k\mathbcal{w}_k+\mathbf{C}_k\mathbf{w}_{k-1}+\mathbcal{v}_k.
\end{equation}

Similarly to Eq.~\eqref{eq16}, $\mathbf{w}_{k-1}$ can be extracted from $\mathbcal{w}_k$ using:
\begin{equation} \label{eq26}
    \mathbf{w}_{k-1}=\bm{\mathbf{\varepsilon}}_n\mathbcal{w}_k,
\end{equation}
\noindent where
\begin{equation} \label{eq27}
    \bm{\mathbf{\varepsilon}}_n = \begin{bmatrix}\mathbf{I}_n & \bm{\mathbf{0}}_{n\times Nn}\\ \end{bmatrix}.
\end{equation}

By applying Eqs.~\eqref{eq16} and \eqref{eq26}, Eq.~\eqref{eq25} can be expressed by:
\begin{equation} \label{eq28}\tilde{\mathbcal{y}}_k=\breve{\mathbf{H}}_k\mathbcal{p}_k+\bm{\mathbf{\Gamma}}_k\tilde{\mathbf{x}}_{k-1}+\breve{\mathbf{D}}_k\mathbcal{w}_k+\mathbcal{v}_k,
\end{equation}
\noindent where 
\begin{equation} \label{eq29} \breve{\mathbf{H}}_k\triangleq\mathbcal{H}_k+\mathbcal{C}_k\mathbf{G}_{k-1}\begin{bmatrix}\mathbf{I}_m & \bm{\mathbf{0}}_{m\times N m}\\ \end{bmatrix},
\end{equation}
\vspace{-0.5cm}
\begin{equation} \label{eq30} \breve{\mathbf{D}}_k\triangleq\mathbcal{D}_k+\begin{bmatrix}\mathbcal{C}_k & \bm{\mathbf{0}}_{\left(N+1\right)d\times N n}\\ \end{bmatrix}.
\end{equation}

An error term denoted by $\mathbf{e}_k$ can be extracted from Eq.~\eqref{eq28}:
\begin{equation} \label{eq31}
    \mathbf{e}_k\triangleq\bm{\mathbf{\Gamma}}_k\tilde{\mathbf{x}}_{k-1}+\breve{\mathbf{D}}_k\mathbcal{w}_k+\mathbcal{v}_k.
\end{equation}

As a result of Eqs.~\eqref{eq28} to~\eqref{eq31}, the innovation $\tilde{\mathbcal{y}}_k$ can be finally simplified as follows:
\begin{equation} \label{eq32}
\tilde{\mathbcal{y}}_k=\breve{\mathbf{H}}_k\mathbcal{p}_k+\mathbf{e}_k.
\end{equation}

Note that $\mathbb{E}\left[\tilde{\mathbf{x}}_{k-1}\right]=\bm{\mathbf{0}}$ if $\hat{\mathbf{x}}_{k-1}$ is unbiased. Since $\mathbf{w}_k$ and $\mathbf{v}_k$ are zero-mean white noise, $\mathbb{E}\left[\mathbf{w}_k\right]=\bm{\mathbf{0}}$ and $\mathbb{E}\left[\mathbf{v}_k\right]=\bm{\mathbf{0}}$, therefore, $\mathbb{E}\left[\mathbf{e}_k\right]=\bm{\mathbf{0}}$. This leads to $\mathbb{E}\left[\tilde{\mathbcal{y}}_k\right]=\breve{\mathbf{H}}_k\mathbcal{p}_k$. Given Eqs.~\eqref{eq14}, \eqref{eq15} and \eqref{eq32}, $\mathbb{E}\left[\hat{\mathbcal{p}}_k\right]=\mathbb{E}\left[\mathbf{M}_k\tilde{\mathbcal{y}}_k\right]=\mathbf{M}_k\breve{\mathbf{H}}_k\mathbcal{p}_k$. Therefore, to ensure that $\hat{\mathbcal{p}}_k$ is an unbiased estimate, i.e., $\mathbb{E}\left[\hat{\mathbcal{p}}_k\right]=\mathbcal{p}_k$, the following requirement must be satisfied: 
\begin{equation} \label{eq33}
    \mathbf{M}_k\breve{\mathbf{H}}_k=\mathbf{I}_{(N+1)m}.
\end{equation}

The weighted least square is herein adopted to achieve a minimum variance estimation, with the weighting being the covariance of the error term $\mathbf{e}_k$, i.e.: 
\begin{equation} \label{eq34}\tilde{\mathbf{R}}_k\triangleq\mathbb{E}\left[\mathbf{e}_k\mathbf{e}_k^T\right]=\bm{\mathbf{\Sigma}}_k\bm{\mathbf{\Lambda}}_k\bm{\mathbf{\Sigma}}_k^T,
\end{equation}
\noindent where
\begin{equation} \label{eq35}
    \mathbf{\Lambda}_k\triangleq
    \begin{bmatrix}\mathbf{P}_{k-1} & \mathbf{P}_{k-1}^{\mathbf{xw}} & \mathbf{P}_{k-1}^{\mathbf{xv}}\\
    \mathbf{P}_{k-1}^{\mathbf{wx}} & \mathbcal{Q}_{k,k} & \mathbf{P}_k^{\mathbf{wv}}\\
    \mathbf{P}_{k-1}^{\mathbf{vx}} & \mathbf{P}_k^{\mathbf{vw}} & \mathbcal{R}_{k,k}\\ \end{bmatrix},
\end{equation}
\vspace{-0.5cm}
\begin{equation} \label{eq36}
    \bm{\mathbf{\Sigma}}_k\triangleq\begin{bmatrix}\bm{\mathbf{\Gamma}}_k & \breve{\mathbf{D}}_k & \mathbf{I}_{(N+1)d}\\ \end{bmatrix}.
\end{equation}

In the above equations, matrix $\mathbf{\Lambda}_k$ contains all error covariance matrices involved in the estimation process. The error covariance $\mathbf{P}_k\triangleq\mathbb{E}\left[\tilde{\mathbf{x}}_k\tilde{\mathbf{x}}_k^T\right]$ is linked to the state estimation, for which the derivation will be presented in Section~\ref{section2.4}. $\mathbf{P}_k^{\mathbf{xw}}\triangleq\mathbb{E}\left[\tilde{\mathbf{x}}_k\mathbcal{w}_{k+1}^T\right]$ and $\mathbf{P}_k^{\mathbf{xv}}\triangleq\mathbb{E}\left[\tilde{\mathbf{x}}_k\mathbcal{v}_{k+1}^T\right]$ are zero in the MVUFs as the state estimation error is uncorrelated with both the modelling error and observation noise, which considerably reduces the complexity of the filters. However, for smoothing methods, $\mathbf{P}_k^{\mathbf{xw}}$ and $\mathbf{P}_k^{\mathbf{xv}}$ are nonzero as the modelling error and the observation noise are correlated to the state estimation due to the extended observation equation, i.e., Eq.~\eqref{eq3}. In \cite{Maes2018}, $\mathbf{P}_k^{\mathbf{xw}}$ and $\mathbf{P}_k^{\mathbf{xv}}$ were derived in conventional batch forms. Although a straightforward derivation, a high computational cost may be required for the processing of the batch forms. Therefore, the recursive approach proposed in \cite{EbrahimzadehHassanabadi2022} is adopted instead to improve computational efficiency. The derivation of $\mathbf{P}_k^{\mathbf{xw}}$ and $\mathbf{P}_k^{\mathbf{xv}}$ will be presented in Section~\ref{section2.5}. It should be noted that $\mathbf{P}_k^{\mathbf{xw}}={\mathbf{P}_k^{\mathbf{wx}}}^T$ and $\mathbf{P}_k^{\mathbf{xv}}={\mathbf{P}_k^{\mathbf{vx}}}^T$. In addition, $\mathbf{P}_k^{\mathbf{wv}}\triangleq\mathbb{E}\left[\mathbcal{w}_k\mathbcal{v}_k^T\right]={\mathbf{P}_k^{\mathbf{vw}}}^T$ is the cross-covariance of the extended modelling error and the extended observation noise. It is a common assumption that the modelling error and the observation noise are taken as uncorrelated~\cite{LourensReyndersEATAL2012, Gillijns2007a, Gillijns2007b, EbrahimzadehHassanabadi2022}. In this paper, however and for the sake of generality, $\mathbf{P}_k^{\mathbf{wv}}$ and $\mathbf{P}_k^{\mathbf{vw}}$ are treated as nonzero in the derivation of the new smoothing method for potential broader applications. Furthermore, the covariances of the extended observation noise $\mathbcal{R}_{l,k}$ and the extended modelling error $\mathbcal{Q}_{l,k}$ are: 
\begin{equation} \label{eq37}
\mathbcal{R}_{l,k}\triangleq\mathbb{E}\left[\mathbcal{v}_l\mathbcal{v}_k^T\right],
\end{equation}
\vspace{-0.5cm}
\begin{equation} \label{eq38}    \mathbcal{Q}_{l,k}\triangleq\mathbb{E}\left[\mathbcal{w}_l\mathbcal{w}_k^T\right].
\end{equation}

Considering Eq.~\eqref{eq32}, applying weighted least square with the weight matrix $\tilde{\mathbf{R}}_k$ yields:
\begin{equation} \label{eq39}
\hat{\mathbcal{p}}_k=\arg\min_{\mathbcal{p}} \left(\tilde{\mathbcal{y}}_k-\breve{\mathbf{H}}_k\mathbcal{p}\right)^T
    \tilde{\mathbf{R}}_k^{-1}
    \left(\tilde{\mathbcal{y}}_k-\breve{\mathbf{H}}_k\mathbcal{p}\right)=\left(\breve{\mathbf{H}}_k^T\tilde{\mathbf{R}}_k^{-1}\breve{\mathbf{H}}_k\right)^{-1}\breve{\mathbf{H}}_k^T\tilde{\mathbf{R}}_k^{-1}\tilde{\mathbcal{y}}_k.
\end{equation}

The input estimation error $\tilde{\mathbf{p}}_k$ and the extended input estimation error $\tilde{\mathbcal{p}}_k$ are defined as:
\vspace{-0.5cm}
\begin{equation} \label{eq40}
    \tilde{\mathbf{p}}_k\triangleq\mathbf{p}_k-\hat{\mathbf{p}}_k,
\end{equation}
\vspace{-0.5cm}
\begin{equation} \label{eq41}
    \tilde{\mathbcal{p}}_k\triangleq\mathbcal{p}_k-\hat{\mathbcal{p}}_k.
\end{equation}

By applying the principle of best linear unbiased estimation~\cite{Verhagen2017}, the error covariance matrix of $\tilde{\mathbcal{p}}_k$ can be obtained as:
\begin{equation} \label{eq42}
\mathbf{P}_k^\mathbcal{p}\triangleq\mathbb{E}\left[\tilde{\mathbcal{p}}_k\tilde{\mathbcal{p}}_k^T\right]=\left(\breve{\mathbf{H}}_k^T\tilde{\mathbf{R}}_k^{-1}\breve{\mathbf{H}}_k\right)^{-1},
\end{equation}

Next, given Eq.~\ref{eq42}, since $\mathbf{p}_k$ can be extracted from $\mathbcal{p}_k$ using Eq.~\eqref{eq17}, the error covariance matrix of $\tilde{\mathbf{p}}_k$ can be obtained:
\begin{equation} \label{eq43}
    \mathbf{P}_k^\mathbf{p} \triangleq \mathbb{E} \left[\tilde{\mathbf{p}}_k \tilde{\mathbf{p}}_k^T\right]
    = {\mathbb{E}} \left[ \mathbf{\bm{\varepsilon}}_m \tilde{\mathbcal{p}}_k \left(\mathbf{\bm{\varepsilon}}_m  \tilde{\mathbcal{p}}_k\right)^T \right]
    =\mathbf{\bm{\varepsilon}}_m\mathbf{P}_k^\mathbcal{p}\mathbf{\bm{\varepsilon}}_m^T.
\end{equation}

With Eqs.~\eqref{eq14},~\eqref{eq15},~\eqref{eq39} and~\eqref{eq42}, the gain matrix can be obtained as follows:
\begin{equation} \label{eq44}
\mathbf{M}_k=\mathbf{P}_k^\mathbcal{p}\breve{\mathbf{H}}_k^T\tilde{\mathbf{R}}_k^{-1}.
\end{equation}

\subsection{State estimation} 
\label{section2.4}
\noindent The state estimation and the state gain $\mathbf{K}_k$ are derived herein. The error of a priori state estimation $\hat{\mathbcal{x}}_k$ can be defined and written in an expression below by substituting Eqs.~\eqref{eq1}, \eqref{eq13} and~\eqref{eq18}:
\begin{equation} \label{eq45}
    \tilde{\mathbcal{x}}_k \triangleq \mathbf{x}_k-\hat{\mathbcal{x}}_k = \mathbf{A}_{k-1}\tilde{\mathbf{x}}_{k-1} + \mathbf{G}_{k-1}\tilde{\mathbf{p}}_k+\mathbf{w}_{k-1}.
\end{equation}

As $\hat{\mathbf{x}}_{k-1}$ and $\hat{\mathbf{p}}_k$ have been proven to be unbiased estimates, i.e., $\mathbb{E}\left[\tilde{\mathbf{x}}_{k-1}\right]=\mathbf{\bm{0}}$ and $\mathbb{E}\left[\tilde{\mathbf{p}}_k\right]=\mathbf{\bm{0}}$, and the modelling error $\mathbf{w}_{k-1}$ is a zero-mean white noise, it is straightforward to prove that $\mathbb{E}\left[\tilde{\mathbcal{x}}_k\right]=\mathbf{\bm{0}}$. This means that the a priori state estimate $\hat{\mathbcal{x}}_k$ is an unbiased estimate as well. For the extended input estimation error $\tilde{\mathbcal{p}}_k$ defined in Eq.~\eqref{eq41}, and using Eqs.~\eqref{eq14},~\eqref{eq15},~\eqref{eq32} and~\eqref{eq33}, one can write: 
\begin{equation} \label{eq46}
    \tilde{\mathbcal{p}}_k=-\mathbf{M}_k\mathbf{e}_k.
\end{equation}

Similarly to Eq.~\eqref{eq16}, extracting $\tilde{\mathbf{p}}_k$ from $\tilde{\mathbcal{p}}_k$ using $\mathbf{\bm{\varepsilon}}_m$ gives:
\begin{equation} \label{eq47}
    \tilde{\mathbf{p}}_k=-\mathbf{\bm{\varepsilon}}_m\mathbf{M}_k\mathbf{e}_k.
\end{equation}

Substituting Eq.~\eqref{eq3} into Eq.~\eqref{eq19} while recalling the definition of $\tilde{\mathbcal{x}}_k$ and Eq.~\eqref{eq46} yields:
\begin{equation} \label{eq48}
    \hat{\mathbf{x}}_k=\hat{\mathbcal{x}}_k+\mathbf{K}_k\left( \mathbcal{C}_k\tilde{\mathbcal{x}}_k-\mathbf{\bm{\Theta}}_k\mathbf{e}_k+\mathbcal{D}_k\mathbcal{w}_k+\mathbcal{v}_k\right),
\end{equation}

\noindent where
\begin{equation} \label{eq49}
    \mathbf{\bm{\Theta}}_k=\mathbcal{H}_k\mathbf{M}_k.
\end{equation}

Since $\hat{\mathbf{x}}_k$ is a linear combination of an unbiased estimate $\hat{\mathbcal{x}}_k$ and zero-mean white noises, it can be shown that $\hat{\mathbf{x}}_k$  is also an unbiased estimate.

To obtain the error covariance of the a priori state estimation error $\tilde{\mathbcal{x}}_k$, Eq.~\eqref{eq47} can be replaced into Eq.~\eqref{eq45} while recalling Eqs.~\eqref{eq26} and~\eqref{eq31} to obtain:
\begin{equation} \label{eq50}
    \tilde{\mathbcal{x}}_k=\breve{\mathbf{A}}_{k-1}\tilde{\mathbf{x}}_{k-1}+\mathbf{W}_k\mathbcal{w}_k-\mathbf{V}_k\mathbcal{v}_k,
\end{equation}
\noindent in which $\mathbf{V}_k\triangleq\mathbf{G}_{k-1}\mathbf{\bm{\varepsilon}}_m\mathbf{M}_k$, $\breve{\mathbf{A}}_{k-1}\triangleq\mathbf{A}_{k-1}-\mathbf{V}_k\mathbf{\bm{\Gamma}}_k$ and $\mathbf{W}_k\triangleq-\mathbf{V}_k\breve{\mathbf{D}}_k+\mathbf{\bm{\varepsilon}}_n$. Next, the error covariance of $\tilde{\mathbcal{x}}_k$ follows as:
\begin{equation} \label{eq51}   \mathbf{P}_k^\mathbcal{x}\triangleq\mathbb{E}\left[\tilde{\mathbcal{x}}_k\tilde{\mathbcal{x}}_k^T\right]=\mathbf{\bm{\Pi}}_k\mathbf{\bm{\Lambda}}_k\mathbf{\bm{\Pi}}_k^T,
\end{equation}
\noindent in which
\begin{equation} \label{eq52}
    \mathbf{\bm{\Pi}}_k\triangleq\begin{bmatrix}\breve{\mathbf{A}}_{k-1} & \mathbf{W}_k & -\mathbf{V}_k\\ \end{bmatrix}.
\end{equation}

Next, with the definition of the error of a posteriori state estimate, i.e., Eq.~\eqref{eq20}, replacing $\hat{\mathbf{x}}_k$ with Eq.~\eqref{eq48}, the expression of $\tilde{\mathbf{x}}_k$ can be obtained as:
\begin{equation} \label{eq53}
    \tilde{\mathbf{x}}_k=\tilde{\mathbcal{x}}_k-\mathbf{K}_k\left(\mathbcal{C}_k\tilde{\mathbcal{x}}_k-\mathbf{\bm{\Theta}}_k\mathbf{e}_k+\mathbcal{D}_k\mathbcal{w}_k+\mathbcal{v}_k\right).
\end{equation}

Following the previous equation, the error covariance of state estimation $\mathbf{P}_k$ reads as:
\begin{equation} \label{eq54}\mathbf{P}_k\triangleq\mathbb{E}\left[\tilde{\mathbf{x}}_k\tilde{\mathbf{x}}_k^T\right]=\mathbf{P}_k^\mathbcal{x}+\mathbf{K}_k\mathbf{\bm{\Upsilon}}_k+\mathbf{\bm{\Upsilon}}_k^T\mathbf{K}_k^T+\mathbf{K}_k\mathbf{\bm{\Phi}}_k\mathbf{K}_k^T,
\end{equation}
\noindent where
\begin{equation} \label{eq55}
    \mathbf{\bm{\Upsilon}}_k\triangleq\mathbb{E}\left[-\mathbf{\bm{\Omega}}_k\mathbf{\bm{\Lambda}}_k\mathbf{\bm{\Pi}}_k^T\right]
    =-\mathbf{\bm{\Omega}}_k\mathbf{\bm{\Lambda}}_k\mathbf{\bm{\Pi}}_k^T,
\end{equation}
\vspace{-0.5cm}
\begin{equation} \label{eq56}   \mathbf{\bm{\Phi}}_k\triangleq\mathbb{E}\left[\mathbf{\bm{\Omega}}_k\mathbf{\bm{\Lambda}}_k\mathbf{\bm{\Omega}}_k^T\right]
    =\mathbf{\bm{\Omega}}_k\mathbf{\bm{\Lambda}}_k\mathbf{\bm{\Omega}}_k^T,
\end{equation}
\vspace{-0.5cm}
\begin{equation} \label{eq57}
    \mathbf{\Omega}_k\triangleq
    \begin{bmatrix} \mathbcal{C}_k\breve{\mathbf{A}}_{k-1}-\mathbf{\bm{\Theta}}_k\mathbf{\bm{\Gamma}}_k
    &\mathbcal{C}_k\mathbf{W}_k-\mathbf{\bm{\Theta}}_k\breve{\mathbf{D}}_k+\mathbcal{D}_k
    &-\mathbcal{C}_k\mathbf{V}_k-\mathbf{\bm{\Theta}}_k+\mathbf{I}_{\left(N+1\right)d\times n}\\ \end{bmatrix}.
\end{equation}

Finally, the gain $\mathbf{K}_k$ can be directly calculated from Eq.~\eqref{eq54} by minimising the trace of the error covariance matrix $\mathbf{P}_k$. To do so, a decomposition of $\mathbf{K}_k$ is firstly assumed:
\begin{equation} \label{eq58}
    \mathbf{K}_k=\bar{\mathbf{K}}_k\mathbf{U}_k.
\end{equation}

$\mathbf{U}_k$ is chosen such that $ \mathbf{U}_k\mathbf{\bm{\Phi}}_k\mathbf{U}_k^T$ has a full rank, which can be achieved by eliminating all of the singular vectors with zero singular values; this enables the matrix inversion necessary for the state gain calculation in the next step. By solving an optimisation problem, $\bar{\mathbf{K}}_k$ can be obtained: 
\begin{equation}  \label{eq59}
\begin{aligned}
    \bar{\mathbf{K}}_k &= \arg{\min_{\bar{\mathbf{K}}}{\left(tr\left(\mathbf{P}_k^\mathbcal{x}+\bar{\mathbf{K}}\mathbf{U}_k\mathbf{\bm{\Upsilon}}_k+\mathbf{\bm{\Upsilon}}_k^T\mathbf{U}_k^T\bar{\mathbf{K}}^T+\bar{\mathbf{K}}\mathbf{U}_k\mathbf{\bm{\Phi}}_k\mathbf{U}_k^T\bar{\mathbf{K}}^T\right)\right)}}\\
   &=-\mathbf{\bm{\Upsilon}}_k^T\mathbf{U}_k^T\left(\mathbf{U}_k\mathbf{\bm{\Phi}}_k\mathbf{U}_k^T\right)^{-1}.
\end{aligned}
\end{equation}

Based on Eqs.~\eqref{eq58} and~\eqref{eq59}, the gain $\mathbf{K}_k$ that minimise the variance of state estimation is: 
\begin{equation} \label{eq60}
    \mathbf{K}_k=-\mathbf{\bm{\Upsilon}}_k^T\mathbf{U}_k^T\left(\mathbf{U}_k\mathbf{\bm{\Phi}}_k\mathbf{U}_k^T\right)^{-1}\mathbf{U}_k.
\end{equation}

\subsection{Error covariance matrices linked to input estimation}
\label{section2.5}
\noindent The conventional computation of $\mathbf{P}_k^{\mathbf{xw}}$ and $\mathbf{P}_k^{\mathbf{xv}}$ in batch format is computationally intensity. Therefore, the recursive approach in~\cite{EbrahimzadehHassanabadi2022} is employed herein. By substituting Eq.~\eqref{eq50} into Eq.~\eqref{eq53}, and with Eq.~\eqref{eq31}, a recursive form of state estimation error ${\tilde{\mathbf{x}}}_k$ can be obtained: 
\begin{equation} \label{eq61}
    \tilde{\mathbf{x}}_k=\mathbcal{A}_k\tilde{\mathbf{x}}_{k-1}+\mathbcal{W}_k\mathbcal{w}_k+\mathbcal{V}_k\mathbcal{v}_k,
\end{equation}

\noindent where
\begin{equation} \label{eq62}
    \mathbcal{A}_k = \mathbf{T}_k\breve{\mathbf{A}}_{k-1}+\mathbf{K}_k\mathbf{\bm{\Theta}}_k\mathbf{\bm{\Gamma}}_k,
\end{equation}
\vspace{-0.5cm}
\begin{equation} \label{eq63}
    \mathbcal{W}_k = \mathbf{T}_k\mathbf{W}_k+\mathbf{K}_k\mathbf{\bm{\Theta}}_k\breve{\mathbf{D}}_k-\mathbf{K}_k\mathbcal{D}_k,
\end{equation}
\vspace{-0.5cm}
\begin{equation} \label{eq64}
    \mathbcal{V}_k=-\mathbf{T}_k\mathbf{V}_k+\mathbf{K}_k\mathbf{\bm{\Theta}}_k-\mathbf{K}_k,
\end{equation}
\vspace{-0.5cm}
\begin{equation} \label{eq65}
    \mathbf{T}_k=\mathbf{I}_{\left(N+1\right)d\times n}-\mathbf{K}_k\mathbcal{C}_k,
\end{equation}

The assumption that $\hat{\mathbf{x}}_0$ is independent of the observation noise and process error leads to $\mathbb{E}\left[\ \tilde{\mathbf{x}}_0\mathbcal{w}_j^T\right]=\mathbf{0}$ and $\mathbb{E}\left[\ \tilde{\mathbf{x}}_0\mathbcal{v}_j^T\right]=\mathbf{0}$ with Eq.~\eqref{eq61}, i.e., the recursive expression of the state estimation error, it can be proven by mathematical induction that $\mathbb{E}\left[\ \tilde{\mathbf{x}}_i\mathbcal{w}_j^T\right]=\mathbf{0}$ and $\mathbb{E}\left[\ \tilde{\mathbf{x}}_i\mathbcal{v}_j^T\right]=\mathbf{0}$ for $j\geq i+N+1$. In addition, $\mathbb{E}\left[ \mathbcal{w}_i\mathbcal{w}_j^T \right] =\mathbf{0}$ and $\mathbb{E}\left[ \mathbcal{v}_i\mathbcal{v}_j^T \right] =\mathbf{0}$ for $\left|j-i\right|\geq N+1$.

Next, the entries of $\mathbcal{w}_k$ and $\mathbcal{v}_k$ can be written as: 
\begin{equation} \label{eq66}
    \mathbcal{w}_{k+1}=\breve{\mathbf{\bm{\varepsilon}}}_n\mathbcal{w}_k+\bar{\mathbf{\bm{\varepsilon}}}_n\mathbcal{w}_{k+1},
\end{equation}
\vspace{-0.5cm}
\begin{equation} \label{eq67}
    \mathbcal{v}_{k+1}=\breve{\mathbf{\bm{\varepsilon}}}_d\mathbcal{v}_k+\bar{\mathbf{\bm{\varepsilon}}}_d\mathbcal{v}_{k+1},
\end{equation}

\noindent where
\begin{equation} \label{eq68}
    \breve{\mathbf{\bm{\varepsilon}}}_i\triangleq
    \begin{bmatrix}\mathbf{0}_{iN\times i} & \mathbf{I}_{Ni}\\
    \mathbf{0}_{i\times i} & \mathbf{0}_{i\times i N}\\\end{bmatrix},
\end{equation}
\vspace{-0.5cm}
\begin{equation} \label{eq69}
    \bar{\mathbf{\bm{\varepsilon}}}_i\triangleq
    \begin{bmatrix}\mathbf{0}_{iN\times i N} & \mathbf{0}_{iN\times i}\\
    \mathbf{0}_{i\times i N} & \mathbf{I}_i\\\end{bmatrix}.
\end{equation}

Since Eq.~\eqref{eq61} implies that there is no correlation between $\tilde{\mathbf{x}}_{k-1}$ and $\mathbf{w}_{k+N}$, i.e., $\mathbb{E}\left[\tilde{\mathbf{x}}_{k-1}\left(\bar{\mathbf{\bm{\varepsilon}}}_n\mathbcal{w}_{k+1}\right)^T\right]=\mathbf{0}$, one can write: 
\begin{equation} \label{eq70}
\begin{aligned}
    \mathbf{P}_k^{\mathbf{xw}}\triangleq\mathbb{E}\left[{\tilde{\mathbf{x}}}_k\mathbcal{w}_{k+1}^T\right] 
    &=\mathbcal{A}_k\tilde{\mathbf{x}}_{k-1}\left(\breve{\mathbf{\bm{\varepsilon}}}_n\mathbcal{w}_k+\bar{\mathbcal{\bm{\varepsilon}}}_n\mathbcal{w}_{k+1}\right)^T+\mathbcal{W}_k\mathbcal{Q}_{k,k+1}\\
    &=\mathbcal{A}_k\mathbf{P}_{k-1}^{\mathbf{xw}}\breve{\mathbf{\bm{\varepsilon}}}_n^T+\mathbcal{W}_k\mathbcal{Q}_{k,k+1}.
\end{aligned}
\end{equation}

Similarly, the recursive form of $\mathbf{P}_k^{\mathbf{xv}}$ can be written as: 
\begin{equation} \label{eq71}
\begin{aligned}
    \mathbf{P}_k^{\mathbf{xv}}\triangleq\mathbb{E}\left[\tilde{\mathbf{x}}_k\mathbcal{v}_{k+1}^T\right]
    &= \mathbcal{A}_k\tilde{\mathbf{x}}_{k-1}\left(\breve{\mathbf{\bm{\varepsilon}}}_d\mathbcal{v}_k+\bar{\mathbf{\bm{\varepsilon}}}_d\mathbcal{v}_{k+1}\right)^T+\mathbcal{V}_k\mathbcal{R}_{k,k+1}\\
    &= \mathbcal{A}_k\mathbf{P}_{k-1}^{\mathbf{xv}}\breve{\mathbf{\bm{\varepsilon}}}_d^T+\mathbcal{V}_k\mathbcal{R}_{k,k+1}.
\end{aligned}
\end{equation}

\clearpage
\subsection{Summary of the Universal Smoothing}
\label{section2.6}
\noindent The US is summarised in Table \ref{tab2}.

{
\footnotesize
\begin{longtable}[t]{l}
    \caption{Summary of the Universal Smoothing}
    \label{tab2}\\

    \toprule
    \multicolumn{1}{l}{\textbf{Initialisation}}\\
    \endfirsthead
    \midrule
    \endfoot
    
    \multicolumn{1}{c}{Continuation of Table \ref{tab2}}\\
    \midrule
    \endhead
    \bottomrule
    \endlastfoot

    \hspace{3mm} • Define process and observation equations\\
    \hspace{3mm} • Assign initial values to $\hat{\mathbf{x}}_0$, $\mathbf{P}_0$, $\mathbf{Q}_k$, $\mathbf{R}_k$ and build $\mathbcal{Q}_{k,k}$, $\mathbcal{Q}_{k,k+1}$, $\mathbcal{R}_{k,k}$ and $\mathbcal{R}_{k,k+1}$\\
    \hspace{3mm} • Assign $\mathbf{P}_0^{\mathbf{xw}}=\mathbf{0}$ and $\mathbf{P}_0^{\mathbf{xv}}=\mathbf{0}$\\
    \hspace{3mm} • Build time-independent matrices\\
    \hspace{6mm} $\mathbf{\bm{\varepsilon}}_m = \begin{bmatrix}\mathbf{I}_m & \bm{\mathbf{0}}_{m\times Nm}\\ \end{bmatrix}$, $\mathbf{\bm{\varepsilon}}_n = \begin{bmatrix}\mathbf{I}_n & \bm{\mathbf{0}}_{n\times Nn}\\ \end{bmatrix}$, $\breve{\mathbf{\bm{\varepsilon}}}_i\triangleq
                     \begin{bmatrix}\mathbf{0}_{iN\times i} & \mathbf{I}_{Ni}\\
                     \mathbf{0}_{i\times i} & \mathbf{0}_{i\times i N}\\\end{bmatrix}$, $\bar{\mathbf{\bm{\varepsilon}}}_i\triangleq
                     \begin{bmatrix}\mathbf{0}_{iN\times i N} & \mathbf{0}_{iN\times i}\\
                     \mathbf{0}_{i\times i N} & \mathbf{I}_i\\\end{bmatrix}$\\
    \midrule
    \textbf{Smoothing loop}\\
    \hspace{3mm} • Input estimation\\
    \hspace{6mm} 1. $\mathbf{\bm{\Gamma}}_k=\mathbcal{C}_k\mathbf{A}_{k-1}$ \\
    \hspace{6mm} 2. $\mathbf{\bm{\Sigma}}_k\triangleq
                     \begin{bmatrix}\mathbf{\bm{\Gamma}}_k & \breve{\mathbf{D}}_k & \mathbf{I}_{(N+1)d}\\ \end{bmatrix}$ \\
    \hspace{6mm} 3. $\tilde{\mathbf{R}}_k=\mathbf{\bm{\Sigma}}_k\mathbf{\bm{\Lambda}}_k\mathbf{\bm{\Sigma}}_k^T$\\
    \hspace{6mm} 4. $\mathbf{P}_k^\mathbcal{p}=\left(\breve{\mathbf{H}}_k^T\tilde{\mathbf{R}}_k^{-1}\breve{\mathbf{H}}_k\right)^{-1}$\\
    \hspace{6mm} 5. $\mathbf{P}_k^\mathbf{p}=\mathbf{\bm{\varepsilon}}_m\mathbf{P}_k^\mathbcal{p}\mathbf{\bm{\varepsilon}}_m^T$\\
    \hspace{6mm} 6. $\mathbf{M}_k=\mathbf{P}_k^\mathbcal{p}\breve{\mathbf{H}}_k^T\tilde{\mathbf{R}}_k^{-1}$\\
    \hspace{6mm} 7. $\hat{\bm{\mathbf{\chi}}}_k=\mathbf{A}_{k-1}\hat{\mathbf{x}}_{k-1}$\\
    \hspace{6mm} 8. $\hat{\mathbcal{p}}_k=\mathbf{M}_k\left(\mathbcal{y}_k-\mathbcal{C}_k\hat{\bm{\mathbf{\chi}}}_k\right)$\\
    \hspace{6mm} 9. $\hat{\mathbf{p}}_k=\bm{\mathbf{\varepsilon}}_m\hat{\mathbcal{p}}_k$\\
    \hspace{3mm} • State estimation\\
    \hspace{6mm} 1. $\hat{\mathbcal{x}}_k=\hat{\bm{\mathbf{\chi}}}_k+\mathbf{G}_{k-1}\hat{\mathbf{p}}_k$\\
    \hspace{6mm} 2. $\mathbf{V}_k\triangleq\mathbf{G}_{k-1}\mathbf{\bm{\varepsilon}}_m\mathbf{M}_k$\\
    \hspace{6mm} 3. $\mathbf{W}_k\triangleq-\mathbf{V}_k\breve{\mathbf{D}}_k+\mathbf{\bm{\varepsilon}}_n$\\
    \hspace{6mm} 4. $\breve{\mathbf{A}}_{k-1}\triangleq\mathbf{A}_{k-1}-\mathbf{V}_k\mathbf{\bm{\Gamma}}_k$\\
    \hspace{6mm} 5. $\mathbf{\bm{\Pi}}_k\triangleq\begin{bmatrix}\breve{\mathbf{A}}_{k-1} & \mathbf{W}_k & -\mathbf{V}_k\\ \end{bmatrix}$\\
    \hspace{6mm} 6. $\mathbf{P}_k^\mathbcal{x}=\mathbf{\bm{\Pi}}_k\mathbf{\bm{\Lambda}}_k\mathbf{\bm{\Pi}}_k^T$\\
    \hspace{6mm} 7. $\mathbf{\bm{\Theta}}_k=\mathbcal{H}_k\mathbf{M}_k$\\
    \hspace{6mm} 8. $\mathbf{\Omega}_k\triangleq
    \begin{bmatrix} \mathbcal{C}_k\breve{\mathbf{A}}_{k-1}-\mathbf{\bm{\Theta}}_k\mathbf{\bm{\Gamma}}_k
    &\mathbcal{C}_k\mathbf{W}_k-\mathbf{\bm{\Theta}}_k\breve{\mathbf{D}}_k+\mathbcal{D}_k
    &-\mathbcal{C}_k\mathbf{V}_k-\mathbf{\bm{\Theta}}_k+\mathbf{I}_{\left(N+1\right)d\times n}\\ \end{bmatrix}$\\
    \hspace{6mm} 9. $\mathbf{\bm{\Upsilon}}_k=-\mathbf{\bm{\Omega}}_k\mathbf{\bm{\Lambda}}_k\mathbf{\bm{\Pi}}_k^T$\\
    \hspace{6mm} 10. $\mathbf{\bm{\Phi}}_k=\mathbf{\bm{\Omega}}_k\mathbf{\bm{\Lambda}}_k\mathbf{\bm{\Omega}}_k^T$\\
    \hspace{6mm} 11. $\mathbf{K}_k=-\mathbf{\bm{\Upsilon}}_k^T\mathbf{U}_k^T\left(\mathbf{U}_k\mathbf{\bm{\Phi}}_k\mathbf{U}_k^T\right)^{-1}\mathbf{U}_k$\\
    \hspace{6mm} 12. $\hat{\mathbf{x}}_k=\hat{\mathbcal{x}}_k+\mathbf{K}_k\left(\mathbcal{y}_k-\mathbcal{C}_k\hat{\mathbcal{x}}_k-\mathbcal{H}_k\hat{\mathbcal{p}}_k\right)$\\
    \hspace{6mm} 13. $\mathbf{P}_k=\mathbf{P}_k^\mathbcal{x}+\mathbf{K}_k\mathbf{\bm{\Upsilon}}_k+\mathbf{\bm{\Upsilon}}_k^T\mathbf{K}_k^T+\mathbf{K}_k\mathbf{\bm{\Phi}}_k\mathbf{K}_k^T$\\
    \hspace{6mm} 14. $\mathbf{T}_k=\mathbf{I}_{\left(N+1\right)d\times n}-\mathbf{K}_k\mathbcal{C}_k$\\
    \hspace{6mm} 15. $\mathbcal{W}_k = \mathbf{T}_k\mathbf{W}_k+\mathbf{K}_k\mathbf{\bm{\Theta}}_k\breve{\mathbf{D}}_k-\mathbf{K}_k\mathbcal{D}_k$\\
    \hspace{6mm} 16. $\mathbcal{V}_k=-\mathbf{T}_k\mathbf{V}_k+\mathbf{K}_k\mathbf{\bm{\Theta}}_k-\mathbf{K}_k$\\
    \hspace{6mm} 17. $\mathbcal{A}_k = \mathbf{T}_k\breve{\mathbf{A}}_{k-1}+\mathbf{K}_k\mathbf{\bm{\Theta}}_k\mathbf{\bm{\Gamma}}_k$ \\
    \hspace{6mm} 18. $\mathbf{P}_k^{\mathbf{xw}}=\mathbcal{A}_k\mathbf{P}_{k-1}^{\mathbf{xw}}\breve{\mathbf{\bm{\varepsilon}}}_n^T+\mathbcal{W}_k\mathbcal{Q}_{k,k+1}$\\
    \hspace{6mm} 19. $\mathbf{P}_k^{\mathbf{xv}}=\mathbcal{A}_k\mathbf{P}_{k-1}^{\mathbf{xv}}\breve{\mathbf{\bm{\varepsilon}}}_d^T+\mathbcal{V}_k\mathbcal{R}_{k,k+1}$\\
 \end{longtable}
}

\clearpage
\subsection{Structural system subjected to dynamic loads}
\label{section2.7}
\noindent The equation of motion of an MDOF structural system subjected to dynamic loads can be expressed by the following second-order system of Ordinary Differential Equation (ODE) in the time domain:
\begin{equation} \label{eq_eom}
    \mathbf{M}_F\ddot{\mathbf{u}}(t)+\mathbf{C}_F\dot{\mathbf{u}}(t)+\mathbf{K}_F\mathbf{u}(t)=\mathbf{B}\mathbf{p}(t),
\end{equation}
\noindent in which $\mathbf{M}_F\in\mathbb{R}^{f\times f}$ is the mass matrix, $\mathbf{C}_F\in\mathbb{R}^{f\times f}$ is the damping matrix, $\mathbf{K}_F\in\mathbb{R}^{f\times f}$ is stiffness matrix, and matrix $\mathbf{B}\in\mathbb{R}^{f\times m}$ indicates the distribution of the dynamic loads $\mathbf{p}(t)$. The overdot denotes the derivative with respect to time, therefore, $\mathbf{u}\in\mathbb{R}^f$, $\dot{\mathbf{u}}\in\mathbb{R}^f$ and $\ddot{\mathbf{u}}\in\mathbb{R}^f$ are the displacement, velocity and acceleration vectors, respectively. In the case of ground acceleration, the term $\mathbf{Bp}(t)$ should be replaced by $-\mathbf{M}_F\mathbf{i}\ddot{\mathbf{u}}_g(t)$ where $\ddot{\mathbf{u}}_g(t)$ is the ground acceleration, and $\mathbf{i} \in \mathbb{R}^f$ is a unitary vector indicating the influence of ground acceleration to the structure. The analysis of a large-scale MDOF structure can be computationally demanding. To improve the efficiency, Eq.~\eqref{eq_eom} can be rewritten in a reduced order form by modal orthogonality, which results in: 
\begin{equation} \label{eq73}
    \mathbf{M}_R\ddot{\mathbf{q}}(t)+\mathbf{C}_R\dot{\mathbf{q}}(t)+\mathbf{K}_R\mathbf{q}(t)=\mathbf{B}_R\mathbf{p}(t),
\end{equation}
\noindent where
\begin{equation} \label{eq74}
\begin{aligned}
    &\mathbf{M}_R=\mathbf{Z}^T\mathbf{M}_F\mathbf{Z},\\
    &\mathbf{C}_R=\mathbf{Z}^T\mathbf{C}_F\mathbf{Z},\\
    &\mathbf{K}_R=\mathbf{Z}^T\mathbf{K}_F\mathbf{Z},\\
    & \mathbf{B}_R=\mathbf{Z}^T\mathbf{B}.
\end{aligned}
\end{equation}

In the above equations, $\mathbf{Z}\in\mathbb{R}^{f\times r}$ is the modal matrix containing eigenvectors of each mode. Furthermore, $\mathbf{q}\in\mathbb{R}^r$, $\dot{\mathbf{q}}\in\mathbb{R}^r$ and $\ddot{\mathbf{q}}\in\mathbb{R}^r$ are the dynamic responses in modal coordinates. Next, Eq.~\eqref{eq73} can be written in an equivalent first-order ODE as: 
\begin{equation} \label{eq75}
    \dot{\mathbf{x}}\left(t\right)=\mathbf{\Psi x}\left(t\right)+\mathbf{\Xi}\mathbf{p}(t),
\end{equation}
\noindent where
\begin{equation} \label{eq76}
    \mathbf{x}\left(t\right)=\left\{
    \begin{matrix}
    \mathbf{q}(t)\\
    \dot{\mathbf{q}}(t)\\
    \end{matrix}
    \right\},
\end{equation}
\vspace{-0.5cm}
\begin{equation} \label{eq77}
    \mathbf{\Psi}=
    \begin{bmatrix}
    \mathbf{0}_{r\times r} & \mathbf{I}_r\\
    -{\mathbf{M}_R}^{-1}\mathbf{K}_R & -{\mathbf{M}_R}^{-1}\mathbf{C}_R\\
    \end{bmatrix},
\end{equation}
\vspace{-0.5cm}
\begin{equation} \label{eq78}
    \mathbf{\Xi}=
    \begin{bmatrix}
    \mathbf{0}_{r\times m}\\
    {\mathbf{M}_R}^{-1}\mathbf{B}_R\\
    \end{bmatrix}.
\end{equation}

By implementing the matrix exponential time integrator~\cite{Brogan1991}, a discrete-time recursive solution of Eq.~\eqref{eq75} can be obtained with a uniform time interval ${\Delta}t$ as: 
\begin{equation} \label{eq79}
    \mathbf{x}_k=\mathbf{A}_{k-1}\mathbf{x}_{k-1}+\mathbf{G}_{k-1}{\mathbf{p}}_k,
\end{equation}

\noindent where
\begin{equation} \label{eq80}
    \mathbf{A}_k=\exp{(\mathbf{\Psi}{\Delta}t)},
\end{equation}
\vspace{-0.5cm}
\begin{equation} \label{eq81}
    \mathbf{G}_k=\left(\mathbf{A}_k-\mathbf{I}_n\right)\mathbf{\Psi}^{-1}\mathbf{\Xi}.
\end{equation}

Assuming a total of $d$ quantities are observed, the output matrix $\mathbf{C}_k$ in Eq.~\eqref{eq2} can be obtained for as follows: 
\begin{equation} \label{eq82}
    \mathbf{C}_k=\mathbf{C}_0
    \begin{bmatrix}
    \mathbf{I}_r & \mathbf{0}_{r\times r}\\
    \mathbf{0}_{r\times r} & \mathbf{I}_r\\
    -{\mathbf{M}_R}^{-1}\mathbf{K}_R&-{\mathbf{M}_R}^{-1}\mathbf{C}_R\\
    \end{bmatrix}.
\end{equation}

In the above equation, $\mathbf{C}_0\in\mathbb{R}^{d\times 3r}$ is a Boolean matrix extracting the locations that are observed. Similarly, $\mathbf{H}_k$ is given by: 
\begin{equation} \label{eq83}
    \mathbf{H}_k=\mathbf{C}_0
    \begin{bmatrix}
    \mathbf{0}_{2r\times m}\\
    {\mathbf{M}_R}^{-1}\mathbf{B}_R\\
    \end{bmatrix}.
\end{equation}

\section{Numerical validations}
\label{section3}
\noindent Four numerical case studies are presented in this section. The first consists of an eight-storey shear frame subjected to a sinusoidal force excitation that is used to demonstrate the estimation quality of the US compared to its filtering counterpart, i.e., the UF, using non-collocated observations and to assess the further improvement made by adopting the Moore–Penrose inverse or Pseudo-inverse (PINV) in the input estimation step. Next, the same shear frame but excited by ground motion is used as the second case study to assess the ability of the US to estimate the state and identify the ground motion with various sensor networks containing different combinations of displacement, velocity and acceleration observations. Furthermore, a 3-D finite element model of the Taipei 101 tower is presented next to assess the performance of the US on large-scale structures. In addition, the Taipei 101 tower is excited by two unknown arbitrary loads to recreate a rank-deficient feedforward matrix. Comparisons are made with other popular methods, including the AKF\cite{LourensReyndersEATAL2012}, MVUF-NDF\cite{Gillijns2007a}, MVUF-DF \cite{Gillijns2007b}, MVUS-NDF\cite{EbrahimzadehHassanabadi2022} and MVUS-DF\cite{Maes2018}.

\subsection{Eight-storey shear building subjected to a sinusoidal force}
\label{section3.1}
\begin{figure}[ht]
  \begin{minipage}[c]{.35\textwidth}
    \centering
    \includegraphics[trim={4.5cm 2.7cm 1.3cm 2.5cm}, clip, width=0.8\textwidth]{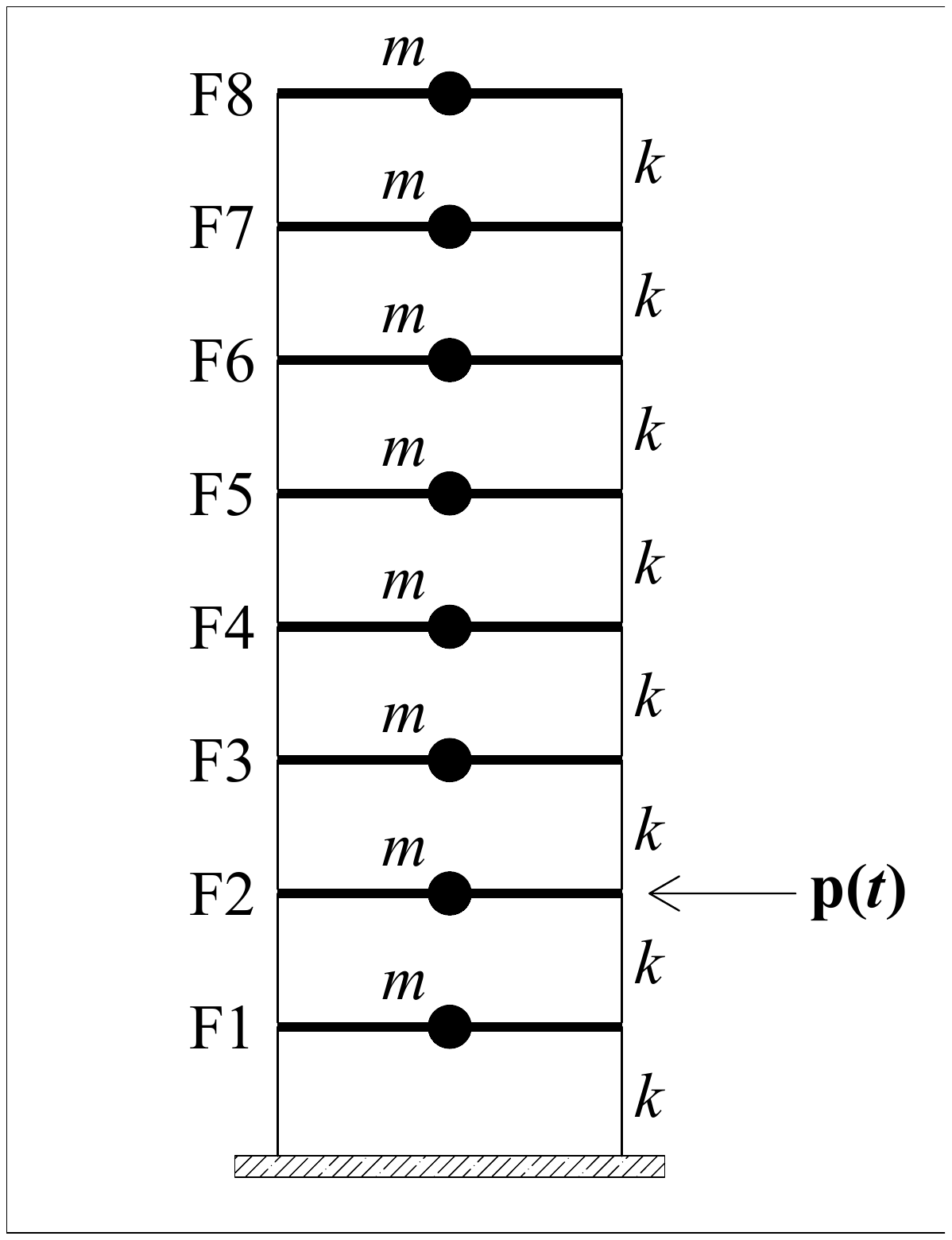}
    \captionof{figure}{Eight-storey shear building.} \label{storey_building_force}
  \end{minipage}\hfill
  \begin{minipage}[c]{.65\textwidth}
    \centering
    \captionsetup{justification=centering}
    \captionof{table}{Observation configurations for the eight-storey shear building subjected to the sinusoidal force.} \label{tab-config-ex1}
    \footnotesize{
    \begin{tabular}{c c c c}
        \toprule
         & \multicolumn{3}{c}{Location of sensors} \\
        \cmidrule{2-4}
        Configurations & Displacement & Velocity & Acceleration \\
        \midrule
        1.1 & F1, F3, F5 and F7 & F1 & - \\
        1.2 & F1, F3, F5 and F7 & - & F1 \\
        \bottomrule
    \end{tabular}}
  \end{minipage}
\end{figure}

\noindent The numerical model of an eight-storey shear building studied by \citet{Callafon2008} is adopted. As shown in Fig.~\ref{storey_building_force}, each floor of the building has a lumped mass $m=625\times{10}^3$~kg and an inter-storey stiffness $k=1\times{10}^9$~N/m. A Rayleigh damping is assumed with factors $\alpha=\beta=0.01$, in which case the damping matrix defined in Eq.~\eqref{eq_eom} can be written as $\mathbf{C}_F=\alpha\mathbf{M}_F+\beta\mathbf{K}_F$. The natural frequencies of the structure range from $7.381$ up to $78.637$ rad/s. The structure is loaded by a sinusoidal force on the second floor, and the load is defined by $\mathbf{p}(t)= \{ 5\times10^3 \times \sin{\left(8t\right)} \}$~N. The load is handled as unknown during the estimation. The time history of the dynamic responses of the structure has a duration of $25$~s and a time interval of ${\Delta}t=0.01$~s, and is obtained using the matrix exponential method described earlier in Section~\ref{section2.7}. The results are artificially contaminated by a zero-mean Gaussian white noise to simulate the observation data. The noise level is $1$\% of the root-mean-square (RMS) of the dynamic responses. Two typical sensor configurations are used in this example: one is the system without direct feedthrough and one is the system with a full-rank direct feedforward matrix, as presented in Table \ref{tab-config-ex1}. Note that the loaded floor is not directly measured.

In this case study, the US is compared with the UF to assess the performance improvement made by the extended observation window. For both methods, the full-order model is adopted in the estimation process, resulting in a zero modelling error covariance, i.e. $\mathbf{Q}=\mathbf{0}$; and the observation error covariance $\mathbf{R}$ is set to the covariance of the simulated added noise. In addition, the initial state $\hat{\mathbf{x}}_0$, initial error covariance matrices $\mathbf{P}_0$, $\mathbf{P}_0^{\mathbf{xw}}$ and $\mathbf{P}_0^{\mathbf{xv}}$ are all set to zero. Furthermore, the observation window for the US is set to $N=25$, which is enough to demonstrate visible improvement. The accuracy of the estimation is evaluated by the dimensionless error $\Sigma\delta_{\alpha}$ defined below:
\begin{equation} 
    \Sigma\delta_{\alpha}=\Sigma(\Delta \alpha_{rms}/\alpha_{max}),
\end{equation}

\noindent in which $\alpha$ can stand for the displacement $u$, velocity $\dot{u}$ and input $p$; $\Delta \alpha_{rms}$ is the RMS of the difference between the estimated and true values, and $\alpha_{max}$ is the maximum absolute value of the true values. The overall dimensionless error can therefore be obtained by $\Sigma\delta_{\alpha} = \mathbf{\sum}_{i=1}^{f}\Delta u_{rms,i}/u_{max,i} +\mathbf{\sum}_{i=1}^{f}\Delta \dot{u}_{rms,i}/\dot{u}_{max,i} + \mathbf{\sum}_{i=1}^{m}\Delta p_{rms,i}/p_{max,i}$.

\begin{figure}[ht]
    \centering
    \begin{subfigure}{0.7\textwidth}
        \caption{} \label{fig3-ex1-config1-a}
        \includegraphics[trim={2cm 7cm 2.5cm 7cm}, clip, width=\textwidth]{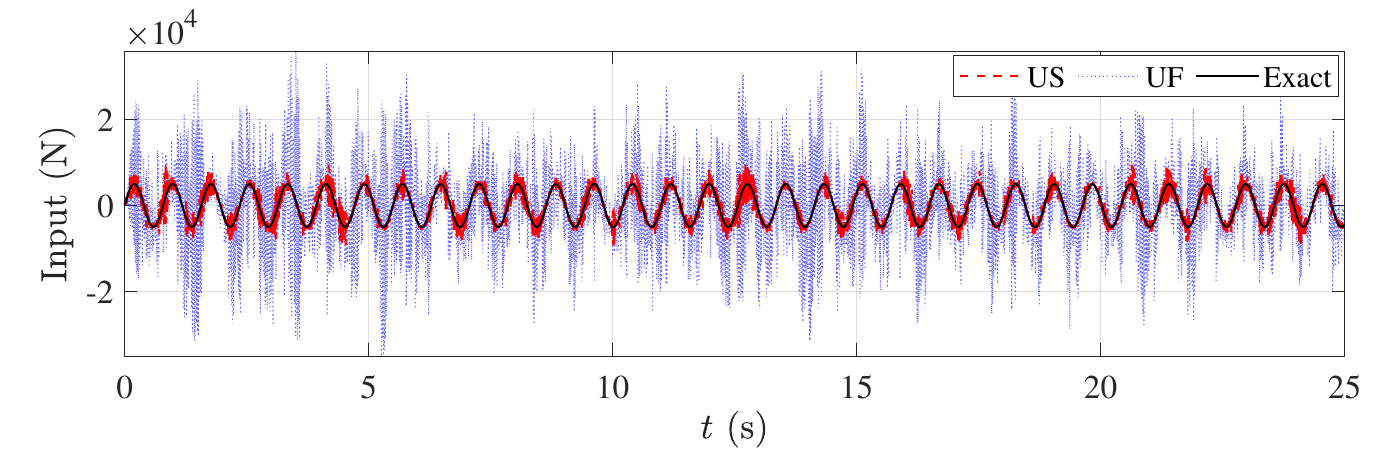}
    \end{subfigure}
    \begin{subfigure}{0.7\textwidth}
        \caption{} \label{fig3-ex1-config1-b}
        \includegraphics[trim={2cm 7cm 2.5cm 7cm}, clip, width=\textwidth]{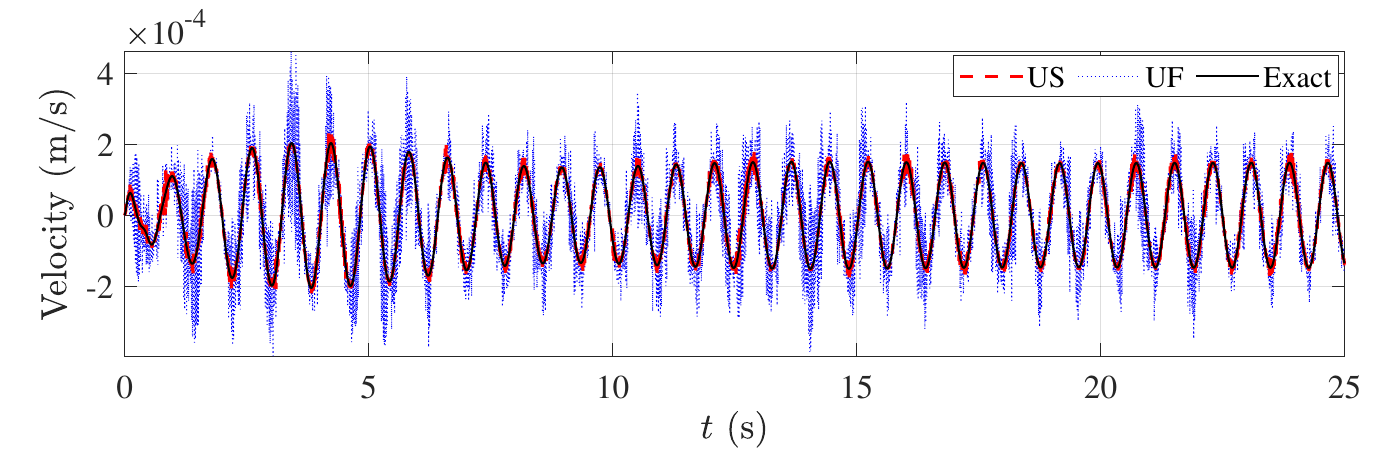}
    \end{subfigure}
    \begin{subfigure}{0.7\textwidth}
        \caption{} \label{fig3-ex1-config1-c}
        \includegraphics[trim={2cm 7cm 2.5cm 7cm}, clip, width=\textwidth]{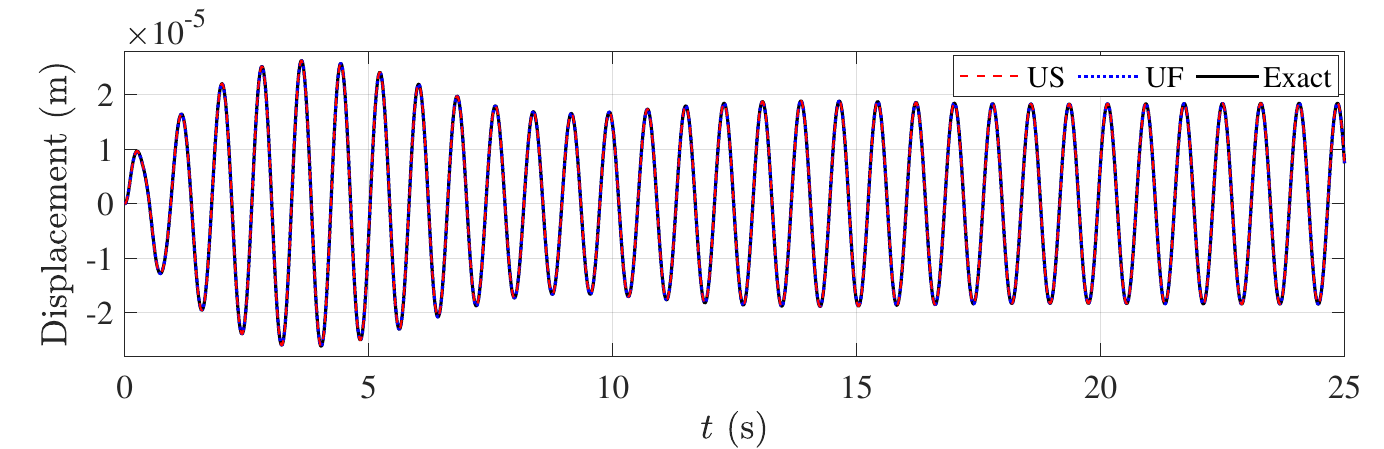}
    \end{subfigure}
    \caption{Configuration 1.1: (a) input estimation, (b) velocity estimation of the 2nd floor; (c) displacement estimation of the 2nd floor.}
    \label{fig3-ex1-config1}
\end{figure}

The estimation results of configuration 1.1 are shown in Fig.~\ref{fig3-ex1-config1}. As shown in Figs.~\ref{fig3-ex1-config1-a} and \ref{fig3-ex1-config1-b}, the noise in the input and velocity estimation is heavily intensified for the UF as impacted by the non-collocated observation. On the contrary, the US can utilise extra sensory information within the extended observation window, resulting in an enhanced estimation quality. Although Fig.~\ref{fig3-ex1-config1-c} shows that both methods have adequate quality for displacement estimation, the US can reach a higher accuracy, evidenced by the dimensionless error of the displacement estimation $\Sigma\delta_{u}$, which is $0.002$ for the US compared to $0.005$ for the UF.

\begin{figure}[!ht]
    \centering
    \begin{subfigure}{0.7\textwidth}
        \caption{} \label{fig-ex1-config2-a}
        \includegraphics[trim={2cm 7cm 2.5cm 7cm}, clip, width=\textwidth]{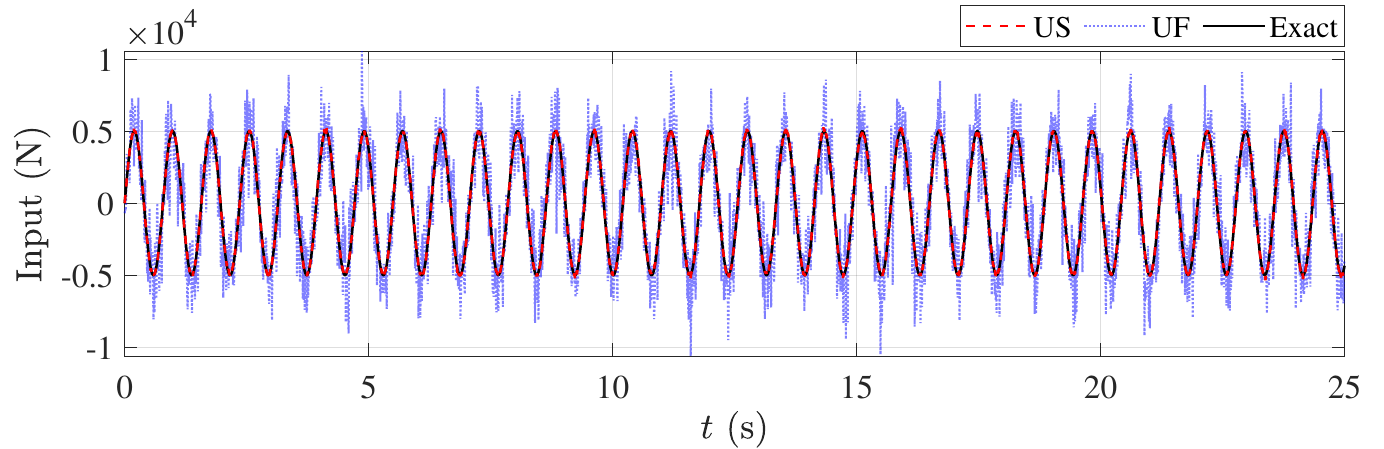}
    \end{subfigure}
    \begin{subfigure}{0.7\textwidth}
        \caption{} \label{fig-ex1-config2-b}
        \includegraphics[trim={2cm 7cm 2.5cm 7cm}, clip, width=\textwidth]{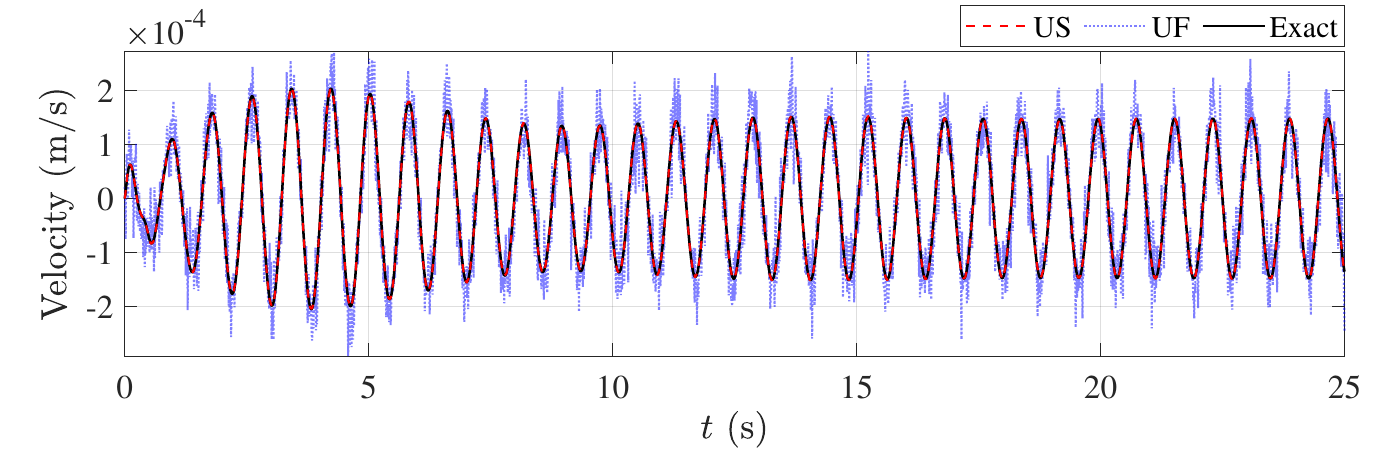}
    \end{subfigure}
    \begin{subfigure}{0.7\textwidth}
        \caption{} \label{fig-ex1-config2-c}
        \includegraphics[trim={2cm 7cm 2.5cm 7cm}, clip, width=\textwidth]{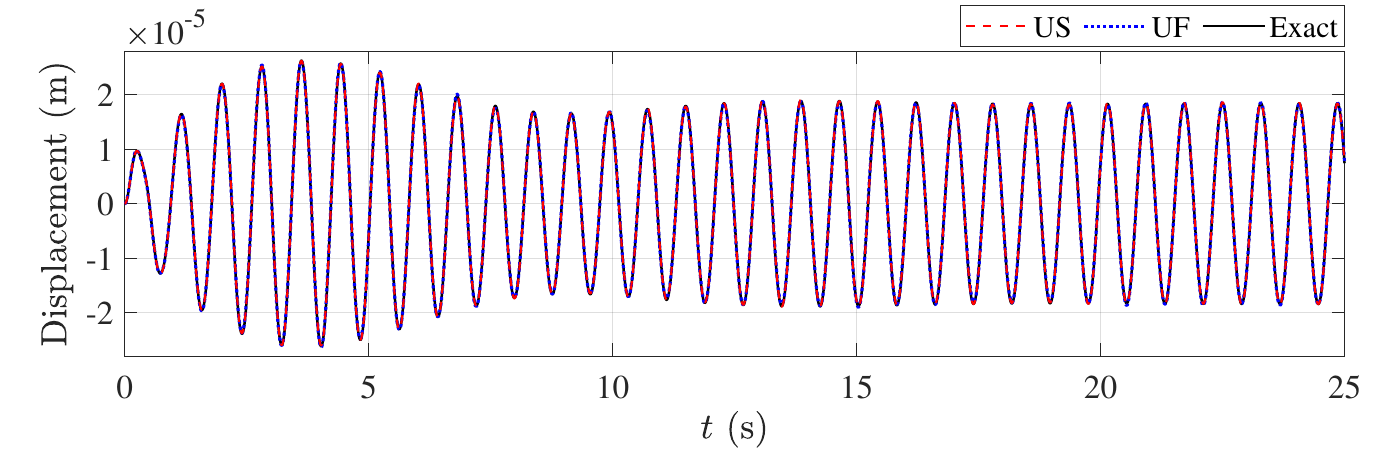}
    \end{subfigure}
    \caption{Configuration 1.2: (a) input estimation, (b) velocity estimation of the 2nd floor; (c) displacement estimation of the 2nd floor.}
    \label{fig-ex1-config2}
\end{figure}

The performance for both methods improves under configuration 1.2 with the contribution of the acceleration measurement on the first floor, as shown in Fig.~\ref{fig-ex1-config2}. However, the input and velocity estimation are yet noisy for the UF despite the improvement compared to the previous configuration. The US, on the other hand, can provide smooth and accurate estimations, and noise in the input and velocity estimates is further reduced when compared with the system without direct feedthrough. The overall dimensionless error $\Sigma\delta_{\alpha}$ for the US is $0.07$ compared to $2.16$ for the UF, confirming a $97$\% improvement.

To further enhance the performance of the proposed US, matrix inversion involved in computing the input error covariance (Eq.~\eqref{eq42}) and the input gain (Eq.~\eqref{eq43}) can be replaced by PINV:
\begin{equation} \label{eq42-2}
\mathbf{P}_k^\mathbcal{p}\triangleq\mathbb{E}\left[\tilde{\mathbcal{p}}_k\tilde{\mathbcal{p}}_k^T\right]=\left(\breve{\mathbf{H}}_k^T\tilde{\mathbf{R}}_k^{\dag}\breve{\mathbf{H}}_k\right)^{\dag},
\end{equation}
\vspace{-0.5cm}
\begin{equation} \label{eq44-2}\mathbf{M}_k=\mathbf{P}_k^\mathbcal{p}\breve{\mathbf{H}}_k^T\tilde{\mathbf{R}}_k^{\dag},
\end{equation}

\noindent where ${\dag}$ stands for PINV, and a tolerance can be selected to truncate the small singular values. The same eight-storey shear frame with the sinusoidal force is repeated to assess the improvement made by adopting PINV in the proposed US. However, the noise level is raised from $1$\% to $5$\% RMS of the dynamic responses. To optimise the estimation accuracy, the PINV tolerance is tuned through a grid search. The tolerance is searched from $1\times10^{-24}$ to $10^{-1}$ with an interval of $10^{0.1}$, and the tuning result is presented in \ref{appendixA1}. Other initialisations remain unchanged.

\begin{figure}[!ht]
    \centering
    \begin{subfigure}{0.7\textwidth}
        \caption{} \label{fig-ex1-config1-PINV-a}
        \includegraphics[trim={2cm 7cm 2.5cm 7cm}, clip, width=\textwidth]{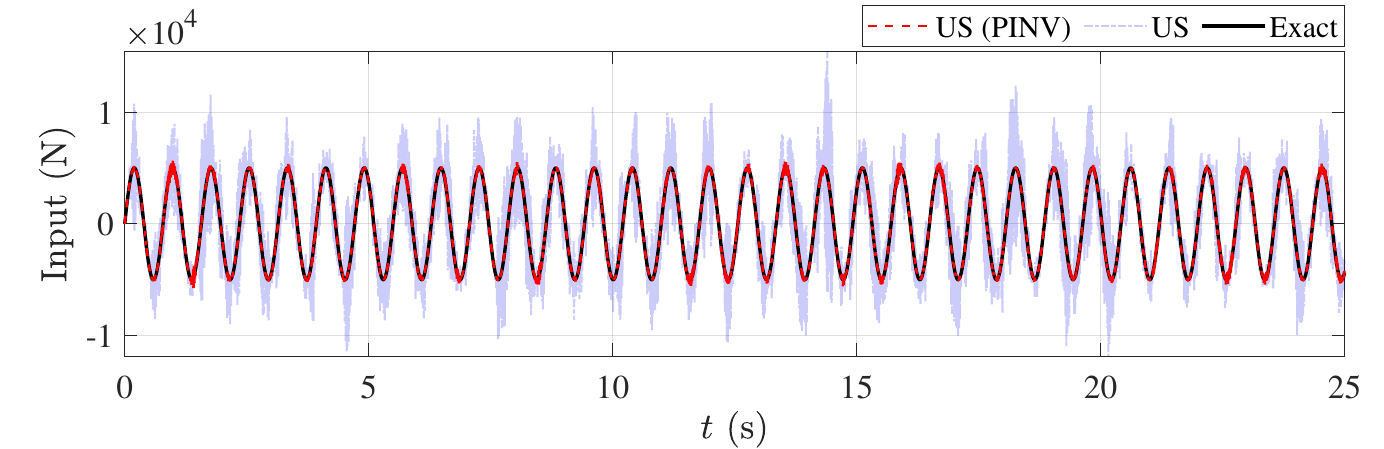}
    \end{subfigure}
    \begin{subfigure}{0.7\textwidth}
        \caption{} \label{fig-ex1-config1-PINV-b}
        \includegraphics[trim={2cm 7cm 2.5cm 7cm}, clip, width=\textwidth]{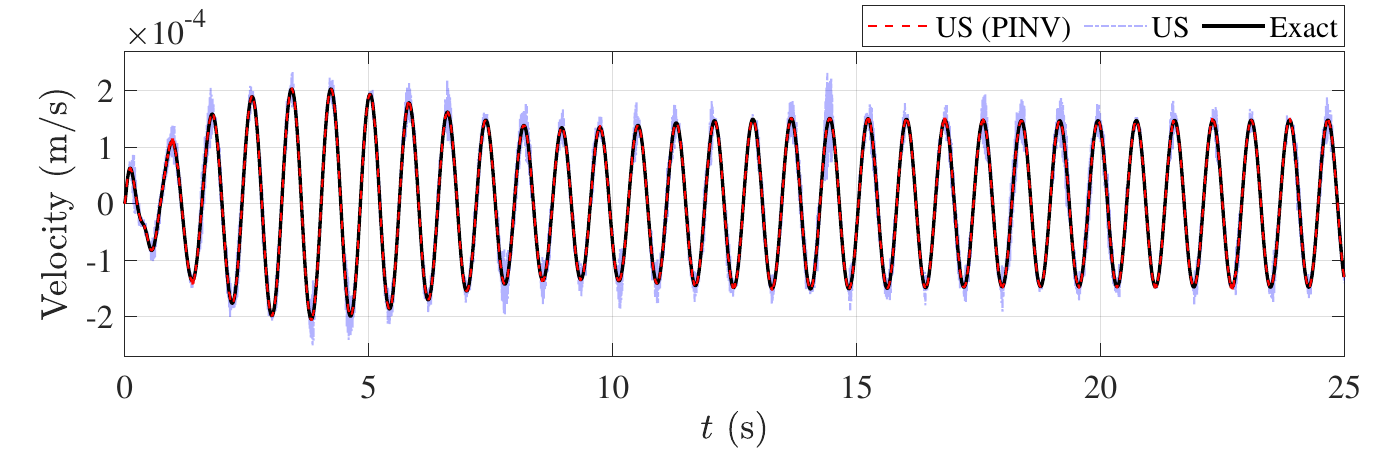}
    \end{subfigure}
    \begin{subfigure}{0.7\textwidth}
        \caption{} \label{fig-ex1-config1-PINV-c}
        \includegraphics[trim={2cm 7cm 2.5cm 7cm}, clip, width=\textwidth]{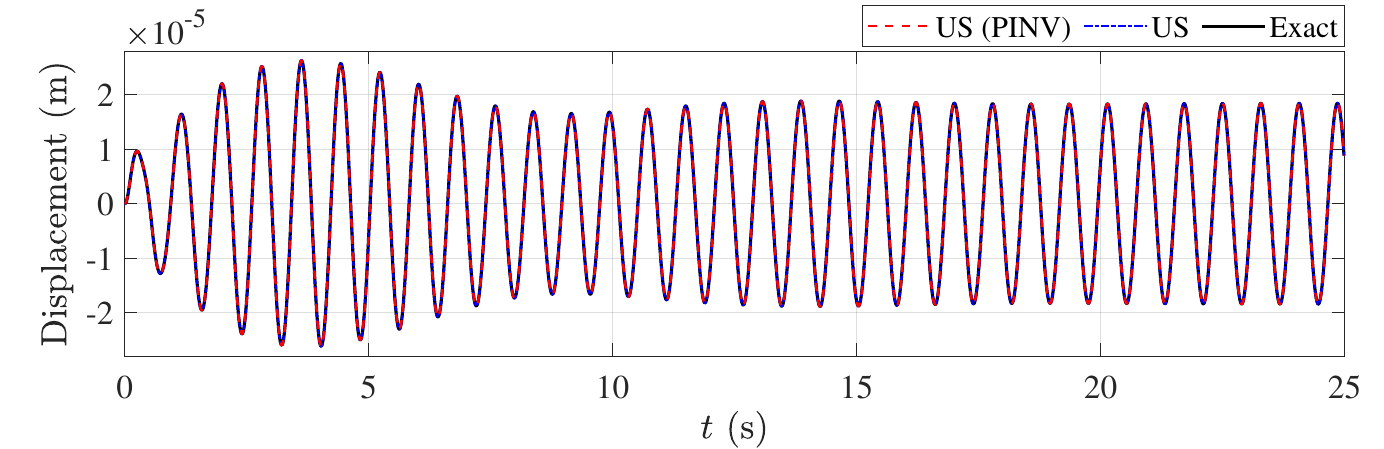}
    \end{subfigure}
    \caption{Configuration 1.1 results: (a) input estimation, (b) velocity estimation of the 2nd floor; (c) displacement estimation of the 2nd floor.}
    \label{fig-ex1-config1-PINV}
\end{figure}

Fig.~\ref{fig-ex1-config1-PINV} shows the results of configuration 1.1 by the US with and without using PINV. As the noise level increases, the accuracy of the US without PINV drops, particularly for the input and velocity estimation shown in Figs.~\ref{fig-ex1-config1-PINV-a} and \ref{fig-ex1-config1-PINV-b}. However, by adopting PINV and properly selecting the tolerance, the US with PINV can maintain a high estimation quality. The overall dimensionless error $\Sigma\delta_{\alpha}$ is reduced from $0.69$ to $0.07$ by adopting PINV.

\begin{figure}[!ht]
    \centering
    \begin{subfigure}{0.7\textwidth}
        \caption{} \label{fig-ex1-config2-PINV-a}
        \includegraphics[trim={1.8cm 7cm 2.5cm 7cm}, clip, width=\textwidth]{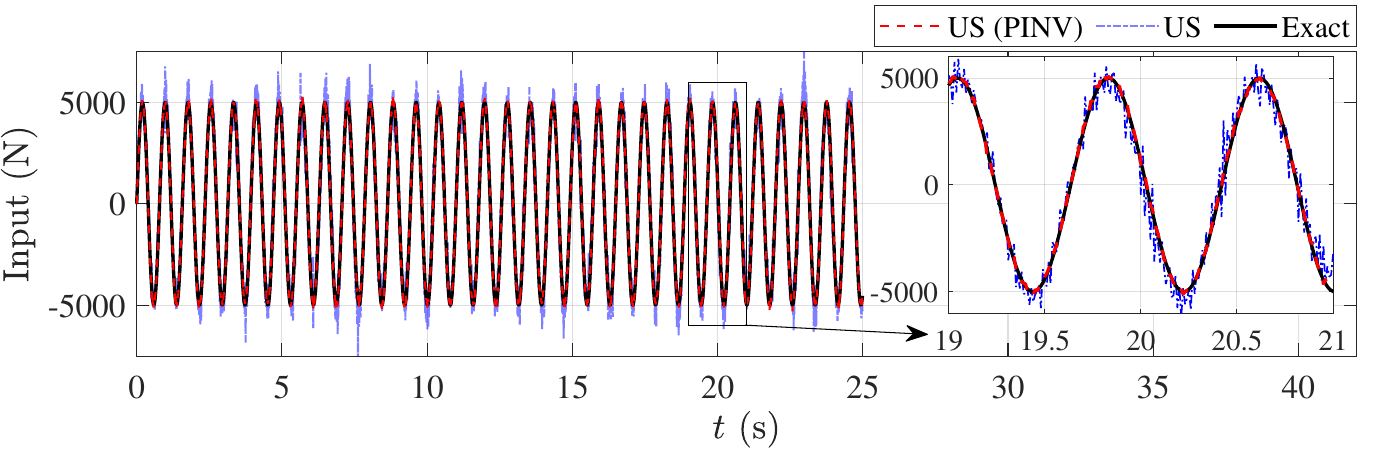}
    \end{subfigure}
    \begin{subfigure}{0.7\textwidth}
        \caption{} \label{fig-ex1-config2-PINV-b}
        \includegraphics[trim={1.8cm 7cm 2.5cm 7cm}, clip, width=\textwidth]{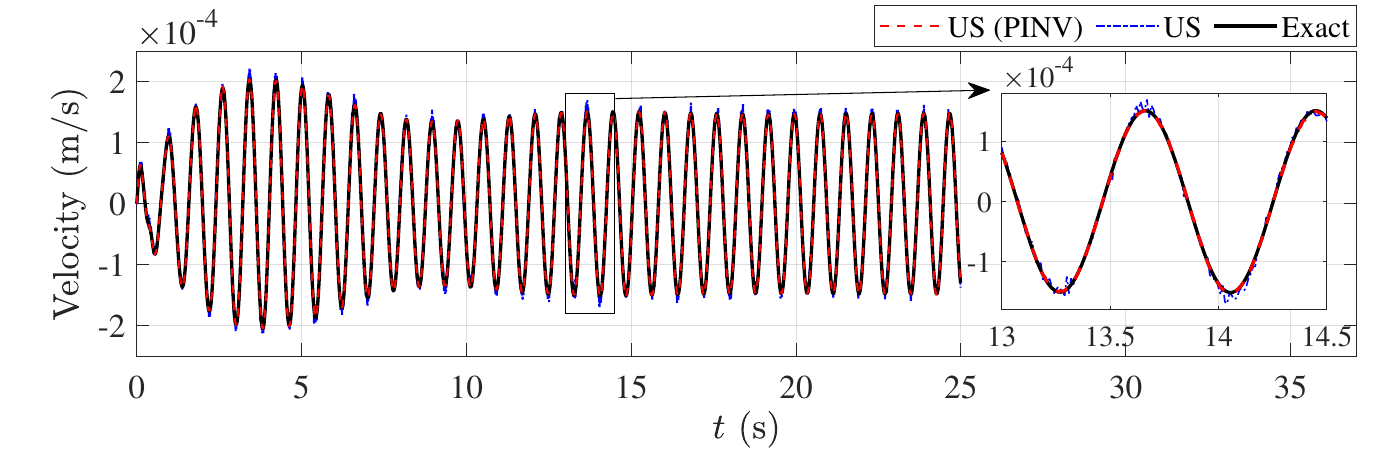}
    \end{subfigure}
    \begin{subfigure}{0.7\textwidth}
        \caption{} \label{fig-ex1-config2-PINV-c}
        \includegraphics[trim={1.8cm 7cm 2.5cm 7cm}, clip, width=\textwidth]{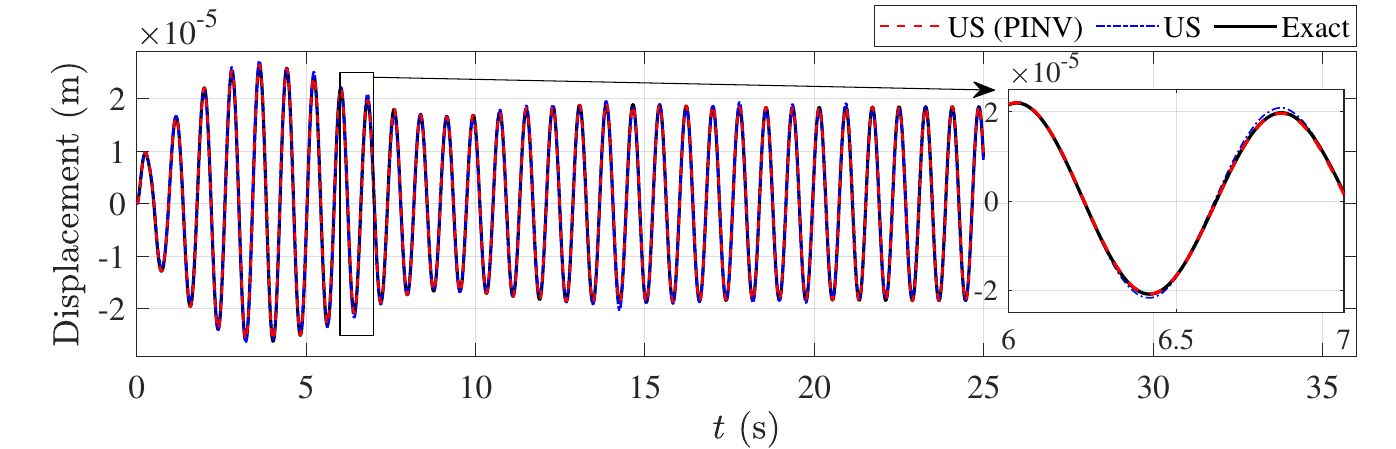}
    \end{subfigure}
    \caption{Configuration 1.2 results: (a) input estimation, (b) velocity estimation of the 2nd floor; (c) displacement estimation of the 2nd floor.}
    \label{fig-ex1-config2-PINV}
\end{figure}

The improvement made to configuration 1.2 is presented in Fig.~\ref{fig-ex1-config2-PINV}. Again, the raised observation noise level impacts the performance of the US when PINV is not in use. The input estimation becomes noisy, fluctuations happen in the velocity estimates, and the peak of the displacement is overestimated. As a result, the overall dimensionless error $\Sigma\delta_{\alpha}$ is $0.41$. In contrast, by using PINV, the estimates perfectly match the exact values, resulting in a lower overall dimensionless error $\Sigma\delta_{\alpha}=0.06$, which is a $85$\% improvement.

\subsection{Eight-storey shear building subjected to ground motion}
\label{section3.2}

\begin{figure}
    \centering
    \includegraphics[trim={3.5cm 9.9cm 4.5cm 10.2cm}, clip, width=0.5\textwidth]{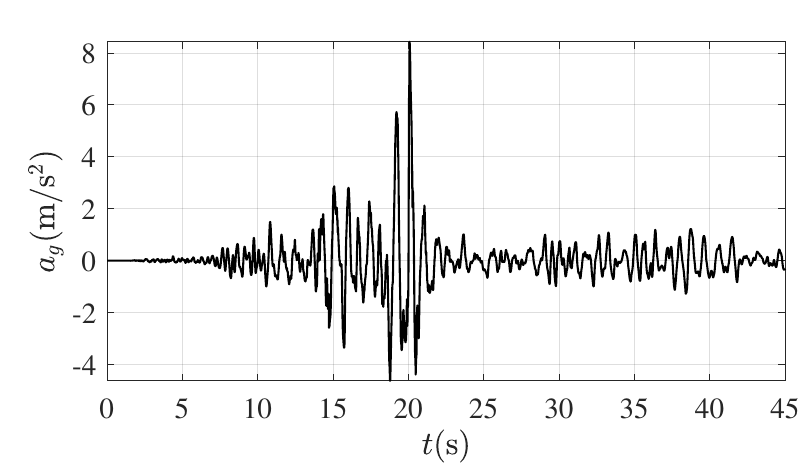}
    \caption{Time history of the 1999 Chi-Chi earthquake ground acceleration}
    \label{fig-ground_acc}
\end{figure}

\noindent In this section, the US is compared with other popular estimators, including AKF, MVU-based filters and smoothing methods. The same eight-storey shear building is adopted to assess the performance of the aforementioned methods focusing on the case of ground motion. The structure is excited by the 1999 Chi-Chi earthquake ground motion -- see the ground acceleration time history represented in Fig.~\ref{fig-ground_acc}. The ground motion is considered unknown during the estimation. The dynamic responses of the structure are obtained by the same matrix exponential method using a full-order model with a duration of $45$~s and a time interval of ${\Delta}t=0.01$~s. The noise level is $5$\% RMS of the dynamic responses for the simulated observation.

For the estimation process, the ROM described in Eqs.~\eqref{eq73} and \eqref{eq74} with the first three natural modes is incorporated to impose nonzero modelling error by creating a mismatch between the model employed in the observer process equation and the model used to generate the observation data. The corresponding natural frequencies are $7.381$, $21.893$, and $35.659$~rad/s. To optimise the estimation accuracy, all hyperparameters involved in the US, AKF and MVU-based methods need to be properly selected. Therefore, the observation error covariance $\mathbf{R}$ is set to the covariance of the simulated added noise; other hyperparameters are tuned thoroughly using comprehensive grid searches case-by-case to minimise the overall dimensionless error $\Sigma\delta_{\alpha}$: the modelling error covariance $\mathbf{Q}$ for all algorithms is assumed to be a diagonal matrix $Q_x\mathbf{I}_n$, and $Q_x$ is searched from $10^{-24}$ to $10^{3}$ with an interval of $10^{0.1}$; the PINV tolerance for the US is searched from $10^{-24}$ to $10^{-1}$ with an interval of $10^{0.1}$; in addition, the AKF requires an input error covariance, which is assumed to be a diagonal matrix $Q_p\mathbf{I}_m$, and $Q_p$ is searched from $10^{-24}$ to $10^{3}$ with an interval of $10^{0.1}$. The grid search results are presented in \ref{appendixA2}.

Considering the feedforward matrix and to assess the universally applicable features of the US, a series of sensor configurations was tested. Table \ref{tab-config-ex2} comprises the types and locations of each sensor network. Configurations 2.1 and 2.2 represent the systems without direct feedthrough, which are very common in computer-vision-based measurements~\cite{Cha2017}. Configuration 2.3 denotes a system with direct feedthrough, and includes displacement and acceleration measurements. Configuration 2.4 is a non-observable system, therefore, the so-called drift effect takes place in the estimated input and displacement~\cite{EftekharAzam2015}. Compared to the sensor placements in Section~\ref{section3.1}, the displacement sensor on the first floor is removed in configurations 2.1 and 2.3, making velocity and acceleration sensors non-collocated with the displacement sensors. It should be mentioned that the MVUF-NDF only works with configurations 2.1 and 2.2, whereas the MVUF-DF can only be applied to configurations 2.3 and 2.4. For each test, the initial state  $\hat{\mathbf{x}}_0$, error covariances $\mathbf{P}_0$, $\mathbf{P}_0^{\mathbf{xw}}$ and $\mathbf{P}_0^{\mathbf{xv}}$ are set to zero. 
\begin{table}[ht!]
    \centering
    \caption{Observation configurations for the eight-storey shear building subjected to ground motion}
    \label{tab-config-ex2}
    \footnotesize{
    \begin{tabular}{c c c c}
    \toprule
     & \multicolumn{3}{c}{Location of sensors} \\
    \cmidrule{2-4}
    Configurations & Displacement & Velocity & Acceleration \\
    \midrule
    2.1 & F3, F5 and F7 & F1 & - \\
    2.2 & F1, F3, F5 and F7 & - & - \\
    2.3 & F3, F5 and F7 & - & F1 \\
    2.4 & - & F4 & F1 \\
    \bottomrule
    \end{tabular}}
\end{table}

\begin{figure}[ht]
    \centering
    \includegraphics[trim={6.5cm 7cm 7cm 7.8cm}, clip, width=0.55\textwidth]{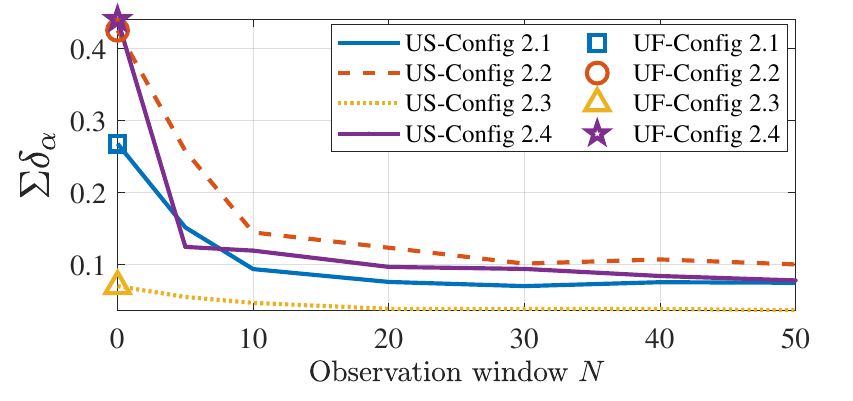}
    \caption{Dimensionless error versus observation window $N$.}
    \label{Ex2_N_vs_error}
\end{figure}

Fig.~\ref{Ex2_N_vs_error} shows the performance of the US with a different observation window $N$ in each configuration. The estimation results of configurations 2.1, 2.2 and 2.4 can worsen if the amount of data is insufficient in the observation window, resulting in a high dimensionless error $\Sigma\delta_{\alpha}$, and the estimation quality is significantly improved by extending the observation window. Configuration 2.3 has the overall best estimation performance due to the contribution of the acceleration observation. Nonetheless, the estimation quality can be further enhanced by the extended observation window. The estimation quality can also improve for configuration 2.4, as the observation window $N$ increases despite being a non-observable system. The results obtained with the UF are also represented in Fig.~\ref{Ex2_N_vs_error}, and they closely match the smoothing algorithm with $N=0$. It is important to denote that, for configuration 2.4, drifts occur in the input and displacement estimates if the observation window is insufficient, regardless of the reduced dimensionless error found in the figure. The performance of the US converges when $N>20$. Therefore, the smoothing results with $N=20$ are selected for comparison. The detailed comparisons are presented next.

\begin{figure}[ht!]
    \centering
    \begin{subfigure}{0.75\textwidth}
        \caption{} \label{fig-ex2-config1-a}
        \includegraphics[trim={2.3cm 7cm 3cm 7cm}, clip, width=\textwidth]{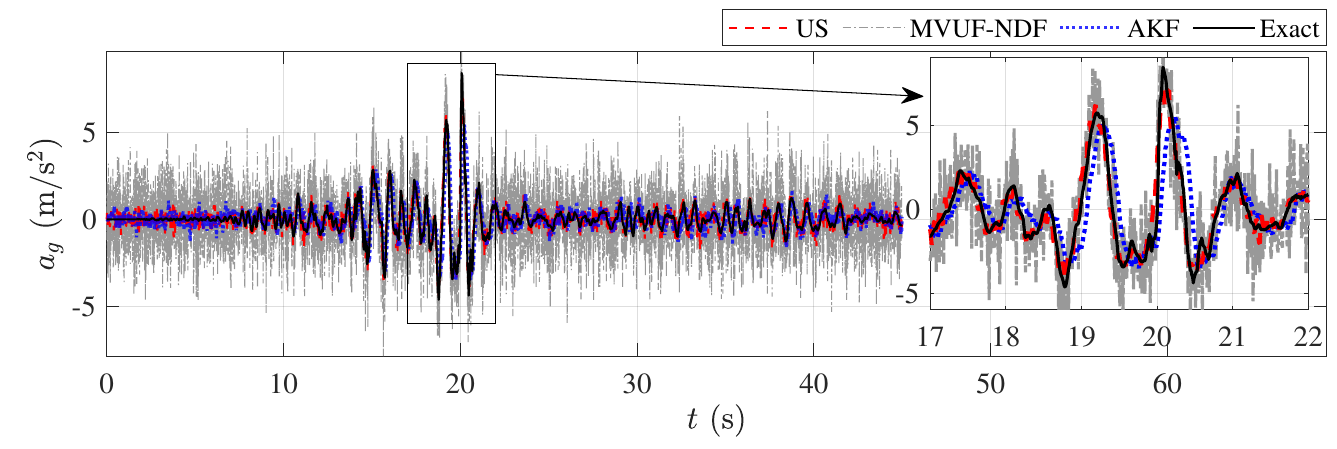}
    \end{subfigure}
    \begin{subfigure}{0.75\textwidth}
        \caption{} \label{fig-ex2-config1-b}
        \includegraphics[trim={2.3cm 7cm 3cm 7cm}, clip, width=\textwidth]{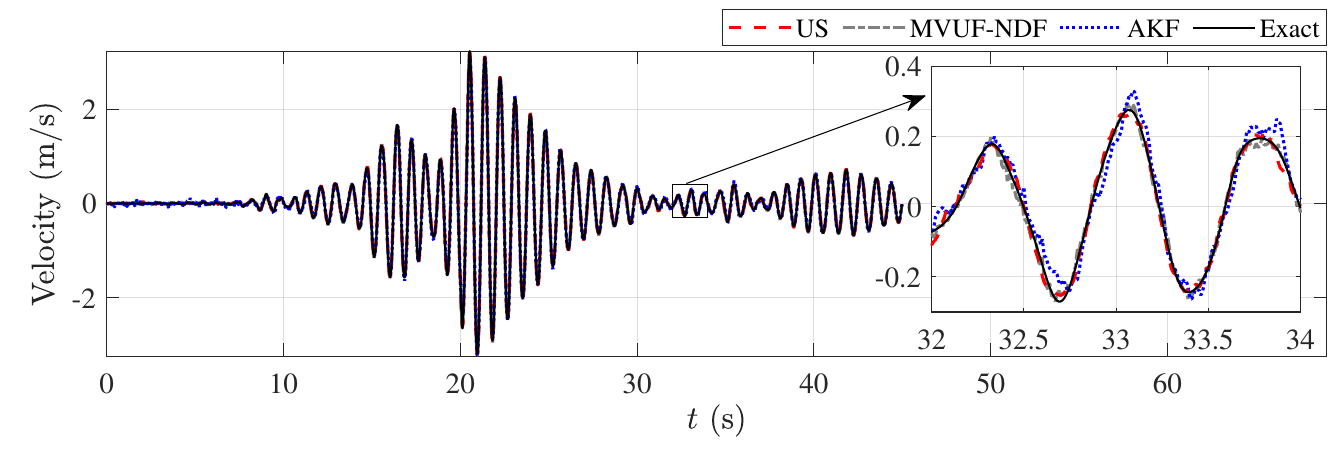}
    \end{subfigure}
    \begin{subfigure}{0.75\textwidth}
        \caption{} \label{fig-ex2-config1-c}
        \includegraphics[trim={2.3cm 7cm 3cm 7cm}, clip, width=\textwidth]{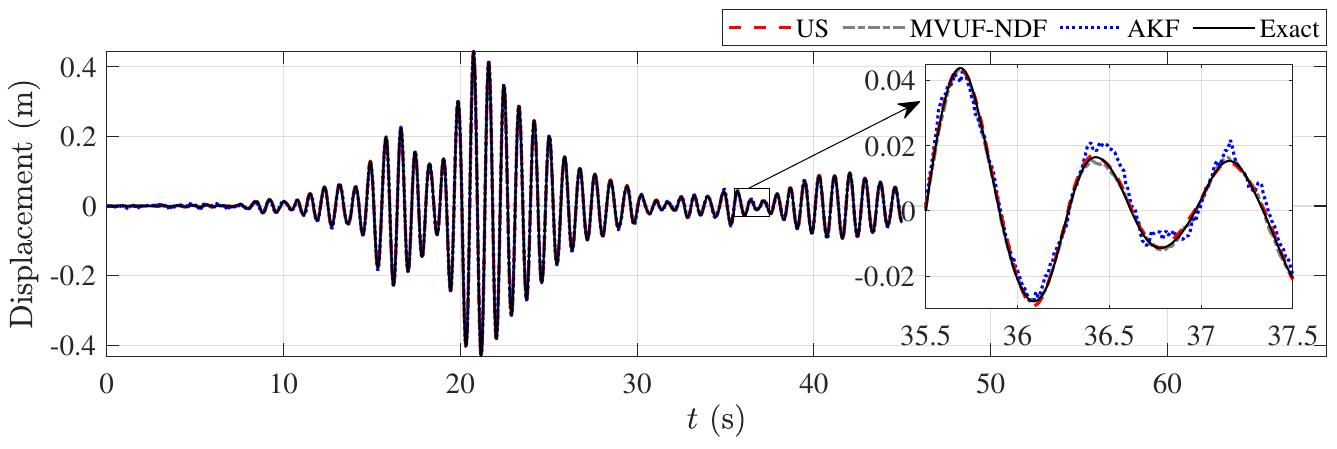}
    \end{subfigure}
    \caption{Configuration 2.1 using the US ($N=20$), MVUF-NDF, and AKF: (a) input estimation; (b) velocity estimation of the 4th floor; and (c) displacement estimation of the 4th floor.}
    \label{fig-ex2-config1}
\end{figure}

The state and input estimates of configurations 2.1 and 2.2 by the US are compared to the MVUF-NDF and the AKF. Fig.~\ref{fig-ex2-config1} shows the input and state estimates of configuration 2.1. In Figs.~\ref{fig-ex2-config1-a}, it is clear that the MVUF-NDF does not provide a quality estimate for the input, whereas the AKF has underestimation and lags; however, by introducing a $20$-timestep observation window, the estimated input by the US has a much improved agreement. The estimated state of the 4th floor is shown in Figs.~\ref{fig-ex2-config1-b} and \ref{fig-ex2-config1-c}; interestingly, although all the algorithms can show an excellent match concerning state estimates, the AKF shows fluctuations at the low-amplitude region in both velocity and displacement estimations. To better present the impact of these fluctuations on the performance of the state estimation, the dimensionless error of state estimates $\Sigma\delta_{x}$ is calculated for the US, MVUF-NDF, and AKF, which are $0.05$, $0.06$ and $0.15$, respectively.

\begin{figure}[ht!]
    \centering
    \begin{subfigure}{0.45\textwidth}
        \caption{} \label{fig-ex2-config2-a}
        \includegraphics[trim={5.1cm 6cm 6.5cm 6.2cm}, clip, width=\textwidth]{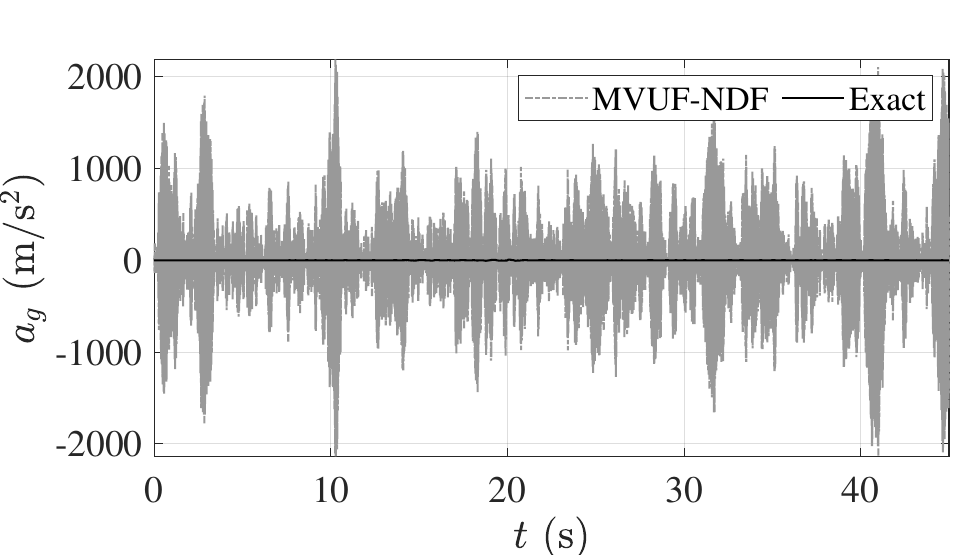}
    \end{subfigure}
    \begin{subfigure}{0.442\textwidth}
        \caption{} \label{fig-ex2-config2-b}
        \includegraphics[trim={5.2cm 6cm 6.5cm 6.2cm}, clip, width=\textwidth]{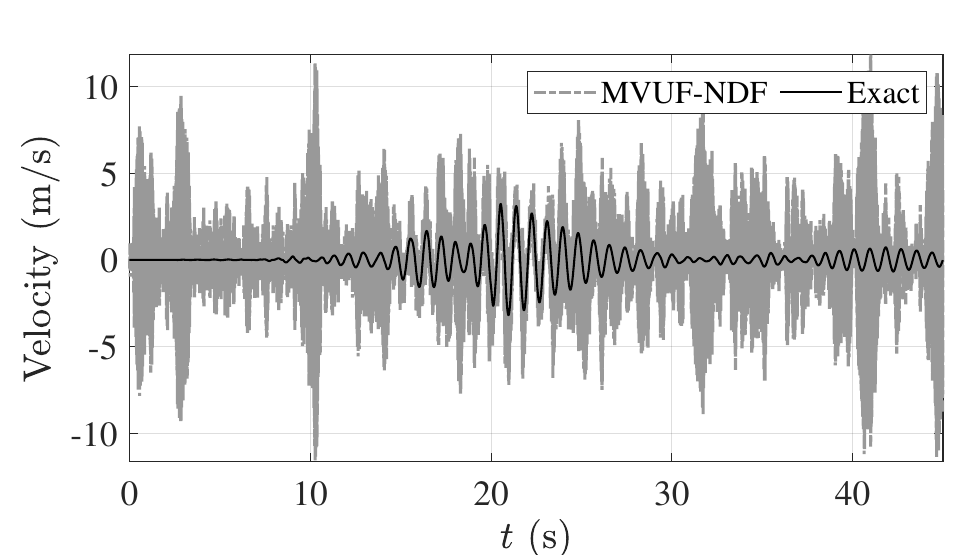}
    \end{subfigure}
    \begin{subfigure}{0.45\textwidth}
        \caption{} \label{fig-ex2-config2-c}
        \includegraphics[trim={5.5cm 5.9cm 6.5cm 6.2cm}, clip, width=\textwidth]{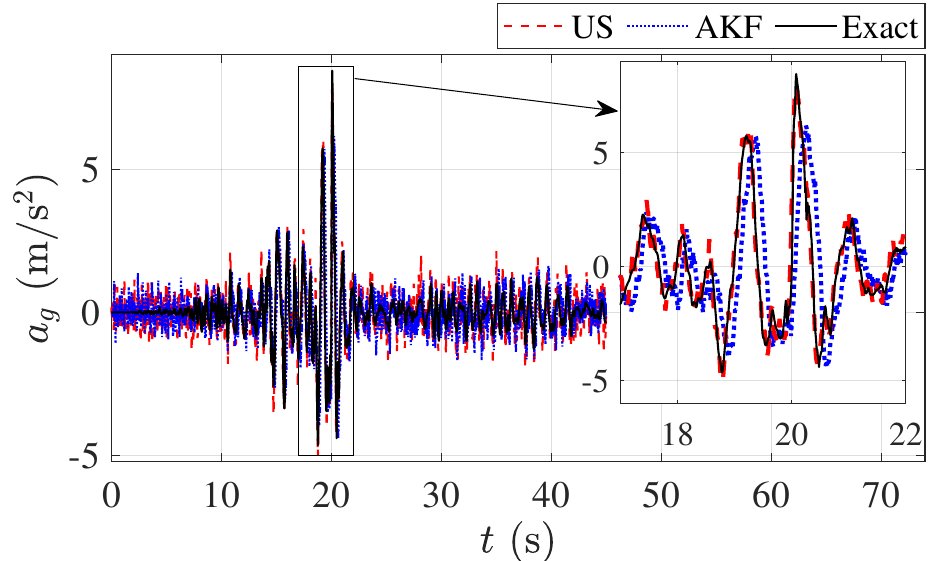}
    \end{subfigure}
    \begin{subfigure}{0.45\textwidth}
        \caption{} \label{fig-ex2-config2-d}
        \includegraphics[trim={5.5cm 5.9cm 6.5cm 6.2cm}, clip, width=\textwidth]{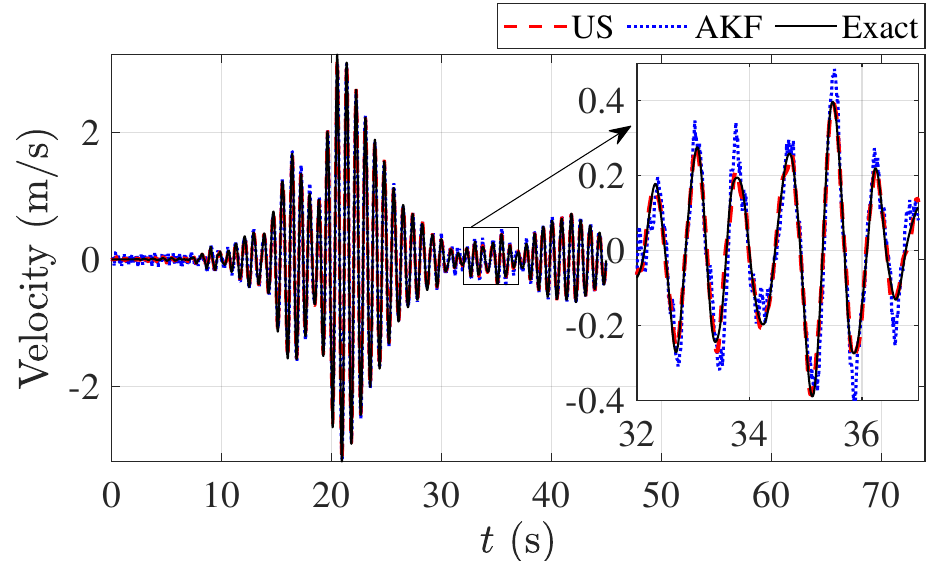}
    \end{subfigure}
    \begin{subfigure}{0.75\textwidth}
        \caption{} \label{fig-ex2-config2-e}
        \includegraphics[trim={2.3cm 7cm 3cm 7cm}, clip, width=\textwidth]{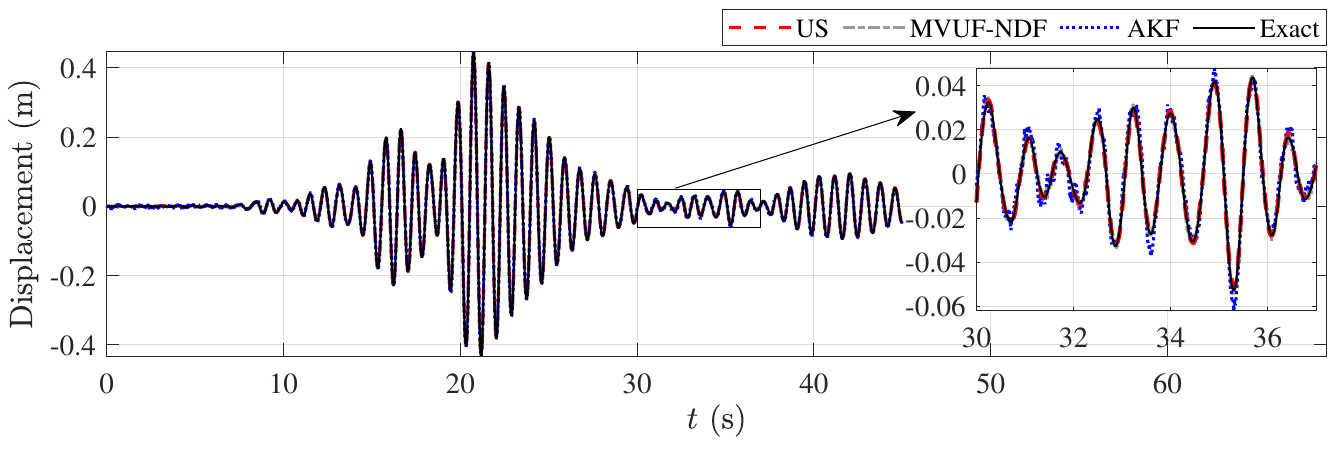}
    \end{subfigure}
    \caption{Configuration 2.2: (a) MVUF-NDF input estimation, (b) MVUF-NDF velocity estimation of the 4th floor; (c) US ($N=20$) and AKF input estimation, (d) US ($N=20$) and AKF velocity estimation of the 4th floor, and (e) displacement estimation of the 4th floor using the US ($N=20$), MVUF-NDF, and AKF.}
    \label{fig-ex2-config2}
\end{figure}

Configuration 2.2, which is a displacement-only observation, represents a case commonly available for industrial applications since computer vision-based structural health monitoring typically provides displacement measurements~\cite{Yu2020}. Indeed, the development of an observer algorithm that can effectively process a limited number of displacement-only measurements could have high practicality. In configuration 2.2, however, the MVUF-NDF is ineffective given that velocity and acceleration observations are not available -- see Figs.~\ref{fig-ex2-config2-a} and \ref{fig-ex2-config2-b} -- in which case the input and velocity of the structure cannot be estimated. Figs.~\ref{fig-ex2-config2-c} and \ref{fig-ex2-config2-d} show that the AKF has underestimation and lags in the input estimation, whereas the velocity is overestimated. In contrast, the US significantly improves the estimation quality of both input and velocity. Furthermore, although the displacement estimation performance of the MVUF-NDF is found to be comparable to the US, the latter can still achieve further enhancement. This can be seen by the dimensionless error $\Sigma\delta_{u}$ for the displacement which is $0.02$ in contrast with $0.03$ for the MVUF-NDF, and $0.07$ for the AKF.
\clearpage

\begin{figure}[ht!]
    \centering
    \begin{subfigure}{0.75\textwidth}
        \caption{} \label{fig-ex2-config3-a}
        \includegraphics[trim={2.3cm 7cm 3cm 7cm}, clip, width=\textwidth]{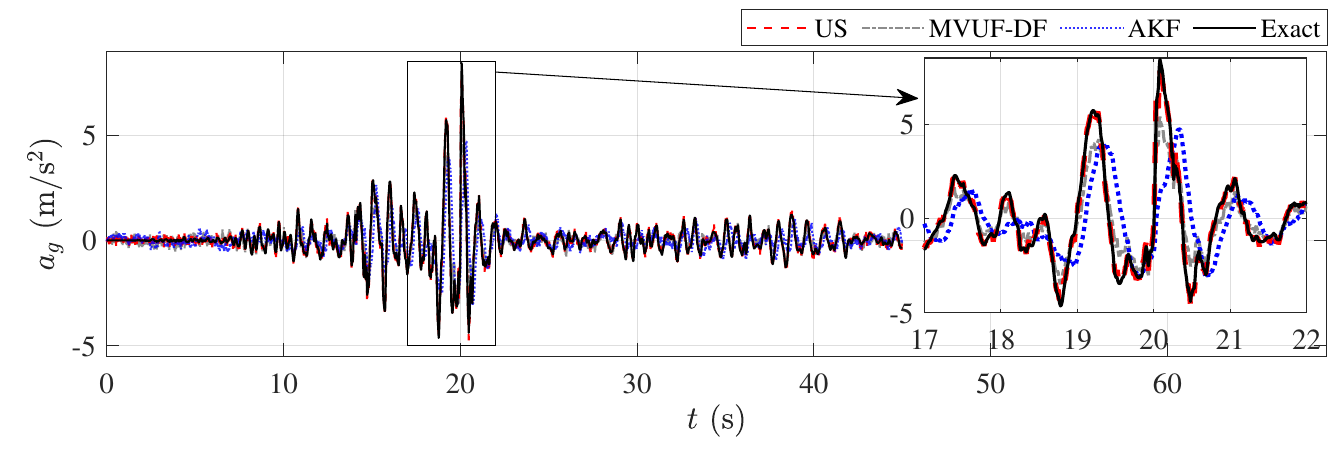}
    \end{subfigure}
    \begin{subfigure}{0.75\textwidth}
        \caption{} \label{fig-ex2-config3-b}
        \includegraphics[trim={2.3cm 7cm 3cm 7cm}, clip, width=\textwidth]{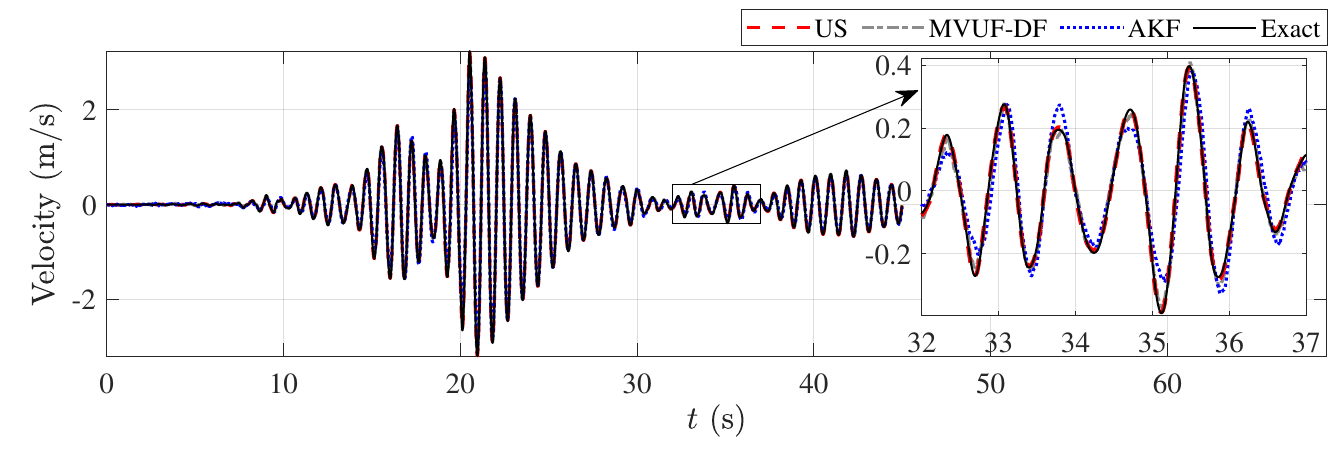}
    \end{subfigure}
    \begin{subfigure}{0.75\textwidth}
        \caption{} \label{fig-ex2-config3-c}
        \includegraphics[trim={2.3cm 7cm 3cm 7cm}, clip, width=\textwidth]{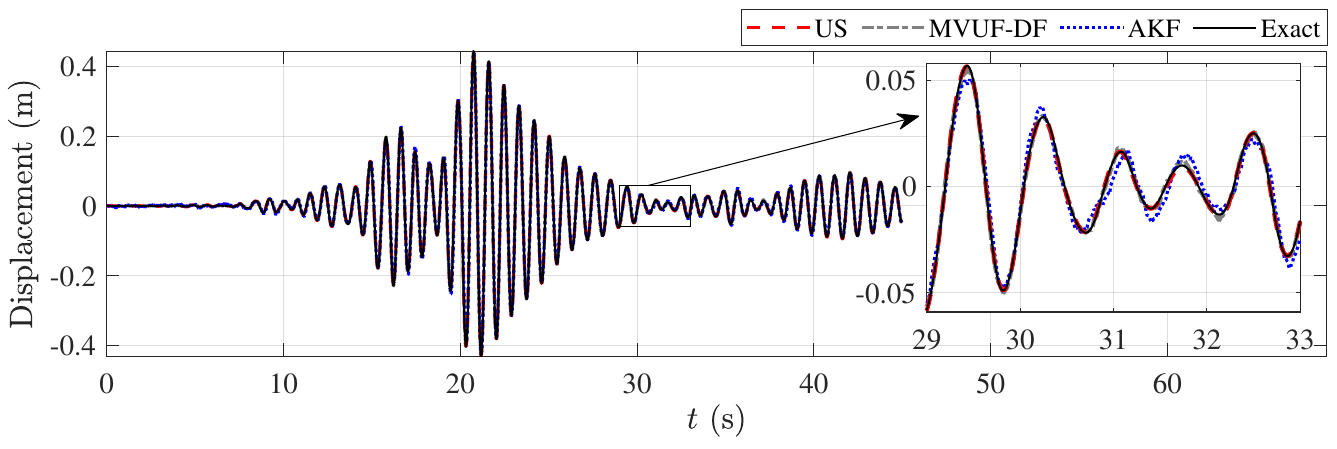}
    \end{subfigure}
    \caption{Configuration 2.3 results using the US ($N=20$), MVUF-DF, and AKF: (a) input estimation, (b) velocity estimation of the 4th floor, and (c) displacement estimation of the 4th floor.}
    \label{fig-ex2-config3}
\end{figure}

For configurations 2.3 and 2.4, the smoothing results with $N=20$ are compared with the MVUF-DF and AKF. As shown in Fig.~\ref{fig-ex2-config3}, with the contribution of the direct feedthrough of the input to the observation, the MVUF-DF reaches a better performance than the one found in configuration 2.1. As illustrated in Fig.~\ref{fig-ex2-config3-a}, although the MVUF-DF yields improved estimation results, the input is still underestimated. Similar to the previous configurations, the AKF not only underestimates and lags the input but also exhibits fluctuations in the estimated state. However, if the observation window is extended to $20$ timesteps, the US yields improved estimates. The overall dimensionless error $\Sigma\delta_{\alpha}$ is $0.04$ for the US, $0.12$ for the MVUF-DF, and $0.35$ for the AKF, which confirms a 66.7\% and 88.6\% reduction in the estimation errors, respectively.

\begin{figure}[ht!]
    \centering
    \begin{subfigure}{0.45\textwidth}
        \caption{} \label{fig-ex2-config4-a}
        \includegraphics[trim={5.5cm 5.9cm 6.5cm 6.2cm}, clip, width=\textwidth]{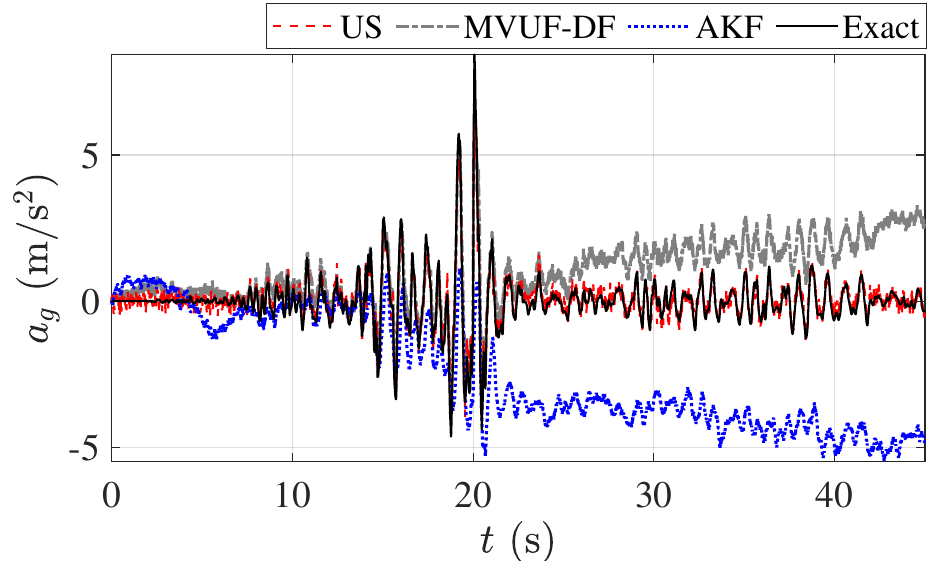}
    \end{subfigure}
    \begin{subfigure}{0.45\textwidth}
        \caption{} \label{fig-ex2-config4-b}
        \includegraphics[trim={5.5cm 5.9cm 6.5cm 6.2cm}, clip, width=\textwidth]{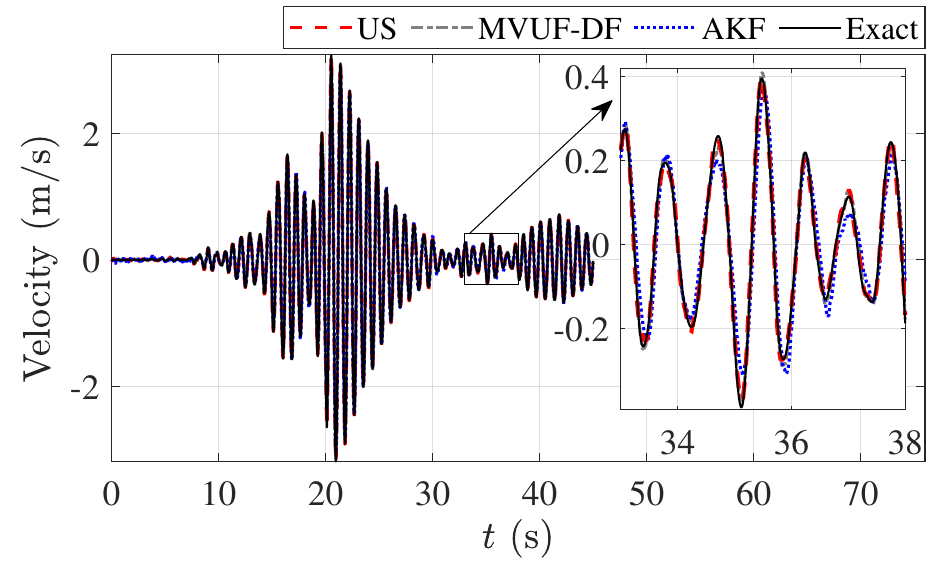}
    \end{subfigure}
    \begin{subfigure}{0.45\textwidth}
        \caption{} \label{fig-ex2-config4-c}
        \includegraphics[trim={5.4cm 6cm 6.5cm 6.2cm}, clip, width=\textwidth]{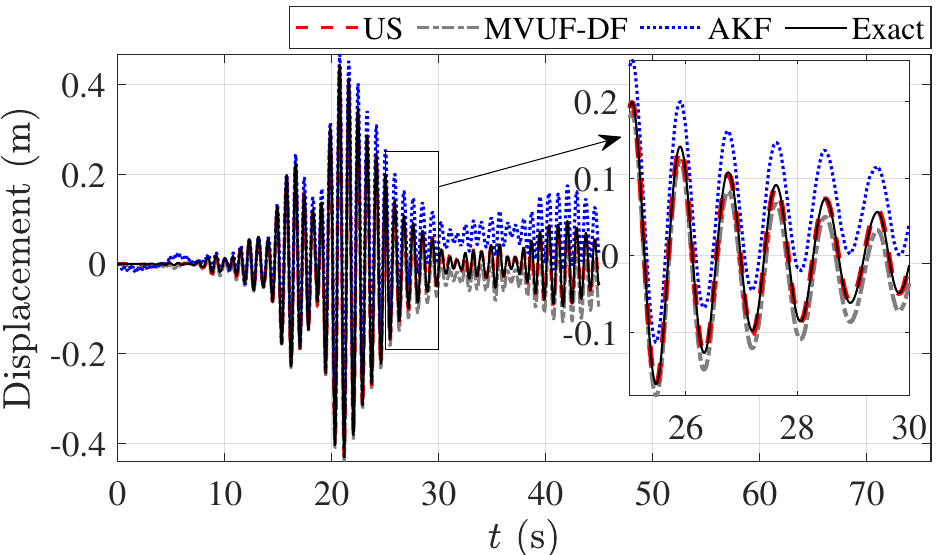}
    \end{subfigure}
    \begin{subfigure}{0.45\textwidth}
        \caption{} \label{fig-ex2-config4-d}
        \includegraphics[trim={5.4cm 6cm 6.5cm 6.2cm}, clip, width=\textwidth]{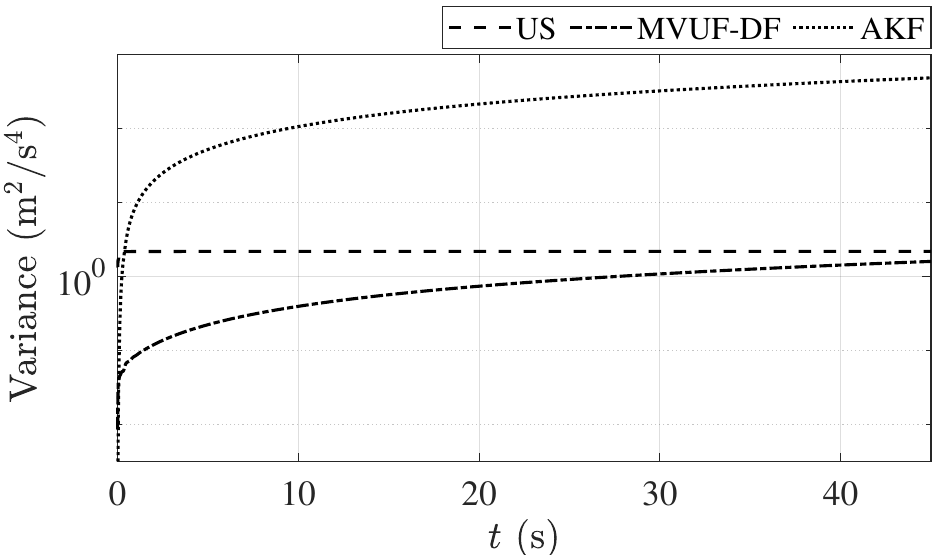}
    \end{subfigure}
    \caption{Configuration 2.4 results using the US ($N=20$), MVUF-DF, and AKF: (a) input estimation, (b) velocity estimation of the 2nd floor, (c) displacement estimation of the 2nd floor, and (d) the error variance of input estimation.}
    \label{fig-ex2-config4}
\end{figure}

The comparison of estimation results for configuration 2.4 is shown in Fig.~\ref{fig-ex2-config4}. The MVUF-DF and AKF become unstable under this sensor arrangement, and the estimated input and displacement drift despite a satisfactory velocity estimation. Unlike the comparative filters, by including extra information in the estimation process, the US can stably estimate the input and state. A stable velocity estimation is also found with the MVUF-DF and AKF, although the US can reach higher accuracy. This is evident in the lower dimensionless error $\Sigma\delta_{\dot{u}}$ for velocity estimation. The US has an error of $0.03$, in comparison to $0.04$ for the MVUF-DF and $0.15$ for the AKF. In Fig.~\ref{fig-ex2-config4-d}, the development of input error variances shows that the US exhibits much higher stability compared with the MVUF-DF and AKF.

\subsection{The Taipei 101 Tower}
\label{section3.3}
\begin{figure}[ht]
    \centering
    \begin{subfigure}{0.24\textwidth}
        \includegraphics[trim={1.8cm 0cm 1.8cm 0cm}, clip, width=\textwidth]{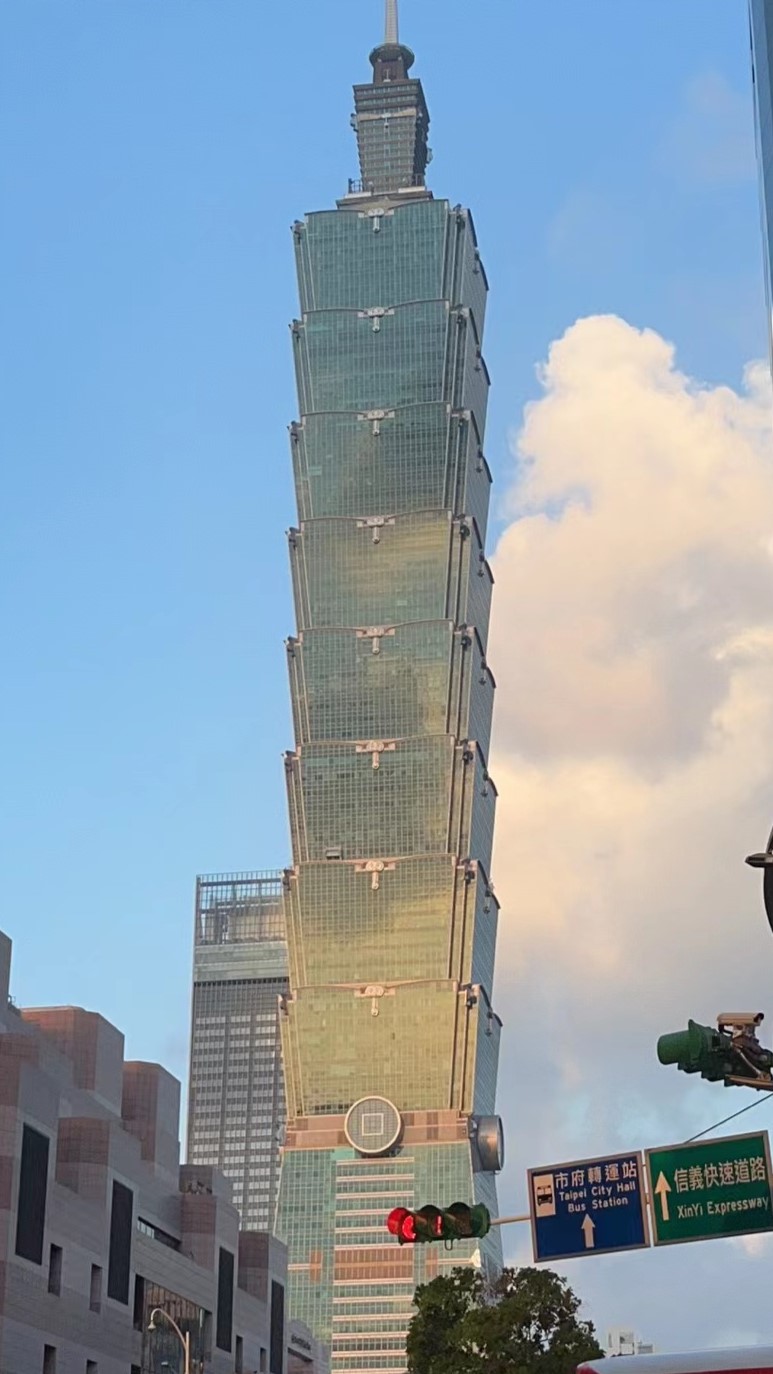}
        \caption{} \label{taipei101}
    \end{subfigure}
    \begin{subfigure}{0.24\textwidth}
        \includegraphics[trim={5cm 2.3cm 5cm 1cm}, clip, width=\textwidth]{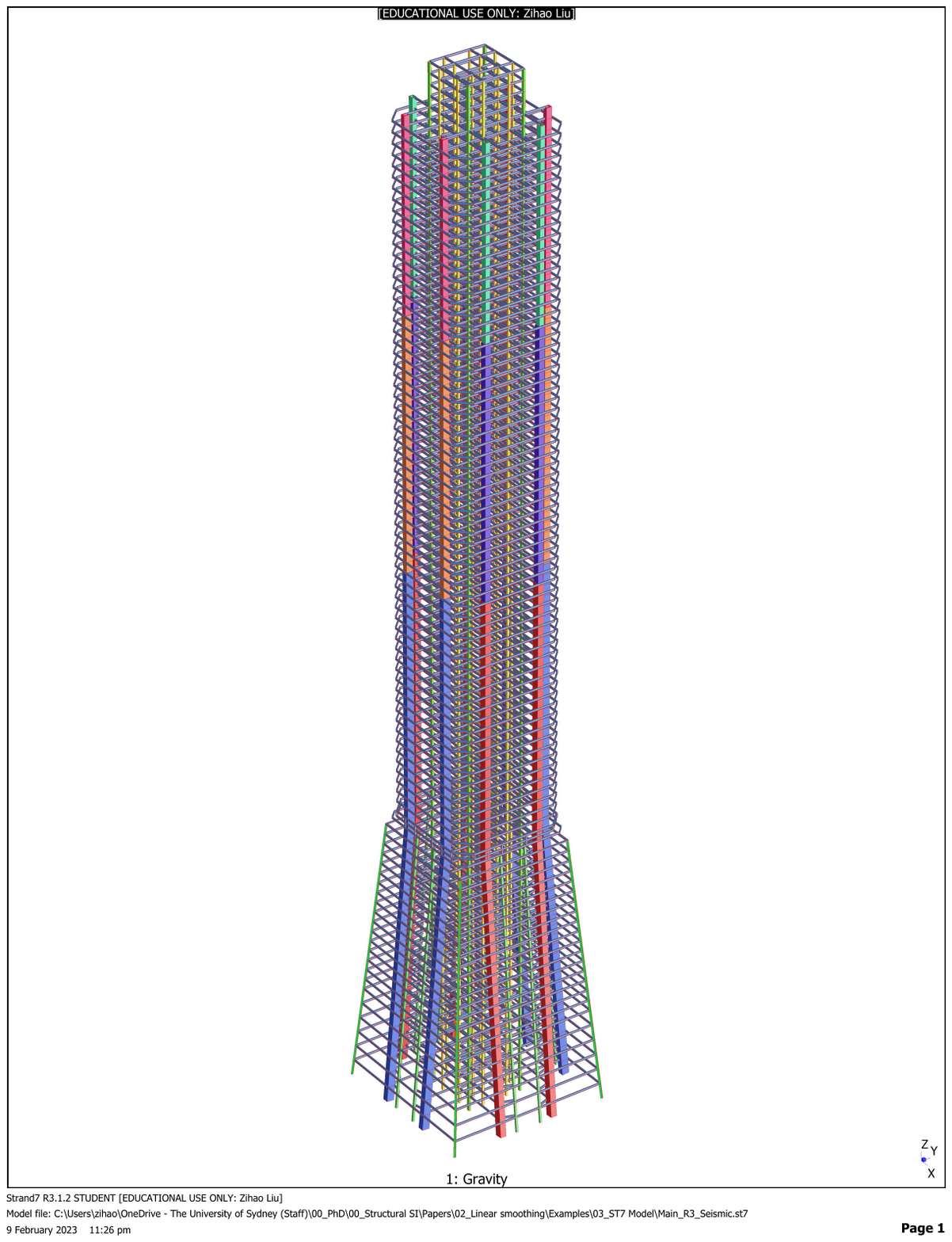}
        \caption{} \label{FEM}
    \end{subfigure}
    \caption{(a) Taipei 101 tower in Taiwan; (b) the finite element model}
    \label{fig-ex3-intro}
\end{figure}

\noindent A high-rise building -- the Taipei World Financial Centre, also known as the Taipei 101 tower -- is shown in Fig.~\ref{taipei101}. The tower has 92 floors and a 9-storey telecommunication tower at the top. There are sixteen columns within a braced core, and eight composite steel-concrete super-columns located on the perimeter. Braced frames are designed in conjunction with the super-columns to resist lateral loads, including wind and seismic forces. For a more detailed analysis of the structural components of the building, readers can refer to the study conducted by \citet{Kourakis2007}. 

The 3-D finite element model in Fig.~\ref{FEM} is deployed to generate the dynamic responses. All the nodes are constrained in the Y- and Z-directions, whereas the Rayleigh damping with factors $\alpha=\beta=0.01$ are taken in the dynamic analysis. An arbitrary excitation is applied at a randomly selected column on the 25th floor, and the load is considered to be unknown during the estimation. To maintain a manageable computational cost despite the dimension and complexity of the structural system, the ROM described in Section~\ref{section2.7} is adopted. The first fifty natural modes, ranging from $9.981$ rad/s up to $30.633$ rad/s, are used to generate the dynamic responses under arbitrary load. These results are then contaminated with a zero-mean Gaussian white noise with a level of 5\% RMS of the dynamic responses to simulate the observation data. 

To assess the performance of the US on a large-scale structural system with a limited number of sensors, two configurations are selected and listed in Table \ref{tab-config-ex3}. Configuration 3.1 represents a typical setup for the system with no direct feedthrough, whereas configuration 3.2 represents a system with direct feedthrough. For each configuration, sensors are set at arbitrary nodes on their designated floor. The loaded floor is not directly measured, and the number of measured nodes is much lower than the total nodes, which is $10$ nodes compared to a total of $5,420$ nodes. The estimation results obtained using the US are then compared with those obtained using the UF and MVUS methods. It should be denoted that the MVUS-NDF is only applicable to configuration 3.1, while the MVUS-DF is only applicable to configuration 3.2.

\begin{table}[h!]
    \centering
    \small
    \caption{Observation configurations for Taipei 101}
    \label{tab-config-ex3}
    \begin{tabular}{c c c c}
    \toprule
     & \multicolumn{3}{c}{Location of sensors} \\
    \cmidrule{2-4}
    Configurations & Displacement & Velocity & Acceleration \\
    \midrule
    3.1 & F12, F24, F36, F48, F60 F72 and F84 & F30, F54 and F78 & - \\
    3.2 & F12, F24, F36, F48, F60 F72 and F84 & F30 and F78 & F54 \\
    \bottomrule
    \end{tabular}
\end{table}

To establish the estimation process, the ROM with forty natural modes is employed to introduce modelling errors. The same algorithm tuning and grid search strategy is adopted for this example: the observation error covariance $\mathbf{R}$ is set to the covariance of the simulated observation noise; the modelling error covariance is assumed to be $\mathbf{Q}=Q_x\mathbf{I}_n$ for all algorithms and $Q_x$ is searched from $10^{-24}$ to $10^{3}$ with an interval of $10^{0.1}$; the PINV tolerance for the US are searched from $10^{-24}$ to $10^{-1}$ with an interval of $10^{0.1}$. The tuning results can be found in \ref{appendixA3}. Other initialisations include $\hat{\mathbf{x}}_0=\mathbf{0}$, and $\mathbf{P}_0=\mathbf{P}_0^{\mathbf{xw}}=\mathbf{P}_0^{\mathbf{xv}}=\mathbf{0}$.

\begin{figure}[ht!]
    \centering
    \includegraphics[trim={6.5cm 5.2cm 7.5cm 6.5cm}, clip, width=0.45\textwidth]{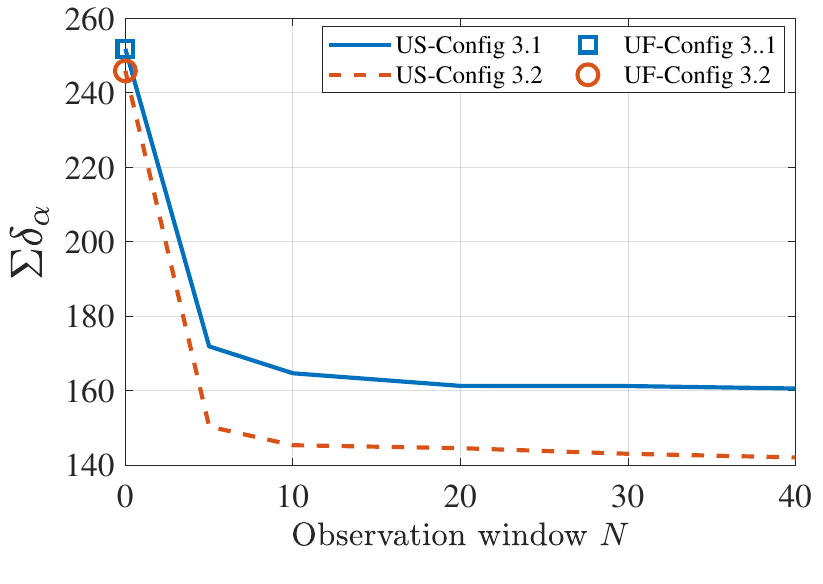}
    \caption{Dimensionless error versus observation window $N$ for the Taipei 101 example.}
    \label{fig-ex3-N_vs_error}
\end{figure}

A parametric study for the observation window is shown in Fig.~\ref{fig-ex3-N_vs_error}. The estimation quality can be seen to improve when the observation window is extended to coverage that can include sufficient information in the smoothing process. Smoothing results with $N=20$ are next selected for further analyses.

\begin{figure}[ht!]
    \centering
    \begin{subfigure}{0.8\textwidth}
        \caption{} \label{fig-ex3-config1-a}
        \includegraphics[trim={2cm 7cm 2.5cm 7cm}, clip, width=\textwidth]{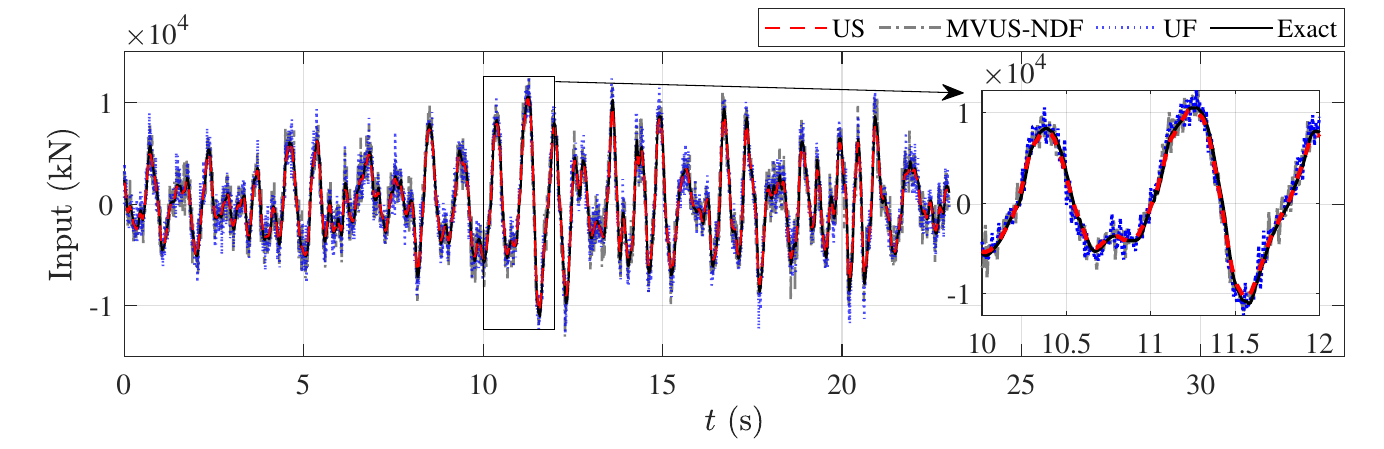}
    \end{subfigure}
    \begin{subfigure}{0.8\textwidth}
        \caption{} \label{fig-ex3-config1-b}
        \includegraphics[trim={2cm 7cm 2.5cm 7cm}, clip, width=\textwidth]{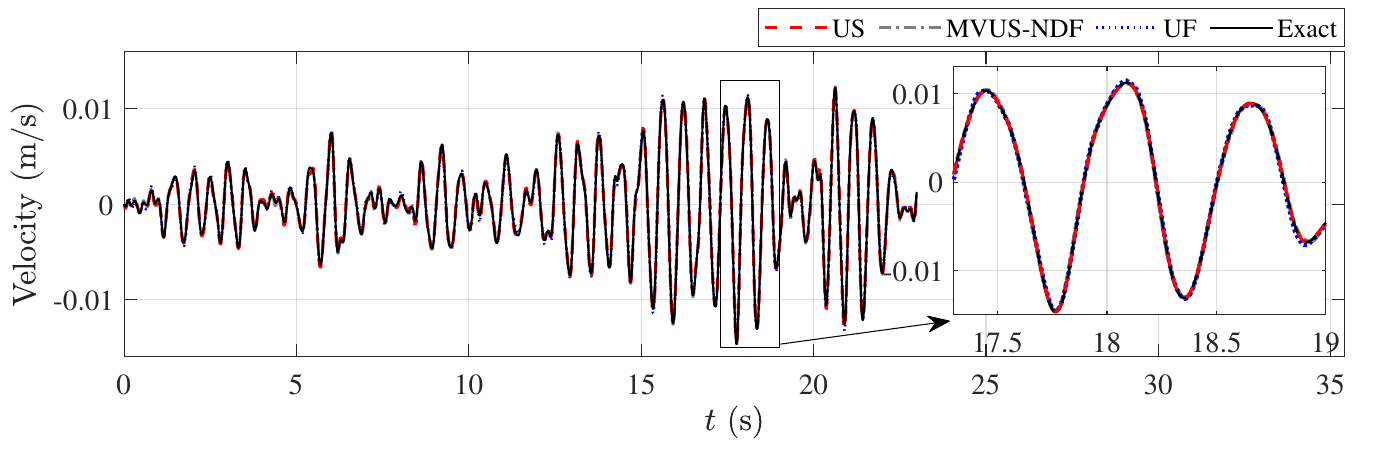}
    \end{subfigure}
    \begin{subfigure}{0.8\textwidth}
        \caption{} \label{fig-ex3-config1-c}
        \includegraphics[trim={2cm 7cm 2.5cm 7cm}, clip, width=\textwidth]{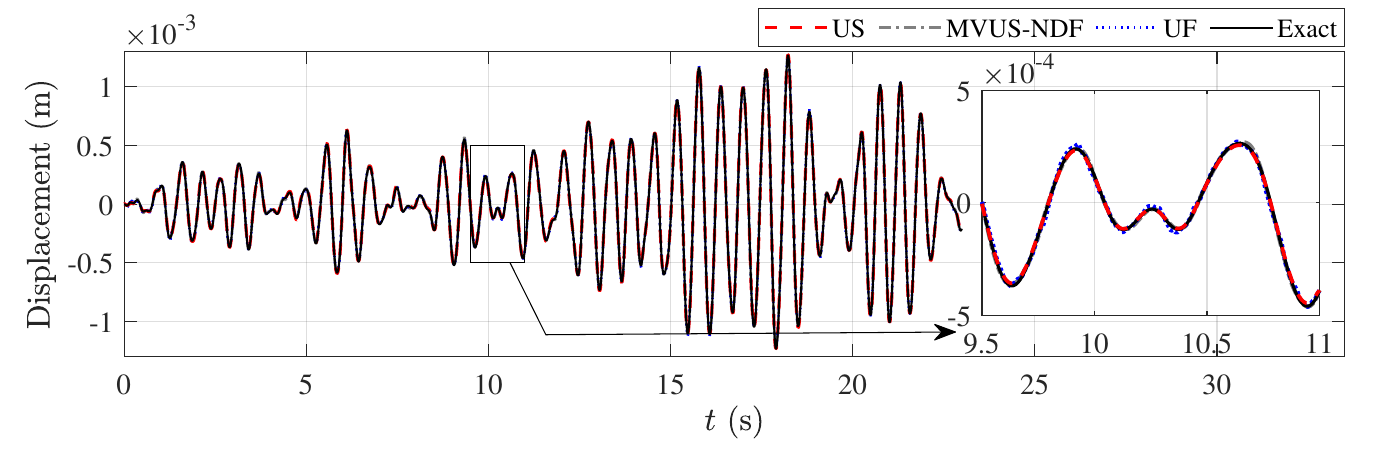}
    \end{subfigure}
    \caption{Configuration 3.1 results using the US ($N=20$), MVUS-NDF ($N=20$), and UF: (a) input estimation; (b) velocity estimation at an arbitrarily selected unmeasured node; and (c) displacement estimation at an arbitrarily selected unmeasured node.}
    \label{fig-ex3-config1}
\end{figure}

Fig.~\ref{fig-ex3-config1} compares the estimation obtained with configuration 3.1 for the US, the UF, and the MVUS-NDF. Since the loaded floor ($25$th floor) is not directly measured, the noise in the input estimation is amplified by the UF and the MVUS-NDF, as shown in Fig.~\ref{fig-ex3-config1-a}. In contrast, the US demonstrates a good agreement for the estimated input. Figs.~\ref{fig-ex3-config1-b} and \ref{fig-ex3-config1-c} display the state estimation results. The non-collocated observations also impact the displacement estimation of the UF, causing fluctuations in the estimates. Despite both the US and the MVUS-NDF can reduce the prediction uncertainty caused by the noncollected observations, the US still has a better performance than the MVUS-NDF. This results in a lower dimensionless error of state estimation which is $\Sigma\delta_{\alpha}=161.3$ when compared to $\Sigma\delta_{\alpha}=251.8$ for the MVUS-NDF, and $231.6$ for the UF. This exhibits a $36$\% and $30$\% improvement compared to the MVUS-NDF and the UF, respectively.
\clearpage

\begin{figure}[ht!]
    \centering
    \begin{subfigure}{0.85\textwidth}
        \caption{} \label{fig-ex3-config2-a}
        \includegraphics[trim={2cm 7cm 2.5cm 7cm}, clip, width=\textwidth]{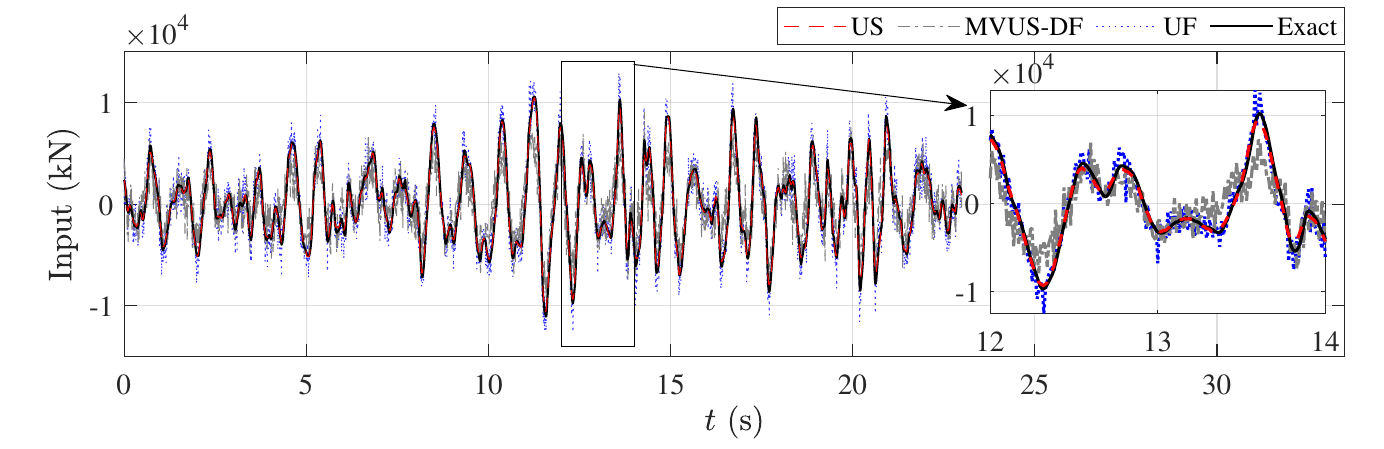}
    \end{subfigure}
    \begin{subfigure}{0.85\textwidth}
        \caption{} \label{fig-ex3-config2-b}
        \includegraphics[trim={2cm 7cm 2.5cm 7cm}, clip, width=\textwidth]{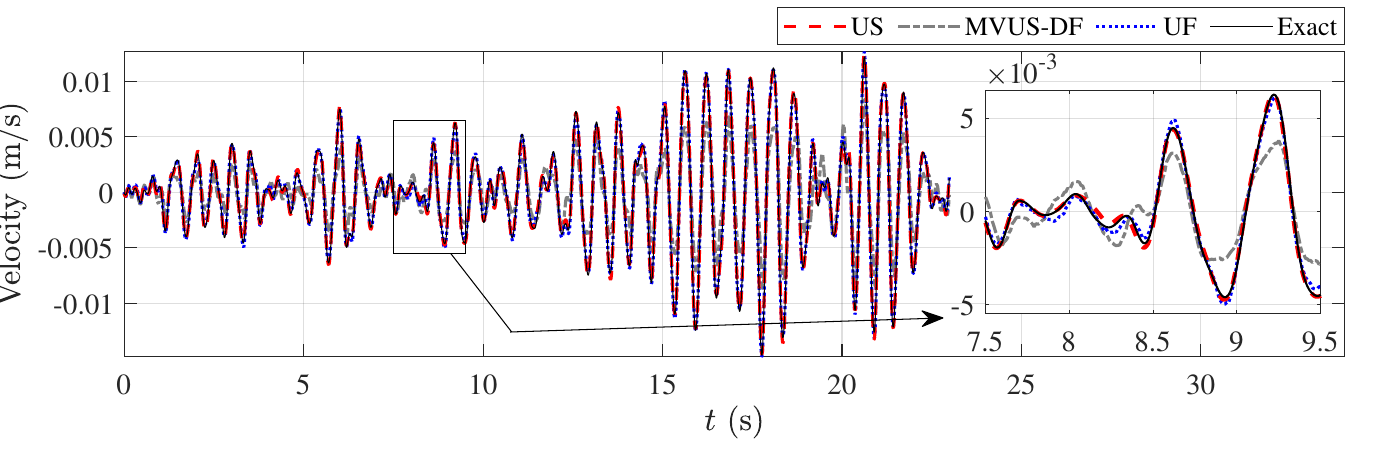}
    \end{subfigure}
    \begin{subfigure}{0.85\textwidth}
        \caption{} \label{fig-ex3-config2-c}
        \includegraphics[trim={2cm 7cm 2.5cm 7cm}, clip, width=\textwidth]{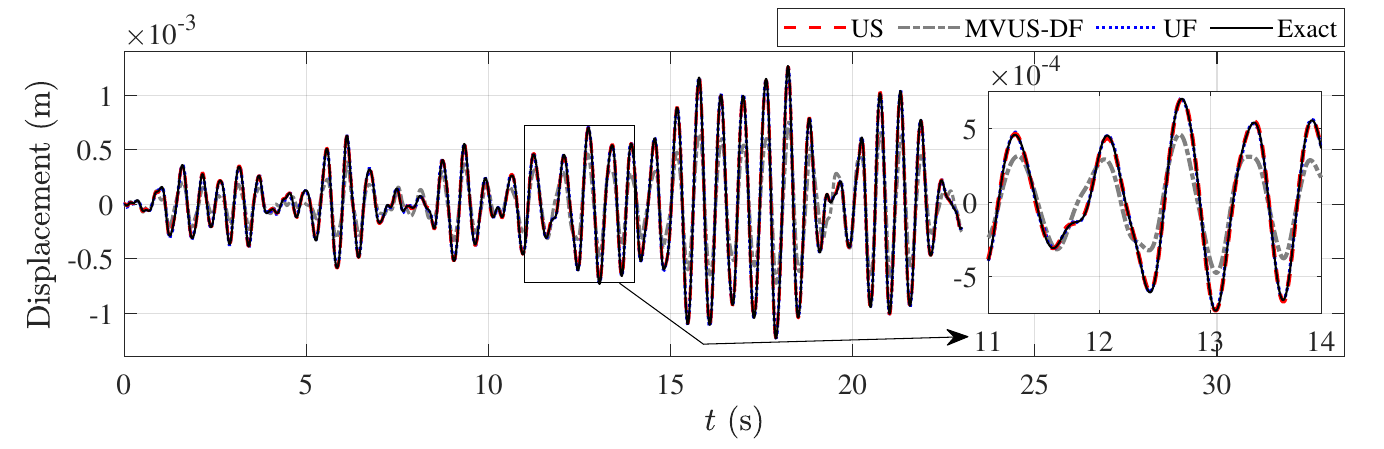}
    \end{subfigure}
    \caption{Configuration 3.2 results using the US ($N=20$), MVUS-DF ($N=20$), and UF: (a) input estimation; (b) velocity estimation at an arbitrarily selected unmeasured node; and (c) displacement estimation at an arbitrarily selected unmeasured node.}
    \label{fig-ex3-config2}
\end{figure}

The estimation results of configuration 3.2 obtained with the US are compared with the UF and MVUS-DF. As noticeable in Fig.~\ref{fig-ex3-config2}, the MVUS-DF benefits from the direct feedthrough of the input, thus resulting in an improved input estimation; However, the MVUS-DF still underestimates both the input and state. In addition, similar to configuration 1, the estimated input is impacted by the non-collocated sensor network for the UF despite the satisfactory state estimation. In contrast, the US keeps a consistent estimation performance, both input and state estimations show a good agreement with the true values. The performance of the US is evidenced by the overall dimensionless error $\Sigma\delta_{\alpha}=145.1$ compared with $\Sigma\delta_{\alpha}=1460.5$ and $245.9$ for the MVUS-DF and UF, respectively.

\subsection{Rank-deficient feedforward}
\label{section3.4}
\begin{figure}[ht]
    \centering
    \includegraphics[trim={4cm 10.8cm 4cm 11cm}, clip, width=0.55\textwidth]{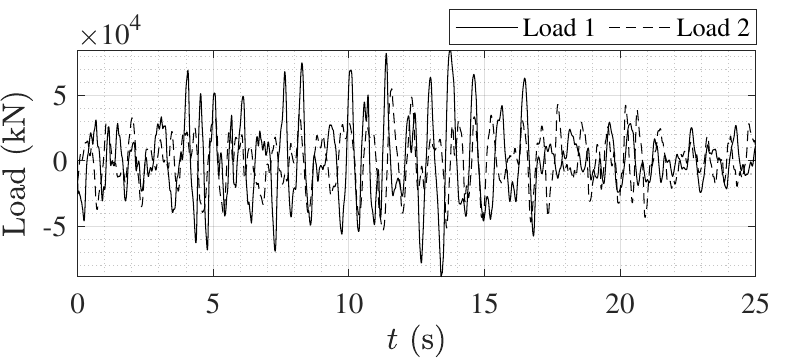}
    \caption{Time history of the two loads used in the rank-deficient feedforward example.}
    \label{fig-ex4_Loads}
\end{figure}

\noindent As discussed in Section~\ref{section1}, the MVUF-DF and the MVUS-DF are only applicable to systems with a full-rank feedforward matrix, i.e., the number of acceleration observations has to be no less than the number of unknown inputs. This restriction can be dropped in the US. To demonstrate the estimation performance of the US with a rank-deficient feedforward matrix, the numerical model of Taipei 101 is loaded with two arbitrary excitations acting simultaneously, as shown in Fig.~\ref{fig-ex4_Loads}; one is located on an arbitrarily selected column on the 20th floor, and the other one is located on an arbitrarily selected column on the 25th floor. Configuration 3.2 defined in Table \ref{tab-config-ex3} is used herein, and other initialisations remain the same as the example in Section~\ref{section3.3}. 

\begin{figure}[!ht]
    \centering
    \begin{subfigure}{0.75\textwidth}
        \caption{} \label{fig-ex4-a}
        \includegraphics[trim={1.8cm 7cm 2.5cm 7cm}, clip, width=\textwidth]{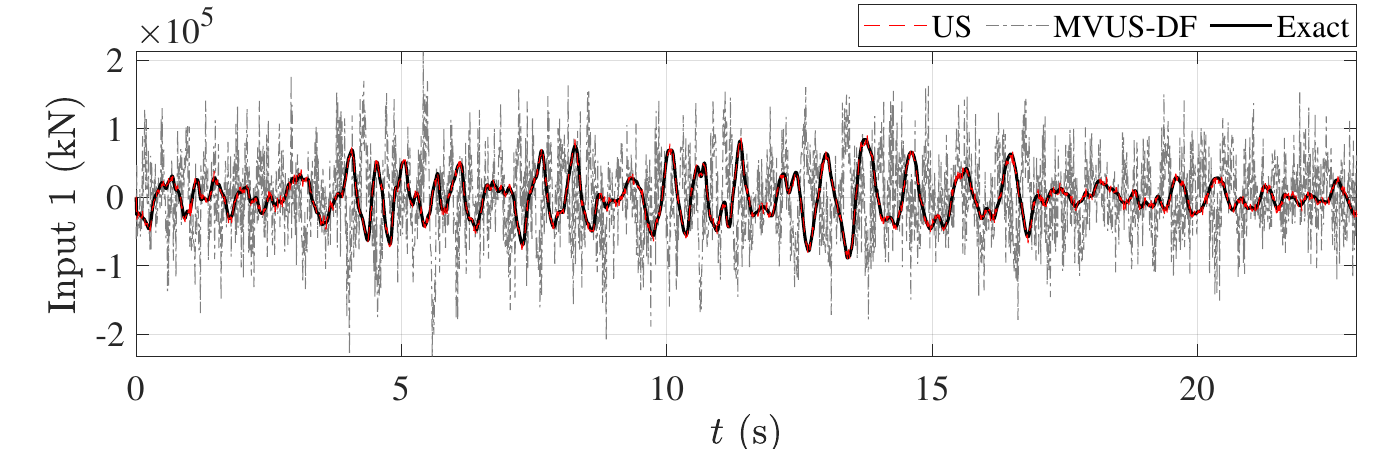}
    \end{subfigure}
    \begin{subfigure}{0.75\textwidth}
        \caption{} \label{fig-ex4-b}
        \includegraphics[trim={1.8cm 7cm 2.5cm 7cm}, clip, width=\textwidth]{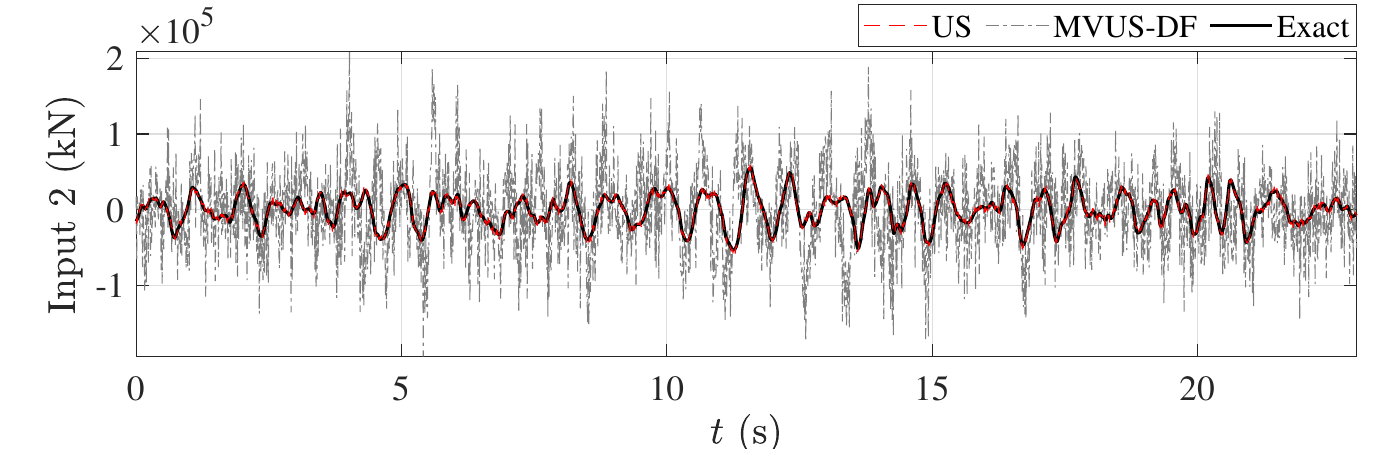}
    \end{subfigure}
    \begin{subfigure}{0.75\textwidth}
        \caption{} \label{fig-ex4-c}
        \includegraphics[trim={1.8cm 7cm 2.5cm 7cm}, clip, width=\textwidth]{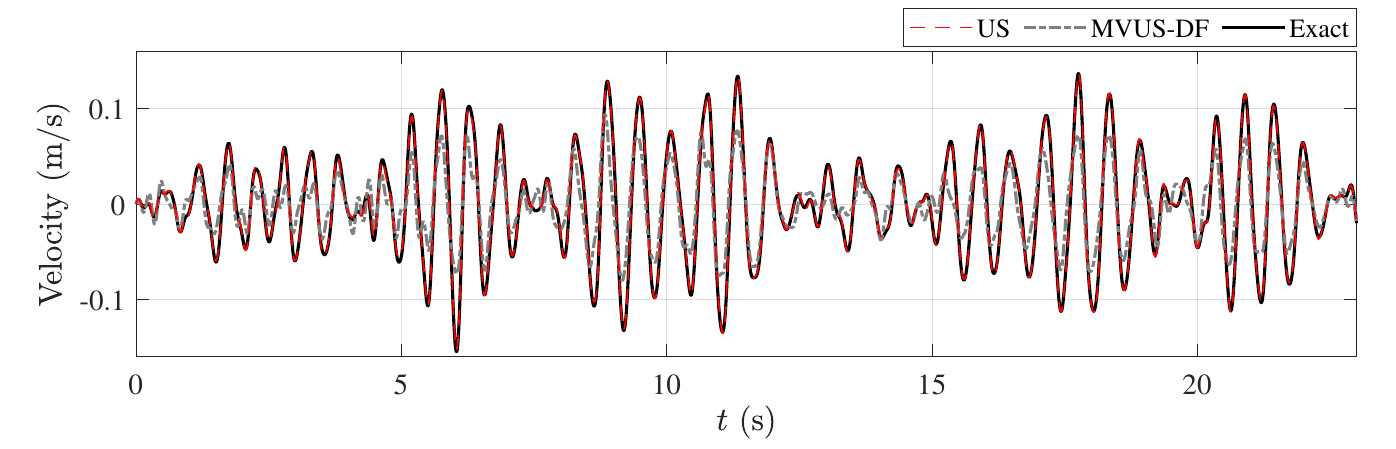}
    \end{subfigure}
    \begin{subfigure}{0.75\textwidth}
        \caption{} \label{fig-ex4-d}
        \includegraphics[trim={1.8cm 7cm 2.5cm 7cm}, clip, width=\textwidth]{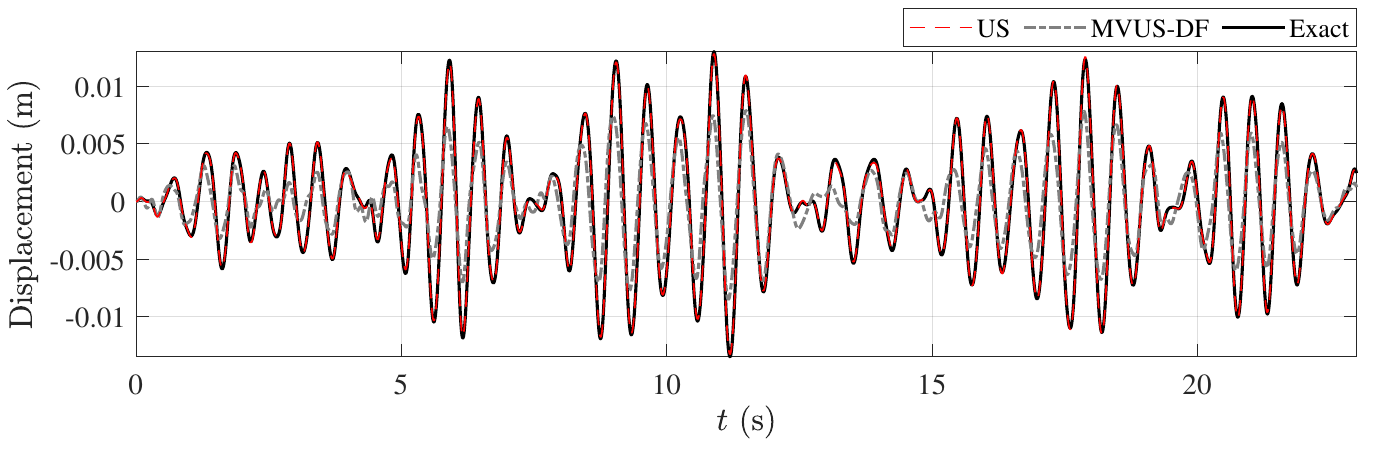}
    \end{subfigure}
    \caption{Estimation results for the rank deficient feedforward: (a) the first excitation, (b) the second excitation, (c) velocity, and (d) displacement.}
    \label{fig-ex4}
\end{figure}

Fig.~\ref{fig-ex4} shows the input and state estimation results using the US and MVUS-DF. It should be noted that the results of MVUF-DF are not included in the plots since it is not possible to run the required computations due to the singular input gain matrix caused by the rank-deficient feedforward matrix. In Figs.~\ref{fig-ex4-a} and \ref{fig-ex4-b}, the MVUS-DF cannot adequately estimate the two unknown inputs, whereas the US exhibits a much-enhanced input identification. In addition, the MVUS-DF underestimates the peak velocity and displacement. The US, on the other hand, can provide consistently accurate state estimation, as shown in Figs.~\ref{fig-ex4-c} and \ref{fig-ex4-d}. As the result, the overall dimensionless error $\Sigma\delta_{\alpha}$ is $134.3$ for the US compared with $1891.7$ for the MVUS-DF.

\section{Conclusions}
\label{section4}
\noindent A novel unbiased recursive smoothing method, the so-called Universal Smoothing, was developed. The smoothing algorithm is universally applicable to structural systems with different sensor networks. The proposed smoothing algorithm can access more observed information than filtering methods by using measured data in an extended observation window, resulting in a better and more consistent estimation quality. A weighted least squares minimised the variance of the input estimation to derive the input gain, and the state gain was obtained by minimising the trace of the error covariance of the state estimation. The covariance propagations were derived in a closed-form solution without neglecting any correlation terms.

Unlike some of the existing Kalman-type estimators, such as the AKF and DKF, the proposed smoothing algorithm does not require any prior knowledge or assumption on the input. The restriction on the rank condition of the feedforward matrix is also relaxed compared to MVUF-DF, MVUF-NDF, MVUS-DF, and MVUS-NDF. In addition, the proposed smoothing algorithm shows numerical stability when deployed to a ROM or a rank-deficient feedforward matrix.

The performance of the proposed smoothing algorithm was assessed using four numerical case studies, and a thorough comparison was made to various existing methods. The following main conclusions can be pointed out:

\begin{itemize}
\item Unlike the filtering methods that rely on the data observed at a single timestep, the proposed smoothing algorithm could adequately reduce the noise in both input and state estimates by utilising the extra sensory information in an extended observation window. It should be noted that for systems in which the sensors are not collocated with the input, the proposed smoothing algorithm could achieve a $97$\% improvement compared to its filtering counterpart, i.e., the UF. Furthermore, in the case of high observation noise, by adopting PINV with a properly selected tolerance in the input estimation step, the US with truncated singular value decomposition can achieve an $87$\% improvement compared with the US without truncated singular value decomposition.

\item The example of the eight-storey shear building with earthquake ground motion showed the proposed smoothing algorithm can accurately identify the seismic acceleration and estimate the state of the structure. This is particularly relevant for displacement-only observations, where filtering errors can be heavily amplified, and for non-displacement observations, where filters are affected by the drift effect. Even with a sensor network with direct feedthrough, the proposed method could show 67\% and 89\% reduction in the estimation errors when compared to MVUF-DF and the AKF, respectively.

\item The Taipei 101 tower study confirmed the robustness of the algorithm when compared to other smoothing methods on a large-scale structural system. Despite the significantly fewer sensors and the lack of collocated sensors, the proposed smoothing algorithm exhibited at least a 36\% and 30\% improvement compared to the MVUS-DF and the UF, respectively. Furthermore, the proposed smoothing algorithm produced good state estimates of the structure and the identification of arbitrary excitations with a rank-deficient feedforward. 
\end{itemize}

The case studies highlighted the flexibility of the Universal Smoothing algorithm, suggesting its potential as a powerful tool for further deployment in applications such as structural health monitoring, system identification, and the development of digital twins.
\clearpage

\appendix 
\section{Algorithm tuning and the grid search results} 

\subsection{Eight-storey shear building subjected to a sinusoidal force} \label{appendixA1}
\begin{figure}[!ht]
    \centering
    \begin{subfigure}{0.6\textwidth}
        \caption{} \label{figA1a}
        \includegraphics[trim={3.5cm 9.5cm 4cm 10cm}, clip, width=\textwidth]{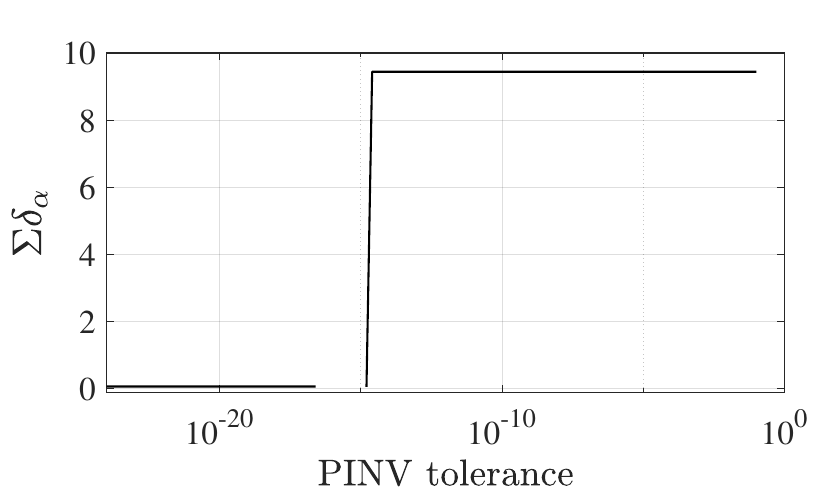}
    \end{subfigure}
    \begin{subfigure}{0.6\textwidth}
        \caption{} \label{figA1b}
        \includegraphics[trim={3.5cm 9.5cm 4cm 10cm}, clip, width=\textwidth]{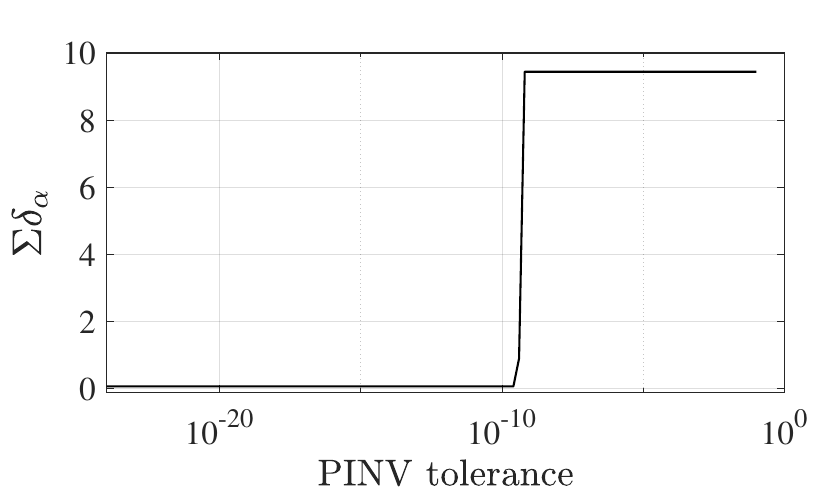}
    \end{subfigure}
    \caption{PINV tolerance tuning results: (a) configuration 1.1, and (b) configuration 1.2.}
    \label{figA1}
\end{figure}

\noindent The PINV tolerance tuning results through the grid search method for the numerical case study in Section~\ref{section3.1} are plotted in Fig.~\ref{figA1}, in which the optimal PINV tolerance is $1\times10^{-14.8}$ for configuration 1.1 and $1\times10^{-13.0}$ for configuration 1.2. Note that the hyperparameters that cause numerical issues or singularities are not plotted. 

\clearpage
\subsection{Eight-storey shear building subjected to ground motion} \label{appendixA2}

\begin{figure}[!ht]
    \centering
    \begin{subfigure}{0.48\textwidth}
        \caption{} \label{figA2a}
        \includegraphics[trim={3.5cm 8.5cm 4cm 9cm}, clip, width=\textwidth]{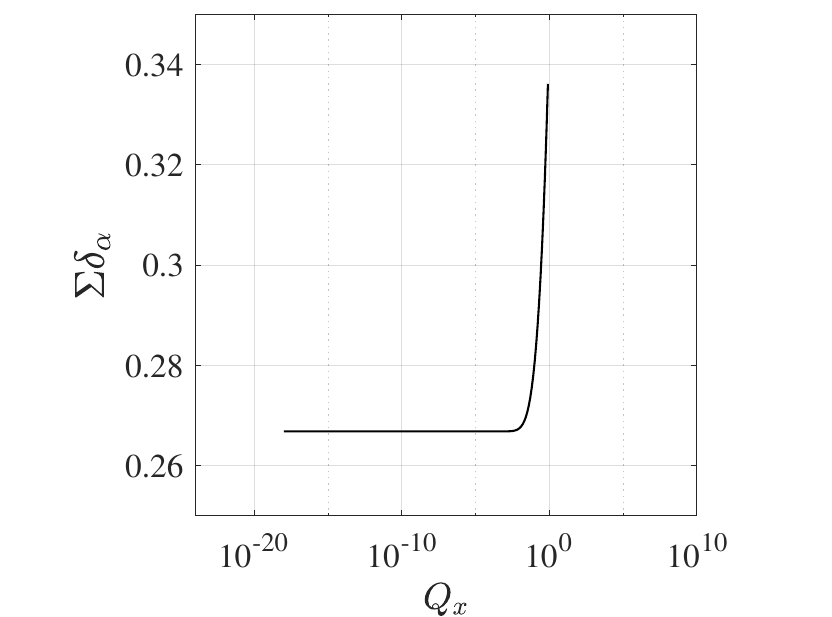}
    \end{subfigure}
    \begin{subfigure}{0.5\textwidth}
        \caption{} \label{figA2b}
        \includegraphics[trim={3.5cm 8.5cm 4cm 9cm}, clip, width=\textwidth]{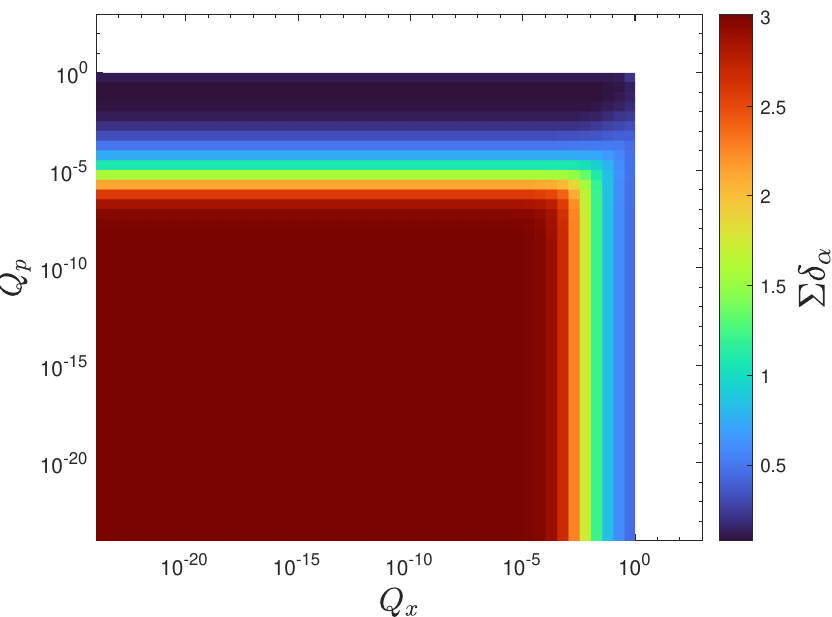}
    \end{subfigure}
    \begin{subfigure}{0.55\textwidth}
        \caption{} \label{figA2c}
        \includegraphics[trim={3.8cm 8.5cm 3.2cm 9cm}, clip, width=\textwidth]{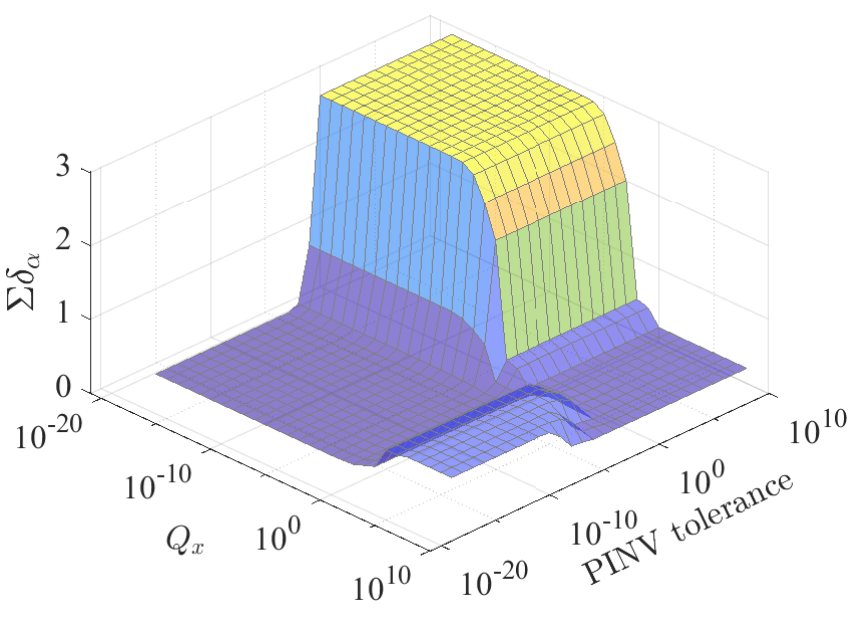}
    \end{subfigure}
    \caption{Tuning results for configuration 2.1: (a) the MVU-NDF, (b) the AKF, and (c) the US.}
    \label{figA2}
\end{figure}

\noindent The hyperparameters tuning results of the eight-storey shear building example in Section~\ref{section3.2} through the grid search method are presented below. The tuning results for configuration 2.1 are shown in Fig.~\ref{figA2}. The best performance can be found at $Q_x=10^{0.7}$ and $\mathrm{PINV \ tolerance = 10^{-3.5} }$ for the US; $Q_x = 10^{-18.0}$ and $Q_p=10^{-1.5}$ for the AKF; and $Q_x=10^{-3.2}$ for the MVUF-NDF. 

\clearpage

\begin{figure}[!ht]
    \centering
    \begin{subfigure}{0.52\textwidth}
        \caption{} \label{figA3a}
        \includegraphics[trim={3.8cm 10.5cm 4.5cm 11.5cm}, clip, width=\textwidth]{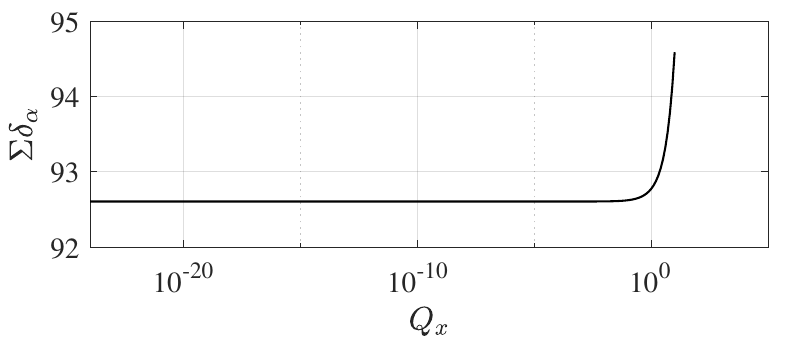}
    \end{subfigure}\\
    \begin{subfigure}{0.5\textwidth}
        \caption{} \label{figA3b}
        \includegraphics[trim={3.5cm 8.5cm 4cm 9cm}, clip, width=\textwidth]{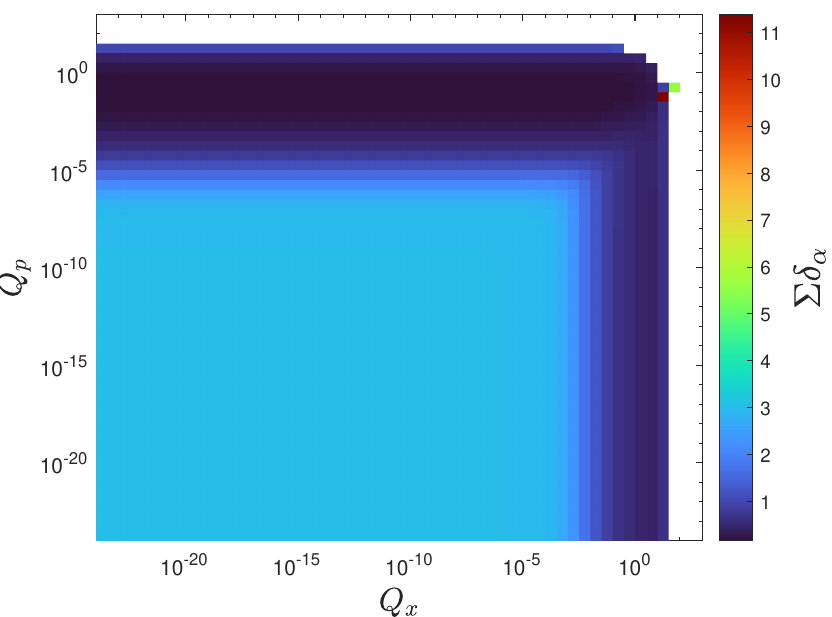}
    \end{subfigure}\\
    \begin{subfigure}{0.55\textwidth}
        \caption{} \label{figA3c}
        \includegraphics[trim={3.5cm 9cm 3.5cm 9cm}, clip, width=\textwidth]{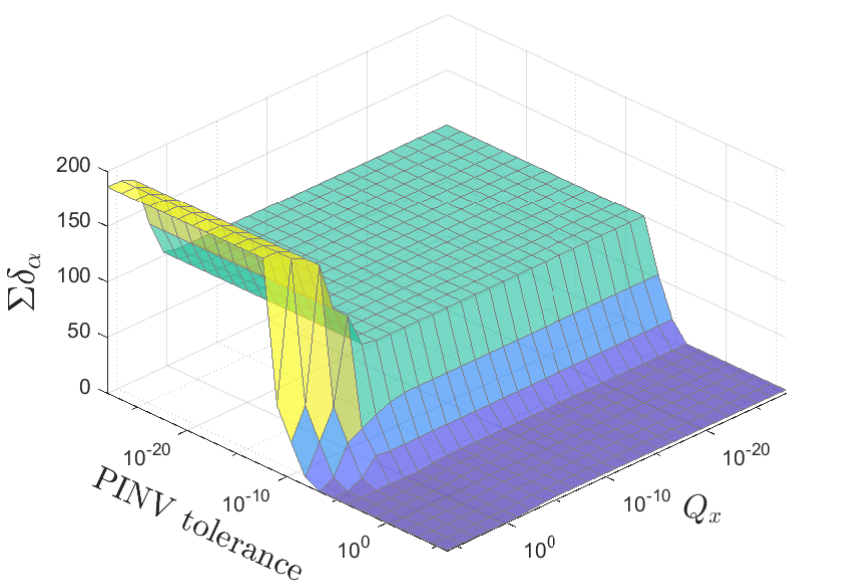}
    \end{subfigure}
    \caption{Tuning results for configuration 2.2: (a) the MVU-NDF, (b) the AKF, and (c) the US.}
    \label{figA3}
\end{figure}

Fig~\ref{figA3} shows the tuning results for configuration 2.2. The best performance can be found at $Q_x=10^{3.0}$ and $\mathrm{PINV \ tolerance = 10^{-2.2} }$ for the US; $Q_x = 10^{-19.5}$ and $Q_p=10^{-1.0}$ for the AKF; and $Q_x=10^{-13.7}$ for the MVUF-NDF.

\clearpage

\begin{figure}[!ht]
    \centering
    \begin{subfigure}{0.52\textwidth}
        \caption{} \label{figA4a}
        \includegraphics[trim={3.8cm 10.5cm 4.5cm 11.5cm}, clip, width=\textwidth]{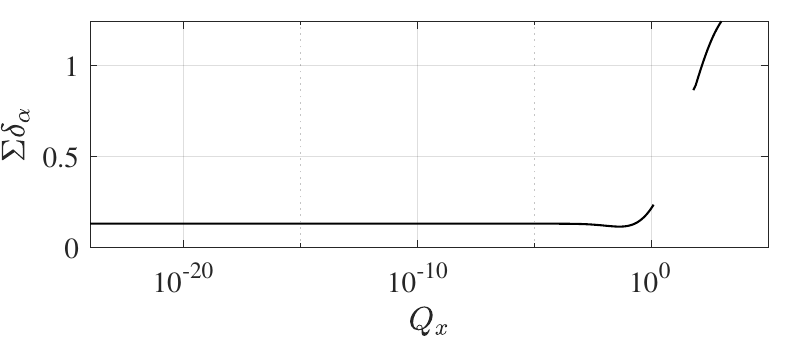}
    \end{subfigure}\\
    \begin{subfigure}{0.48\textwidth}
        \caption{} \label{figA4b}
        \includegraphics[trim={3.2cm 8.5cm 3.9cm 9cm}, clip, width=\textwidth]{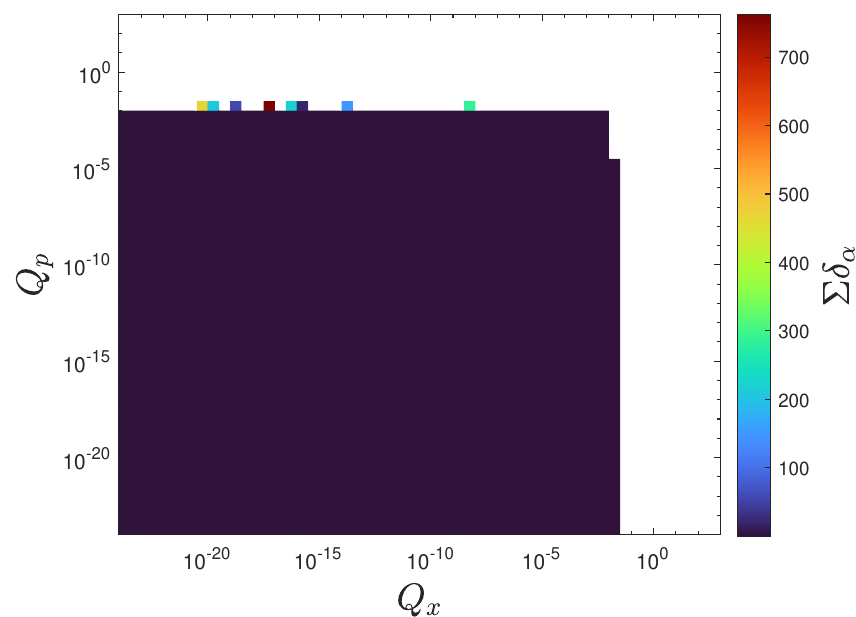}
    \end{subfigure}\\
    \begin{subfigure}{0.55\textwidth}
        \caption{} \label{figA4c}
        \includegraphics[trim={4cm 8.5cm 3.2cm 9cm}, clip, width=\textwidth]{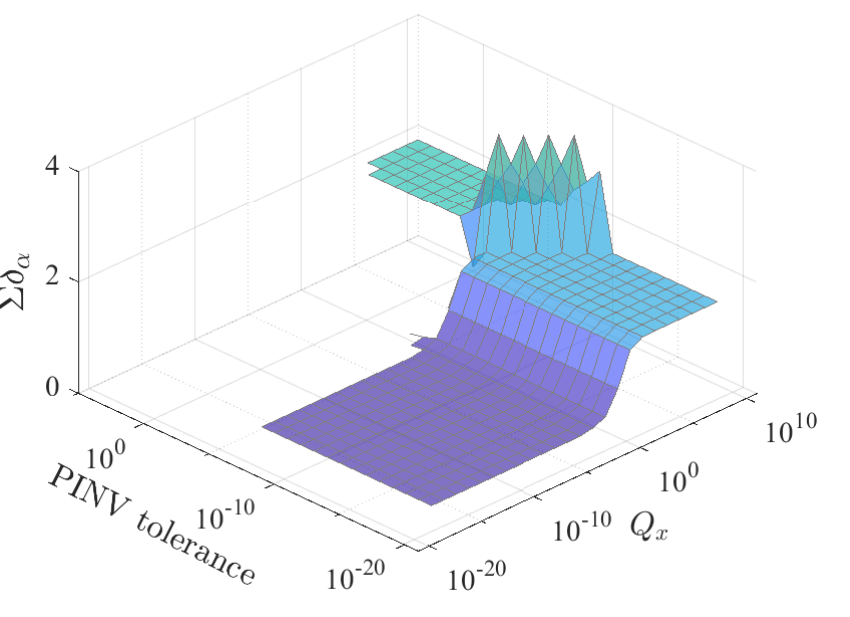}
    \end{subfigure}
    \caption{Tuning results for configuration 2.3: (a) the MVU-DF, (b) the AKF, and (c) the US.}
    \label{figA4}
\end{figure}

The tuning results for configuration 2.3 are presented in Fig~\ref{figA4}. The best performance can be found at $Q_x=10^{-3.1}$ and $\mathrm{PINV \ tolerance = 10^{-18.0} }$ for the US; $Q_x = 10^{-19.0}$ and $Q_p=10^{-2.5}$ for the AKF; and $Q_x=10^{-1.4}$ for the MVUF-DF.

\clearpage

\begin{figure}[!ht]
    \centering
    \begin{subfigure}{0.52\textwidth}
        \caption{} \label{figA5a}
        \includegraphics[trim={4cm 10.5cm 4.5cm 11cm}, clip, width=\textwidth]{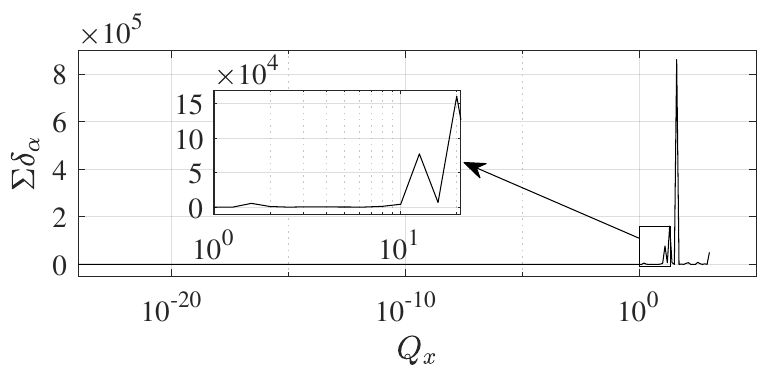}
    \end{subfigure}\\
    \begin{subfigure}{0.5\textwidth}
        \caption{} \label{figA5b}
        \includegraphics[trim={3.2cm 8cm 3.7cm 9cm}, clip, width=\textwidth]{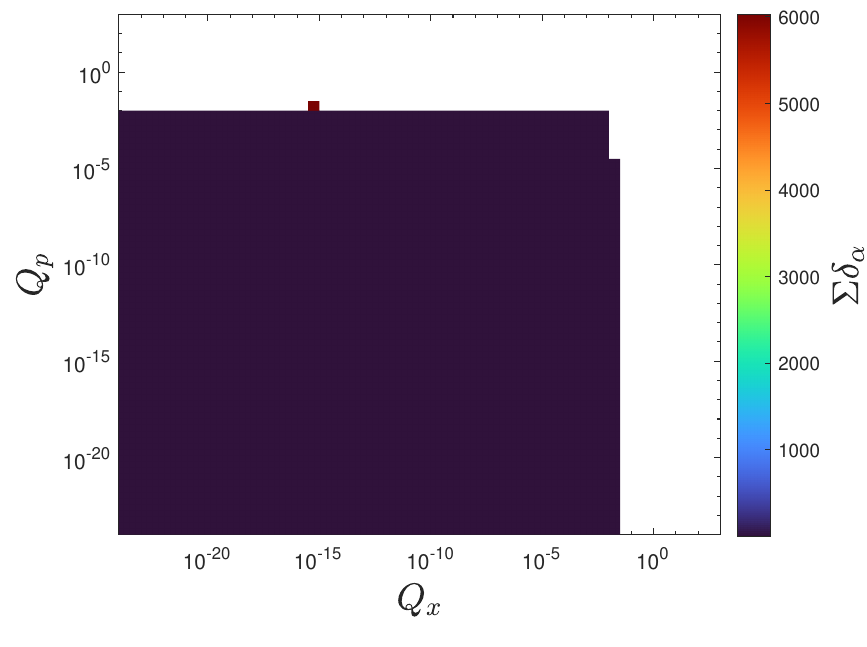}
    \end{subfigure}\\
    \begin{subfigure}{0.55\textwidth}
        \caption{} \label{figA5c}
        \includegraphics[trim={3.8cm 8.5cm 3cm 9cm}, clip, width=\textwidth]{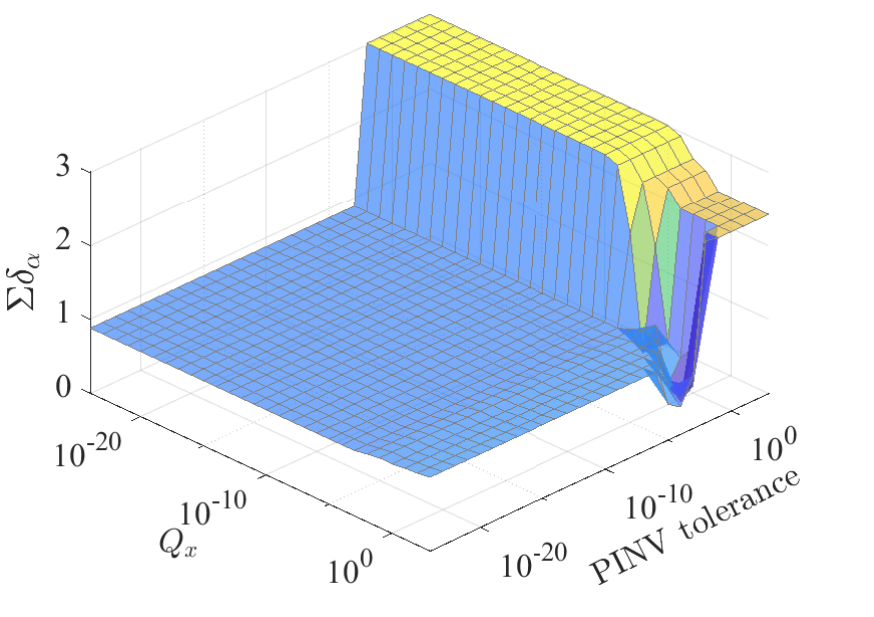}
    \end{subfigure}
    \caption{Tuning results for configuration 2.4: (a) the MVU-DF, (b) the AKF, and (c) the US.}
    \label{figA5}
\end{figure}

Finally, the tuning results for configuration 2.4 are presented in Fig~\ref{figA5}. The best performance can be found at $Q_x=10^{1.0}$ and $\mathrm{PINV \ tolerance = 10^{-2.0} }$ for the US; $Q_x = 10^{-17.5}$ and $Q_p=10^{-4.0}$ for the AKF; and $Q_x=10^{-1.2}$ for the MVUF-DF. 

\clearpage

\subsection{The Taipei 101 example} \label{appendixA3}
\noindent The tuning results of the Taipei 101 case study in Section~\ref{section3.3} and the rank-deficient feedforward example in Section~\ref{section3.4} are presented below. Fig.~\ref{figA6} shows the grid search results for configuration 3.1. The best performance can be found at  $Q_x=10^{-1.9}$ and $\mathrm{PINV \ tolerance = 10^{-13.5} }$ for the US; $Q_x = 10^{-3.0}$ for the MVUS-NDF; and $Q_x=10^{-2.7}$ for the UF.

\begin{figure}[!htbp]
    \centering
    \begin{subfigure}{0.45\textwidth}
        \caption{} \label{figA6a}
        \includegraphics[trim={3.5cm 8.5cm 4cm 8.5cm}, clip, width=\textwidth]{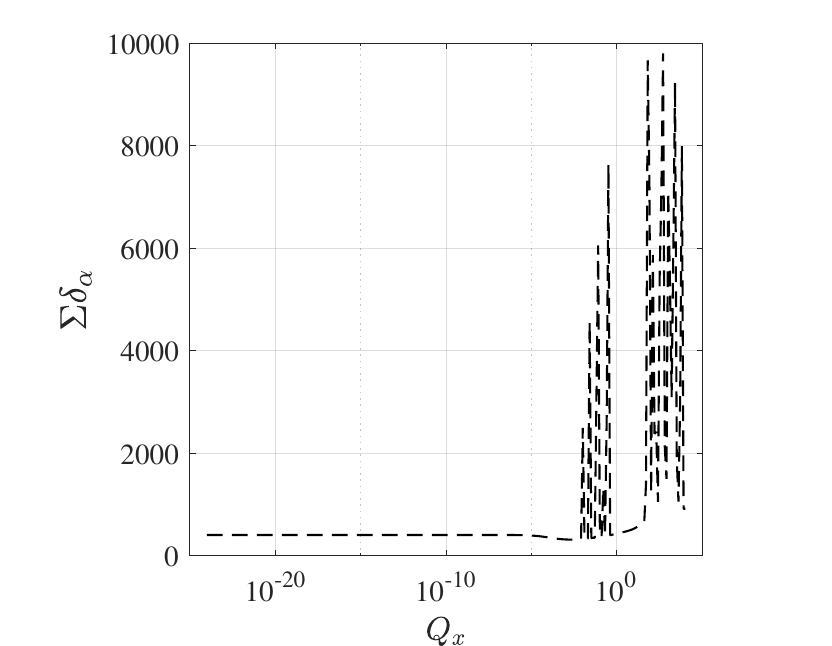}
    \end{subfigure}
    \begin{subfigure}{0.45\textwidth}
        \caption{} \label{figA6b}
        \includegraphics[trim={3.5cm 8.5cm 4cm 9cm}, clip, width=\textwidth]{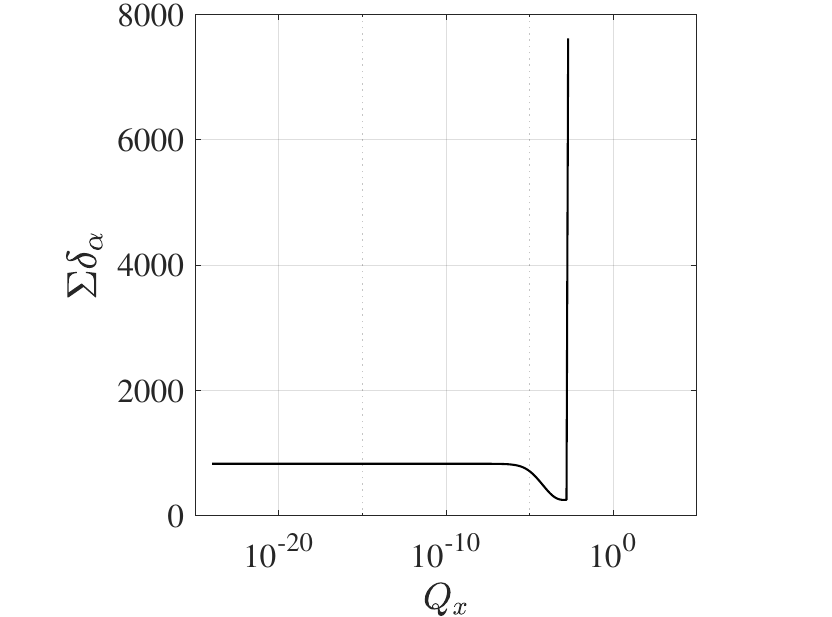}
    \end{subfigure}
    \begin{subfigure}{0.55\textwidth}
        \caption{} \label{figA6c}
        \includegraphics[trim={3.5cm 9cm 3.5cm 9cm}, clip, width=\textwidth]{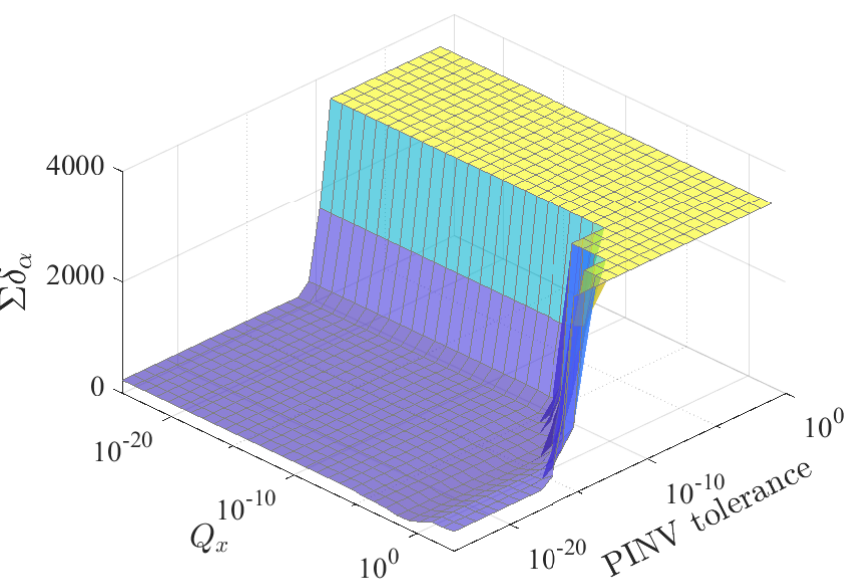}
    \end{subfigure}
    \caption{Tuning results for configuration 3.1: (a) the UF, (b) the MVUS-NDF, and (c) the US.}
    \label{figA6}
\end{figure}

\clearpage

\begin{figure}[!htbp]
    \centering
    \begin{subfigure}{0.45\textwidth}
        \caption{} \label{figA7a}
        \includegraphics[trim={3.5cm 8.5cm 4cm 8.5cm}, clip, width=\textwidth]{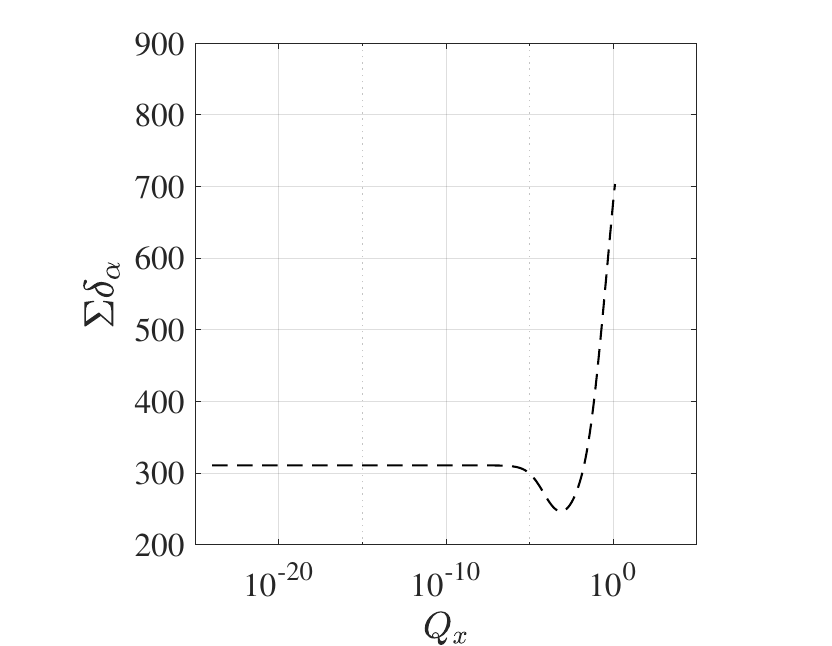}
    \end{subfigure}
    \begin{subfigure}{0.45\textwidth}
        \caption{} \label{figA7b}
        \includegraphics[trim={3.5cm 8.5cm 4cm 9cm}, clip, width=\textwidth]{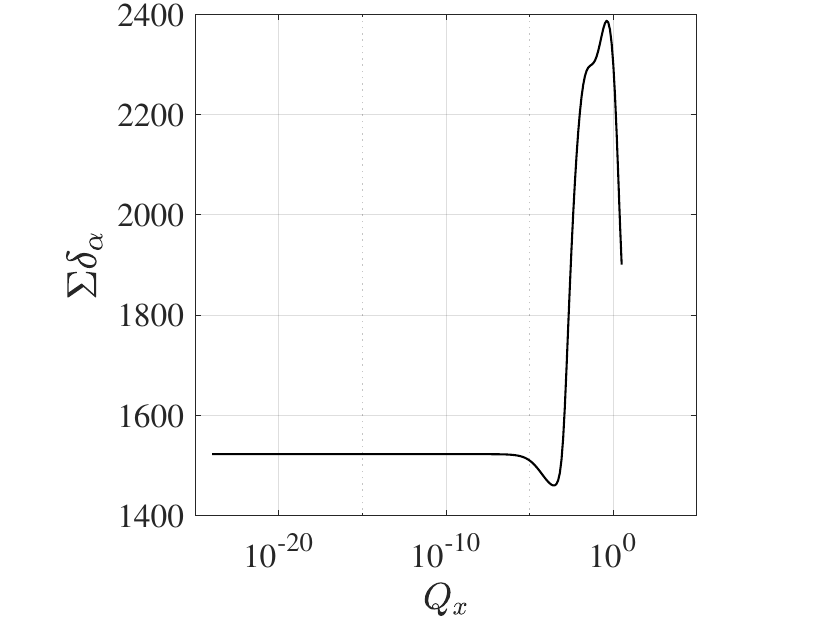}
    \end{subfigure}
    \begin{subfigure}{0.55\textwidth}
        \caption{} \label{figA7c}
        \includegraphics[trim={3.5cm 8.5cm 3.5cm 9cm}, clip, width=\textwidth]{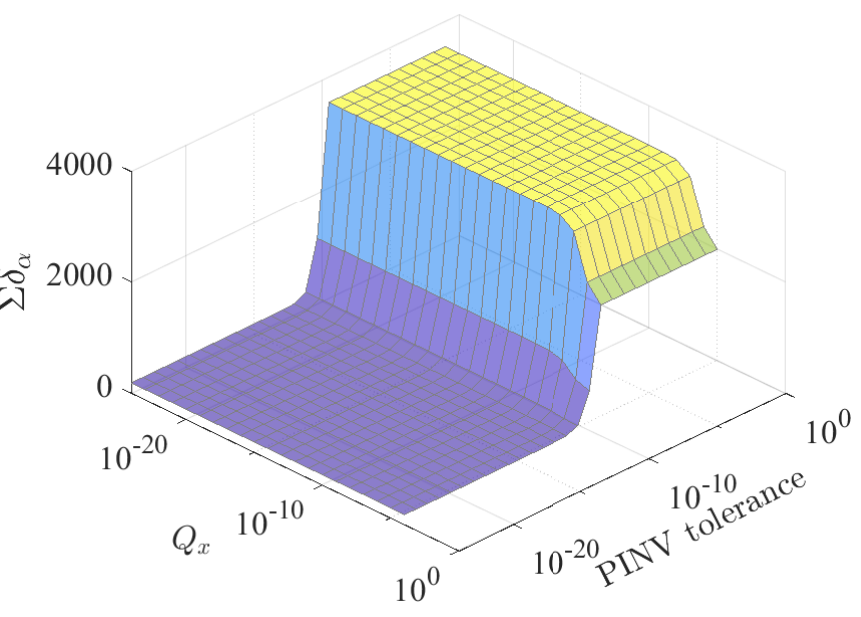}
    \end{subfigure}
    \caption{Tuning results for configuration 3.2: (a) the UF, (b) the MVUS-DF, and (c) the US.}
    \label{figA7}
\end{figure}

The grid search results for configuration 3.2 are presented in Fig.~\ref{figA7}. The best performance can be found at  $Q_x=10^{-3.2}$ and $\mathrm{PINV \ tolerance = 10^{-13.4} }$ for the US; $Q_x = 10^{-3.5}$ for the MVUS-DF; and $Q_x=10^{-2.8}$ for the UF.

\clearpage

\begin{figure}[!htbp]
    \centering
    \begin{subfigure}{0.6\textwidth}
        \caption{} \label{figA8a}
        \includegraphics[trim={3.2cm 10.5cm 4.5cm 11.5cm}, clip, width=\textwidth]{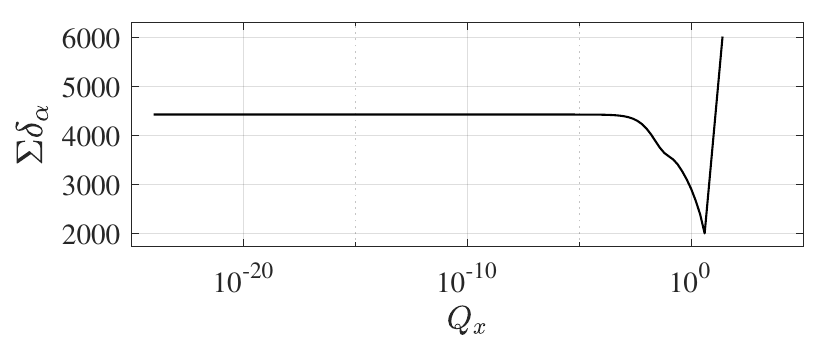}
    \end{subfigure}\\
    \begin{subfigure}{0.55\textwidth}
        \caption{} \label{figA8b}
        \includegraphics[trim={3.5cm 8.5cm 3.5cm 9cm}, clip, width=\textwidth]{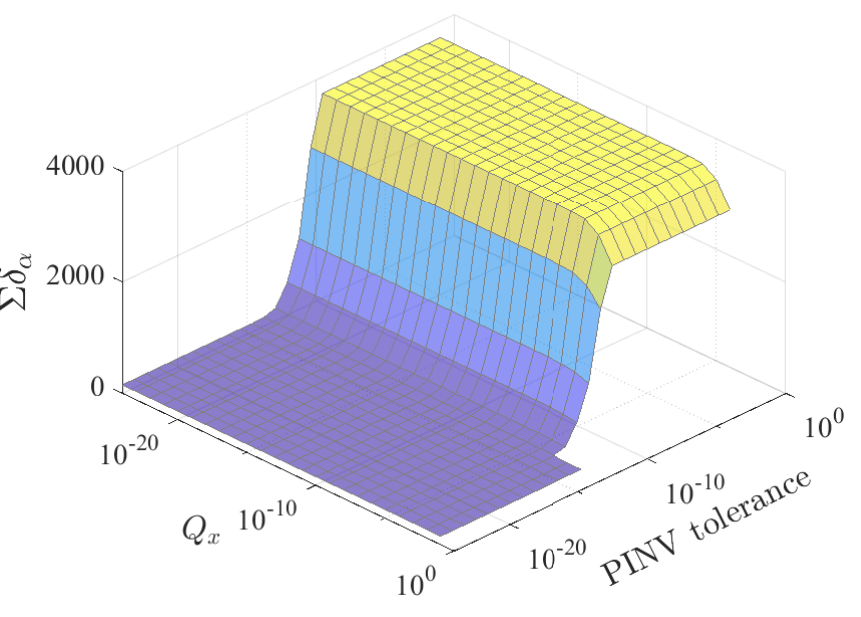}
    \end{subfigure}
    \caption{Tuning results for the rank-deficient feedforward example: (a) the MVUS-DF, and (b) the US.}
    \label{figA8}
\end{figure}

The grid search results for the rank-deficient feedforward example in Section \ref{section3.4} are presented in Fig.~\ref{figA8}. The best performance can be found at  $Q_x=10^{-1.3}$ and $\mathrm{PINV \ tolerance = 10^{-15.5} }$ for the US; and $Q_x = 10^{0.6}$ for the MVUS-DF.

\bibliography{ref} 

\begin{thebibliography}{30}
\expandafter\ifx\csname natexlab\endcsname\relax\def\natexlab#1{#1}\fi
\providecommand{\url}[1]{\texttt{#1}}
\providecommand{\href}[2]{#2}
\providecommand{\path}[1]{#1}
\providecommand{\DOIprefix}{doi:}
\providecommand{\ArXivprefix}{arXiv:}
\providecommand{\URLprefix}{URL: }
\providecommand{\Pubmedprefix}{pmid:}
\providecommand{\doi}[1]{\href{http://dx.doi.org/#1}{\path{#1}}}
\providecommand{\Pubmed}[1]{\href{pmid:#1}{\path{#1}}}
\providecommand{\bibinfo}[2]{#2}
\ifx\xfnm\relax \def\xfnm[#1]{\unskip,\space#1}\fi
\bibitem[{Kalman(1960)}]{Kalman1960}
\bibinfo{author}{R.~E. Kalman},
\newblock \bibinfo{title}{A new approach to linear filtering and prediction problems},
\newblock \bibinfo{journal}{Journal of Basic Engineering} \bibinfo{volume}{82} (\bibinfo{year}{1960}) \bibinfo{pages}{35--45}. \DOIprefix\doi{10.1115/1.3662552}.
\bibitem[{Zhu et~al.(2023)Zhu, Lu, and Zhu}]{Zhu2023}
\bibinfo{author}{Z.~Zhu}, \bibinfo{author}{J.~Lu}, \bibinfo{author}{S.~Zhu},
\newblock \bibinfo{title}{Multi-rate kalman filtering for structural dynamic response reconstruction by fusing multi-type sensor data with different sampling frequencies},
\newblock \bibinfo{journal}{Engineering Structures} \bibinfo{volume}{293} (\bibinfo{year}{2023}) \bibinfo{pages}{116573}. \DOIprefix\doi{10.1016/j.engstruct.2023.116573}.
\bibitem[{Papadimitriou et~al.(2011)Papadimitriou, Fritzen, Kraemer, and Ntotsios}]{Papadimitriou2011}
\bibinfo{author}{C.~Papadimitriou}, \bibinfo{author}{C.-P. Fritzen}, \bibinfo{author}{P.~Kraemer}, \bibinfo{author}{E.~Ntotsios},
\newblock \bibinfo{title}{Fatigue predictions in entire body of metallic structures from a limited number of vibration sensors using kalman filtering},
\newblock \bibinfo{journal}{Structural Control and Health Monitoring} \bibinfo{volume}{18} (\bibinfo{year}{2011}) \bibinfo{pages}{554--573}. \DOIprefix\doi{10.1002/stc.395}.
\bibitem[{Lourens et~al.(2012)Lourens, Reynders, {De Roeck}, Degrande, and Lombaert}]{LourensReyndersEATAL2012}
\bibinfo{author}{E.~Lourens}, \bibinfo{author}{E.~Reynders}, \bibinfo{author}{G.~{De Roeck}}, \bibinfo{author}{G.~Degrande}, \bibinfo{author}{G.~Lombaert},
\newblock \bibinfo{title}{An augmented kalman filter for force identification in structural dynamics},
\newblock \bibinfo{journal}{Mechanical Systems and Signal Processing} \bibinfo{volume}{27} (\bibinfo{year}{2012}) \bibinfo{pages}{446--460}. \DOIprefix\doi{10.1016/j.ymssp.2011.09.025}.
\bibitem[{Saleem and Jo(2019)}]{Saleem2019}
\bibinfo{author}{M.~M. Saleem}, \bibinfo{author}{H.~Jo},
\newblock \bibinfo{title}{Impact force localization for civil infrastructure using augmented kalman filter optimization},
\newblock \bibinfo{journal}{Smart Structures and Systems} \bibinfo{volume}{23} (\bibinfo{year}{2019}) \bibinfo{pages}{123 – 139}. \DOIprefix\doi{10.12989/sss.2019.23.2.123}.
\bibitem[{Wang et~al.(2021)Wang, Guo, and Takewaki}]{Wang2021}
\bibinfo{author}{L.~Wang}, \bibinfo{author}{J.~Guo}, \bibinfo{author}{I.~Takewaki},
\newblock \bibinfo{title}{Real-time hysteresis identification in structures based on restoring force reconstruction and kalman filter},
\newblock \bibinfo{journal}{Mechanical Systems and Signal Processing} \bibinfo{volume}{150} (\bibinfo{year}{2021}). \DOIprefix\doi{10.1016/j.ymssp.2020.107297}.
\bibitem[{{Eftekhar Azam} et~al.(2015){Eftekhar Azam}, Chatzi, and Papadimitriou}]{EftekharAzam2015}
\bibinfo{author}{S.~{Eftekhar Azam}}, \bibinfo{author}{E.~Chatzi}, \bibinfo{author}{C.~Papadimitriou},
\newblock \bibinfo{title}{A dual kalman filter approach for state estimation via output-only acceleration measurements},
\newblock \bibinfo{journal}{Mechanical Systems and Signal Processing} \bibinfo{volume}{60-61} (\bibinfo{year}{2015}) \bibinfo{pages}{866--886}. \DOIprefix\doi{10.1016/j.ymssp.2015.02.001}.
\bibitem[{Cazzulani et~al.(2013)Cazzulani, Moschini, Resta, and Ripamonti}]{Cazzulani2013}
\bibinfo{author}{G.~Cazzulani}, \bibinfo{author}{S.~Moschini}, \bibinfo{author}{F.~Resta}, \bibinfo{author}{F.~Ripamonti},
\newblock \bibinfo{title}{A diagnostic logic for preventing structural failure in concrete displacing booms},
\newblock \bibinfo{journal}{Automation in Construction} \bibinfo{volume}{35} (\bibinfo{year}{2013}) \bibinfo{pages}{499--506}. \DOIprefix\doi{10.1016/j.autcon.2013.06.004}.
\bibitem[{Roffel and Narasimhan(2014)}]{Roffel2014}
\bibinfo{author}{A.~Roffel}, \bibinfo{author}{S.~Narasimhan},
\newblock \bibinfo{title}{Extended kalman filter for modal identification of structures equipped with a pendulum tuned mass damper},
\newblock \bibinfo{journal}{Journal of Sound and Vibration} \bibinfo{volume}{333} (\bibinfo{year}{2014}) \bibinfo{pages}{6038--6056}. \DOIprefix\doi{10.1016/j.jsv.2014.06.030}.
\bibitem[{Naets et~al.(2015)Naets, Croes, and Desmet}]{Naets2015}
\bibinfo{author}{F.~Naets}, \bibinfo{author}{J.~Croes}, \bibinfo{author}{W.~Desmet},
\newblock \bibinfo{title}{An online coupled state/input/parameter estimation approach for structural dynamics},
\newblock \bibinfo{journal}{Computer Methods in Applied Mechanics and Engineering} \bibinfo{volume}{283} (\bibinfo{year}{2015}) \bibinfo{pages}{1167--1188}. \DOIprefix\doi{10.1016/j.cma.2014.08.010}.
\bibitem[{{Ebrahimzadeh Hassanabadi} et~al.(2020){Ebrahimzadeh Hassanabadi}, Heidarpour, {Eftekhar Azam}, and Arashpour}]{EbrahimzadehHassanabadi2020}
\bibinfo{author}{M.~{Ebrahimzadeh Hassanabadi}}, \bibinfo{author}{A.~Heidarpour}, \bibinfo{author}{S.~{Eftekhar Azam}}, \bibinfo{author}{M.~Arashpour},
\newblock \bibinfo{title}{Recursive principal component analysis for model order reduction with application in nonlinear bayesian filtering},
\newblock \bibinfo{journal}{Computer Methods in Applied Mechanics and Engineering} \bibinfo{volume}{371} (\bibinfo{year}{2020}) \bibinfo{pages}{113334}. \DOIprefix\doi{10.1016/j.cma.2020.113334}.
\bibitem[{Wu and Smyth(2007)}]{Wu2007}
\bibinfo{author}{M.~Wu}, \bibinfo{author}{A.~W. Smyth},
\newblock \bibinfo{title}{Application of the unscented kalman filter for real-time nonlinear structural system identification},
\newblock \bibinfo{journal}{Structural Control and Health Monitoring} \bibinfo{volume}{14} (\bibinfo{year}{2007}) \bibinfo{pages}{971--990}. \DOIprefix\doi{10.1002/stc.186}.
\bibitem[{Dertimanis et~al.(2021)Dertimanis, Chatzi, and Masri}]{Dertimanis2021}
\bibinfo{author}{V.~K. Dertimanis}, \bibinfo{author}{E.~N. Chatzi}, \bibinfo{author}{S.~F. Masri},
\newblock \bibinfo{title}{On the active vibration control of nonlinear uncertain structures},
\newblock \bibinfo{journal}{Journal of Applied and Computational Mechanics} \bibinfo{volume}{7} (\bibinfo{year}{2021}) \bibinfo{pages}{1183 – 1197}. \DOIprefix\doi{10.22055/JACM.2020.34007.2322}.
\bibitem[{Cha et~al.(2017)Cha, Chen, and Büyüköztürk}]{Cha2017}
\bibinfo{author}{Y.-J. Cha}, \bibinfo{author}{J.~Chen}, \bibinfo{author}{O.~Büyüköztürk},
\newblock \bibinfo{title}{Output-only computer vision based damage detection using phase-based optical flow and unscented kalman filters},
\newblock \bibinfo{journal}{Engineering Structures} \bibinfo{volume}{132} (\bibinfo{year}{2017}) \bibinfo{pages}{300--313}. \DOIprefix\doi{10.1016/j.engstruct.2016.11.038}.
\bibitem[{Ma et~al.(2022)Ma, Choi, Liu, and Sohn}]{Ma2022}
\bibinfo{author}{Z.~Ma}, \bibinfo{author}{J.~Choi}, \bibinfo{author}{P.~Liu}, \bibinfo{author}{H.~Sohn},
\newblock \bibinfo{title}{Structural displacement estimation by fusing vision camera and accelerometer using hybrid computer vision algorithm and adaptive multi-rate kalman filter},
\newblock \bibinfo{journal}{Automation in Construction} \bibinfo{volume}{140} (\bibinfo{year}{2022}) \bibinfo{pages}{104338}. \DOIprefix\doi{10.1016/j.autcon.2022.104338}.
\bibitem[{Lin et~al.(2022)Lin, Bao, and Li}]{Lin2022}
\bibinfo{author}{Q.~Lin}, \bibinfo{author}{X.~Bao}, \bibinfo{author}{C.~Li},
\newblock \bibinfo{title}{Deep learning based missing data recovery of non-stationary wind velocity},
\newblock \bibinfo{journal}{Journal of Wind Engineering and Industrial Aerodynamics} \bibinfo{volume}{224} (\bibinfo{year}{2022}) \bibinfo{pages}{104962}. \DOIprefix\doi{10.1016/j.jweia.2022.104962}.
\bibitem[{Zhou et~al.(2020)Zhou, Pei, Li, Fang, Zhao, and Yi}]{Zhou2022}
\bibinfo{author}{Y.~Zhou}, \bibinfo{author}{Y.~Pei}, \bibinfo{author}{Z.~Li}, \bibinfo{author}{L.~Fang}, \bibinfo{author}{Y.~Zhao}, \bibinfo{author}{W.~Yi},
\newblock \bibinfo{title}{Vehicle weight identification system for spatiotemporal load distribution on bridges based on non-contact machine vision technology and deep learning algorithms},
\newblock \bibinfo{journal}{Measurement} \bibinfo{volume}{159} (\bibinfo{year}{2020}) \bibinfo{pages}{107801}. \DOIprefix\doi{10.1016/j.measurement.2020.107801}.
\bibitem[{Gillijns and {De Moor}(2007)}]{Gillijns2007a}
\bibinfo{author}{S.~Gillijns}, \bibinfo{author}{B.~{De Moor}},
\newblock \bibinfo{title}{Unbiased minimum-variance input and state estimation for linear discrete-time systems},
\newblock \bibinfo{journal}{Automatica} \bibinfo{volume}{43} (\bibinfo{year}{2007}) \bibinfo{pages}{111--116}. \DOIprefix\doi{10.1016/j.automatica.2006.08.002}.
\bibitem[{Kitanidis(1987)}]{Kitanidis1987}
\bibinfo{author}{P.~K. Kitanidis},
\newblock \bibinfo{title}{Unbiased minimum-variance linear state estimation},
\newblock \bibinfo{journal}{Automatica} \bibinfo{volume}{23} (\bibinfo{year}{1987}) \bibinfo{pages}{775--778}. \DOIprefix\doi{10.1016/0005-1098(87)90037-9}.
\bibitem[{Gillijns and {De Moor}(2007)}]{Gillijns2007b}
\bibinfo{author}{S.~Gillijns}, \bibinfo{author}{B.~{De Moor}},
\newblock \bibinfo{title}{Unbiased minimum-variance input and state estimation for linear discrete-time systems with direct feedthrough},
\newblock \bibinfo{journal}{Automatica} \bibinfo{volume}{43} (\bibinfo{year}{2007}) \bibinfo{pages}{934--937}. \DOIprefix\doi{10.1016/j.automatica.2006.11.016}.
\bibitem[{Lourens et~al.(2012)Lourens, Papadimitriou, Gillijns, Reynders, {De Roeck}, and Lombaert}]{LourensPapadimitriouETAL2012}
\bibinfo{author}{E.~Lourens}, \bibinfo{author}{C.~Papadimitriou}, \bibinfo{author}{S.~Gillijns}, \bibinfo{author}{E.~Reynders}, \bibinfo{author}{G.~{De Roeck}}, \bibinfo{author}{G.~Lombaert},
\newblock \bibinfo{title}{Joint input-response estimation for structural systems based on reduced-order models and vibration data from a limited number of sensors},
\newblock \bibinfo{journal}{Mechanical Systems and Signal Processing} \bibinfo{volume}{29} (\bibinfo{year}{2012}) \bibinfo{pages}{310--327}. \DOIprefix\doi{10.1016/j.ymssp.2012.01.011}.
\bibitem[{Wan et~al.(2018)Wan, Wang, Li, and Xu}]{Wan2018}
\bibinfo{author}{Z.~Wan}, \bibinfo{author}{T.~Wang}, \bibinfo{author}{L.~Li}, \bibinfo{author}{Z.~Xu},
\newblock \bibinfo{title}{A novel coupled state/input/parameter identification method for linear structural systems},
\newblock \bibinfo{journal}{Shock and Vibration} \bibinfo{volume}{2018} (\bibinfo{year}{2018}) \bibinfo{pages}{7691721}. \DOIprefix\doi{10.1155/2018/7691721}.
\bibitem[{Ebrahimzadeh~Hassanabadi et~al.(2023)Ebrahimzadeh~Hassanabadi, Liu, Eftekhar~Azam, and Dias-da Costa}]{EbrahimzadehHassanabadi2023}
\bibinfo{author}{M.~Ebrahimzadeh~Hassanabadi}, \bibinfo{author}{Z.~Liu}, \bibinfo{author}{S.~Eftekhar~Azam}, \bibinfo{author}{D.~Dias-da Costa},
\newblock \bibinfo{title}{A linear bayesian filter for input and state estimation of structural systems},
\newblock \bibinfo{journal}{Computer-Aided Civil and Infrastructure Engineering} \bibinfo{volume}{38} (\bibinfo{year}{2023}) \bibinfo{pages}{1749--1766}. \DOIprefix\doi{10.1111/mice.12973}.
\bibitem[{Maes et~al.(2018)Maes, Gillijns, and Lombaert}]{Maes2018}
\bibinfo{author}{K.~Maes}, \bibinfo{author}{S.~Gillijns}, \bibinfo{author}{G.~Lombaert},
\newblock \bibinfo{title}{A smoothing algorithm for joint input-state estimation in structural dynamics},
\newblock \bibinfo{journal}{Mechanical Systems and Signal Processing} \bibinfo{volume}{98} (\bibinfo{year}{2018}) \bibinfo{pages}{292--309}. \DOIprefix\doi{10.1016/j.ymssp.2017.04.047}.
\bibitem[{Ebrahimzadeh~Hassanabadi et~al.(2022)Ebrahimzadeh~Hassanabadi, Heidarpour, Eftekhar~Azam, and Arashpour}]{EbrahimzadehHassanabadi2022}
\bibinfo{author}{M.~Ebrahimzadeh~Hassanabadi}, \bibinfo{author}{A.~Heidarpour}, \bibinfo{author}{S.~Eftekhar~Azam}, \bibinfo{author}{M.~Arashpour},
\newblock \bibinfo{title}{A bayesian smoothing for input-state estimation of structural systems},
\newblock \bibinfo{journal}{Computer-Aided Civil and Infrastructure Engineering} \bibinfo{volume}{37} (\bibinfo{year}{2022}) \bibinfo{pages}{317--334}. \DOIprefix\doi{10.1111/mice.12733}.
\bibitem[{Verhagen and Teunissen(2017)}]{Verhagen2017}
\bibinfo{author}{S.~Verhagen}, \bibinfo{author}{P.~J. Teunissen},
\newblock \bibinfo{title}{Least-squares estimation and kalman filtering},
\newblock in: \bibinfo{editor}{P.~J. Teunissen}, \bibinfo{editor}{O.~Montenbruck} (Eds.), \bibinfo{booktitle}{Springer Handbook of Global Navigation Satellite Systems}, \bibinfo{publisher}{Springer International Publishing}, \bibinfo{address}{Cham}, \bibinfo{year}{2017}, pp. \bibinfo{pages}{639--660}. \DOIprefix\doi{10.1007/978-3-319-42928-1_22}.
\bibitem[{Brogan(1991)}]{Brogan1991}
\bibinfo{author}{W.~L. Brogan}, \bibinfo{title}{Modern control theory}, \bibinfo{edition}{third edition} ed., \bibinfo{publisher}{Prentice Hall}, \bibinfo{year}{1991}.
\bibitem[{{De Callafon} et~al.(2008){De Callafon}, Moaveni, Conte, He, and Udd}]{Callafon2008}
\bibinfo{author}{R.~A. {De Callafon}}, \bibinfo{author}{B.~Moaveni}, \bibinfo{author}{J.~P. Conte}, \bibinfo{author}{X.~He}, \bibinfo{author}{E.~Udd},
\newblock \bibinfo{title}{General realization algorithm for modal identification of linear dynamic systems},
\newblock \bibinfo{journal}{Journal of Engineering Mechanics} \bibinfo{volume}{134} (\bibinfo{year}{2008}) \bibinfo{pages}{712--722}. \DOIprefix\doi{10.1061/(ASCE)0733-9399(2008)134:9(712)}.
\bibitem[{Yu and Zhang(2020)}]{Yu2020}
\bibinfo{author}{S.~Yu}, \bibinfo{author}{J.~Zhang},
\newblock \bibinfo{title}{Fast bridge deflection monitoring through an improved feature tracing algorithm},
\newblock \bibinfo{journal}{Computer-Aided Civil and Infrastructure Engineering} \bibinfo{volume}{35} (\bibinfo{year}{2020}) \bibinfo{pages}{292--302}. \DOIprefix\doi{10.1111/mice.12499}.
\bibitem[{Kourakis(2007)}]{Kourakis2007}
\bibinfo{author}{I.~Kourakis}, \bibinfo{title}{Structural systems and tuned mass dampers of super-tall buildings: case study of Taipei 101}, Master's thesis, Massachusetts Institute of Technology, Dept. of Civil and Environmental Engineering, \bibinfo{year}{2007}.

\end{thebibliography}




\end{document}